\documentclass[10pt,aps,prd,twocolumn,floatfix,shownopacs,preprintnumbers,nofootinbib,longbibliography]{revtex4-1}
\pdfoutput=1
\usepackage{amsmath,amsfonts,amsthm,bm,bbold,amssymb,mathtools,tabu}
\usepackage{graphicx}
\graphicspath{{attachments/}}
\usepackage{gensymb}
\usepackage[caption=false]{subfig}
\usepackage{epstopdf}
\usepackage{csquotes}
\usepackage[T1]{fontenc}
\usepackage{lmodern}
\usepackage{mathtools}
\usepackage[colorlinks=true,breaklinks=true]{hyperref}
\hypersetup{allcolors=[rgb]{0.0 0.0 0.6},linkcolor=[rgb]{0.8 0.0 0.4}}   
\usepackage{slashed}             
\usepackage{url}
\usepackage[detect-none]{siunitx}
\usepackage{color}  
\usepackage{comment}              
\usepackage[dvipsnames]{xcolor}

\DeclareTextCommand{\textprime}{\encodingdefault}{%
  \mbox{$\m@th'\kern-\scriptspace$}%
}
\DeclarePairedDelimiter\abs{\lvert}{\rvert}%
\DeclarePairedDelimiter\norm{\lVert}{\rVert}%
\makeatletter
\let\oldabs\abs
\def\abs{\@ifstar{\oldabs}{\oldabs*}}
\let\oldnorm\norm
\def\norm{\@ifstar{\oldnorm}{\oldnorm*}}
\makeatother

\allowdisplaybreaks

\bibpunct{[}{]}{,}{n}{}{,}

\providecommand{\abs}[1]{\lvert#1\rvert}
\providecommand{\norm}[1]{\lVert#1\rVert}

\definecolor{blue}{rgb}{0.0,0.0,1.0}

\usepackage{tikz,xcolor,hyperref}

\definecolor{lime}{HTML}{A6CE39}
\DeclareRobustCommand{\orcidicon}{\hspace{-1mm}
	\begin{tikzpicture}
	\draw[lime, fill=lime] (0,0) 
	circle [radius=0.16] 
	node[white] {{\fontfamily{qag}\selectfont \tiny \,ID}};
	\draw[white, fill=white] (-0.0525,0.095) 
	circle [radius=0.007];
	\end{tikzpicture}
	\hspace{-3mm}
}

\foreach \x in {A, ..., Z}{\expandafter\xdef\csname orcid\x\endcsname{\noexpand\href{https://orcid.org/\csname orcidauthor\x\endcsname}
			{\noexpand\orcidicon}}
}

\begin{document}

\title{Elaborating the Depolarization of Fast Collective Neutrino Flavor Oscillations}
\title{Elaborating the Ultimate Fate of Fast Collective Neutrino Flavor Oscillations}

\author{Soumya~Bhattacharyya\orcidA{}}
\email{soumyaquanta@gmail.com}
\affiliation{Tata Institute of Fundamental Research,
	Homi Bhabha Road, Mumbai, 400005, India}
\author{Basudeb~Dasgupta\orcidB{}}
\email{bdasgupta@theory.tifr.res.in}
\affiliation{Tata Institute of Fundamental Research,
             Homi Bhabha Road, Mumbai, 400005, India}

\date{\today}

\begin{abstract} 
Dense clouds of neutrinos and antineutrinos can exhibit fast collective flavor oscillations. Previously, in Phys. Rev. Lett. 126 (2021) 061302, we proposed that such flavor oscillations lead to \emph{depolarization}, i.e., an irreversible mixing of the flavors, whose extent depends on the initial momentum distributions of the different flavors. In this paper,  we elaborate and extend this proposal, and compare it with related results in the literature. We present an accurate analytical estimate for the lower resting point of the fast flavor pendulum and underline the relaxation mechanisms, i.e., transverse relaxation, multipole cascade, and mixing of flavor-waves, that cause it to settle down. We estimate the extent of depolarization, its dependence on momentum and net lepton asymmetry, and its generalization to three flavors. Finally, we prescribe approximate analytical recipes for the depolarized distributions and fluxes that can be used in supernova/nucleosynthesis simulations and supernova neutrino phenomenology.
\end{abstract}
\preprint{TIFR/TH/22-18}

\maketitle

\tableofcontents 
 
\section{Introduction}
\label{sec:intro}
Neutrinos change their flavor with a \emph{time-periodic} probability, e.g., $\sin^2 2\theta\sin^2 \left[{\omega t}/{2}\right]$ in vacuum, due to quantum interference of two eigenstates evolving with a frequency difference $\omega=\Delta m^2/(2E)$~\cite{Pontecorvo:1957qd,Maki:1962mu,Pontecorvo:1967fh}. In ordinary matter, forward-scattering off the background particles modifies the mixing angle $\theta$ and the oscillation rate~$\omega$~\cite{Wolfenstein:1977ue, Mikheev:1987jp}. However, often one is interested in the flavor composition after a sufficiently long time, when the flavor conversion probability is found to become \emph{time-independent}, e.g., $\frac{1}{2}\sin^2 2\theta$ for averaged oscillations in vaccum~\cite{Bethe:1986ej,Parke:1986jy,Bilenky:1987ty}. Generally this is because of \emph{decoherence}, which can occur in two ways: Either the flavor evolution of an individual neutrino becomes stochastic, e.g., due to collisions~\cite{Harris:1978rv, Harris:1980zi}; Or due to observational limitations  -- such as spatial, temporal, or energy resolution -- which result in a pooling together of many neutrinos, each with a slightly different relative phase between its two components~\cite{Stodolsky:1998tc}. See Ref.~\cite{Akhmedov:2009rb} for a clear exposition. It should be noted that the nature and extent of the late-time neutrino mixing, even after the oscillations have ceased,  can have nontrivial dependence on energy/momentum and can encode information about the system~\cite{Bethe:1986ej,Parke:1986jy,Bilenky:1987ty}.

Flavor oscillations of neutrinos from dense astrophysical sources, e.g., deep in a supernova, merging neutron stars, or the early Universe, exhibit an additional novelty. These neutrinos can frequently forward-scatter off other oscillating neutrinos, leading to novel \emph{collective} flavor oscillations~\cite{Capozzi:2022slf}. The effect depends on the neutrino-neutrino forward-scattering rate $\mu=\sqrt{2}G_F n_\nu$~\cite{Pantaleone:1992xh, Pantaleone:1992eq}, which typically exceeds the average oscillation rate $\langle \omega \rangle$ in these environments. Under its influence, neutrinos can collectively oscillate at a synchronized rate~$\langle \omega \rangle$~\cite{Kostelecky:1994dt}, or the  bipolar/slow rate $\sqrt{\mu\langle \omega \rangle}$~\cite{Kostelecky:1993dm}, or the fast rate~$\mu$~\cite{Sawyer:2005jk}. Remarkably, collective oscillations are predicted to occur with large amplitudes even for the matter-suppressed mixing angles expected in dense regions~\cite{Pastor:2001iu, Duan:2005cp, Duan:2006an,Hannestad:2006nj}.

Collective oscillations display a rich phenomenology, but most remarkably they can lead to novel signatures of flavor mixing at late times. 
For slow collective effects a prominent signature is a set of energy-dependent swaps between the flavor spectra~\cite{Raffelt:2007cb, Fogli:2007bk, Dasgupta:2009mg}, with subleading decoherence effects~\cite{Raffelt:2007yz,Esteban-Pretel:2007jwl}. For fast oscillations, the signature is less clear, but it is plausible that the decoherence effects are more important. There are two noteworthy issues: Firstly, one cannot therefore derive the late-time decoherent limit by straight-forwardly averaging over a known coherent oscillation probability. This is because all neutrinos evolve interdependently, with unusually weak dependence on both $\theta$ and $\omega$, and an analytical solution is not available in general;  see however Ref.~\cite{Dasgupta:2017oko}; Secondly, fast oscillations can occur very quickly. While slow instabilities develop over a few $100$\,km, or more, fast oscillations and their associated decoherence effects can occur over much smaller distances $\sim10^{-4}\,{\rm m}$. Thus, their impact can be important already inside the star. E.g., stellar heating and nucleosynthesis could be affected~\cite{Dasgupta:2011jf,Duan:2010af}. See Refs.\,\cite{Wu:2017drk,Capozzi:2018rzl,Zaizen:2019ufj,Morinaga:2019wsv,Stapleford:2019yqg,Xiong:2020ntn,Li:2021vqj} for studies in this direction. As a result, it is both challenging and important to understand the late-time behavior of fast oscillations.

Starting with the first explorations of fast flavor oscillations in the nonlinear regime~\cite{Sawyer:2008zs}, it was seen that the survival probability eventually stops oscillating, and instead approaches a quasi-steady state~\cite{Dasgupta:2016dbv}. The phase space distributions of the different flavors get irreversibly mixed~\cite{Bhattacharyya:2020dhu,Bhattacharyya:2020jpj}. We note that this is because of dephasing, and not because of collisions which help kick-start but do not overwhelm fast oscillations~\cite{Capozzi:2018clo}. To emphasize this distinction, we denote this as \emph{depolarization}~\cite{Bhattacharyya:2020dhu,Bhattacharyya:2020jpj}.  The moniker is borrowed from optics, where it is used to refer to the shrinking of the polarization sphere (i.e., the Stokes parameters Q, U, V get smaller) without dissipation (i.e., loss in intensity I).  We use it also to draw attention to the novel associated flavor conversion --  full/partial equilibration of the flavor spectra, depending on velocity, while conserving the lepton asymmetry.  

In two previous papers, hereafter  B20a~\cite{Bhattacharyya:2020dhu}, and B20b~\cite{Bhattacharyya:2020jpj}, we have explored this phenomenon in detail. The purpose of this work is to elaborate these results, and to compare them with several closely related works. First, we compare our predictions for the behavior of the so-called \emph{fast flavor pendulum}, with those by Johns et al.~\cite{Johns:2019izj}, hereafter J20, and the recent work by Padilla-Gay et al.~\cite{Padilla-Gay:2021haz}, hereafter PG21. Then we compare our \emph{depolarization} proposal with work by Wu et al., hereafter Wu21~\cite{Wu:2021uvt}, and by Richers et al., hereafter R21a~\cite{Richers:2021nbx} and R21b~\cite{Richers:2021xtf}, respectively, which contain closely related results.  We also compare and contrast our results with those by Martin et al., hereafter M20~\cite{Martin:2019gxb} and M21~\cite{Martin:2021xyl}, where they do not find a depolarized steady state. Although our study is not intended to supplant a systematic code-comparison, the comparisons provided here should clarify a number of conceptual issues. See Refs.\,\cite{Abbar:2018beu,Johns:2020qsk,Sigl:2021tmj,Xiong:2021dex,Shalgar:2020wcx,Sasaki:2021zld,Abbar:2021lmm} for related studies of fast oscillations in the nonlinear regime. A separate body of work has focussed on the initial growth of fast instabilities; see Refs.\,\cite{Izaguirre:2016gsx, Capozzi:2017gqd, Abbar:2017pkh, Morinaga:2018aug,Airen:2018nvp, Yi:2019hrp, Doring:2019axc, Capozzi:2019lso,Capozzi:2020kge, Bhattacharyya:2021klg,Morinaga:2021vmc, Dasgupta:2021gfs}.

This paper is structured as follows: We outline our set-up in Sec.\,\ref{sec:not}. Sec.\,\ref{sec:flavor} presents an accurate estimate for the lower resting point of the fast flavor pendulum, and analyses of transverse relaxation, cascading of multipoles, and mixing of flavor waves. Sec.\,\ref{sec:range} gives an estimate of the extent of depolarization and its generalization to three flavors. Sec.\,\ref{Pres} contains recipes for the depolarized distributions and fluxes in a form that is usable for supernova simulations or neutrino phenomenology. Finally, in Sec.\,\ref{Con}, we conclude with a summary and outlook.


\section{Framework and Methods}
\label{sec:not}
We use natural units throughout, with $\hbar=c=1$. In each phase space cell $d^3{\bf p}\,d^3{\bf x}$, one has~\cite{Sigl:1993ctk}
\begin{equation}
i(\partial_t+{\bf v}\cdot{\bf \partial}_{\bf x})\rho_{\bf p}=[{\cal H}_{\bf p},\rho_{\bf p}]\,,
\end{equation}
 where $\rho_{\bf p}$ is the matrix of densities and ${\cal H}_{\bf p}$ is the flavor Hamiltonian matrix. The phase space cells are taken to be sufficiently large, so that position and momentum can be simultaneously specified~\cite{Stirner:2018ojk}. We ignore momentum changing collisions, external forces, and neutrino mass-mixing~\cite{Chakraborty:2016lct}, which are typically negligible on time-scales of the fastest neutrino oscillations. The velocities ${\bf v}={\bf p}/|{\bf p}|$ and energies $E=|{\bf p}|$ do not change and serve as immutable labels. The range of $E$ is from $-\infty$ to $+\infty$, to include antineutrinos of energy $E$ by writing them as if they were neutrinos of energy $-E$. Axisymmetry restricts that the flavor evolution depends on a single spatial coordinate $z$, a single momentum coordinate $v$, and of course on time. This is a simple model for neutrino flavor evolution in a supernova, after it starts free streaming. 

\begin{figure*}[t!]
	\hspace{-0.25cm}\includegraphics[height=0.64\columnwidth]{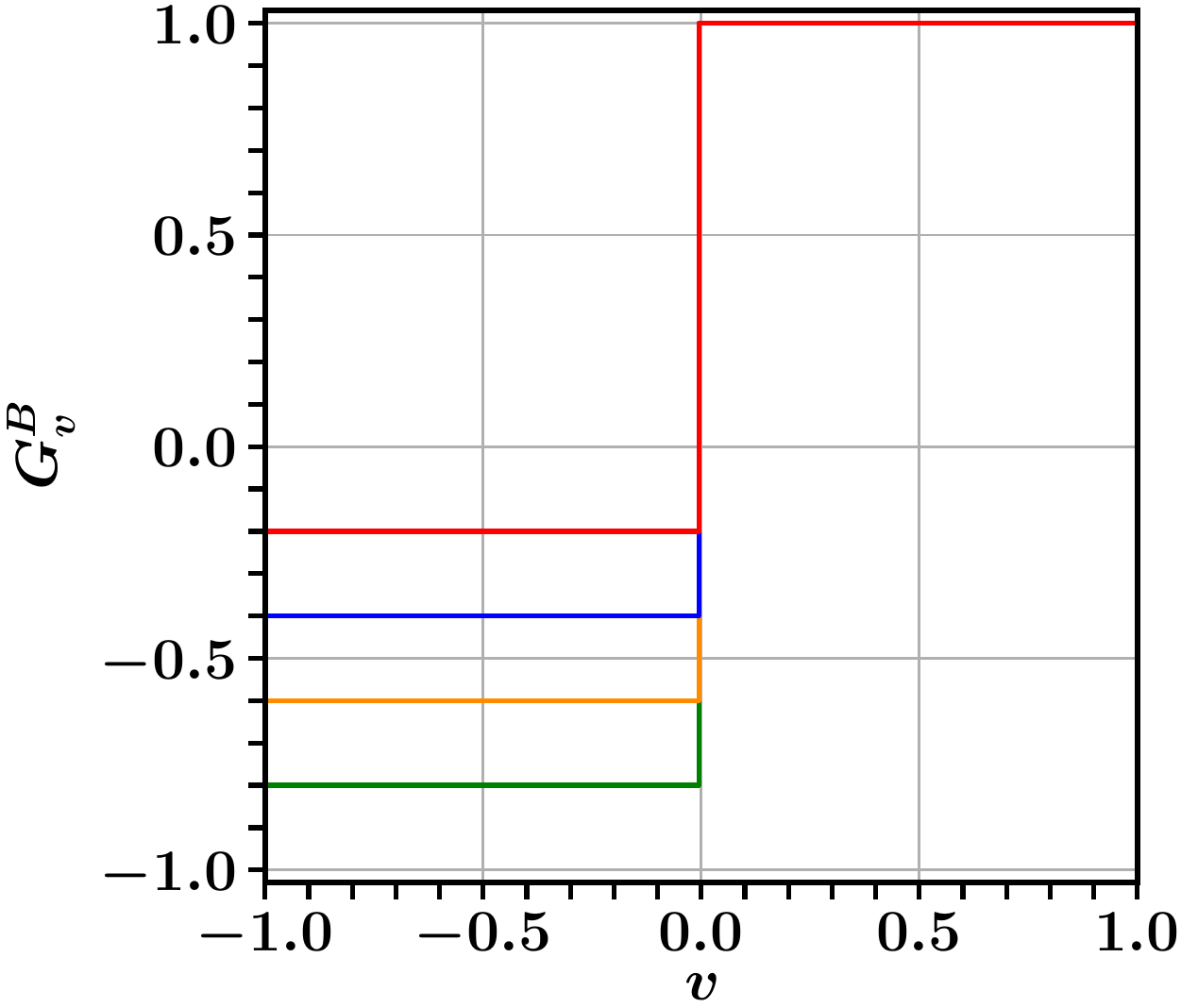}~\quad\qquad\,\,
	\includegraphics[height=0.65\columnwidth]{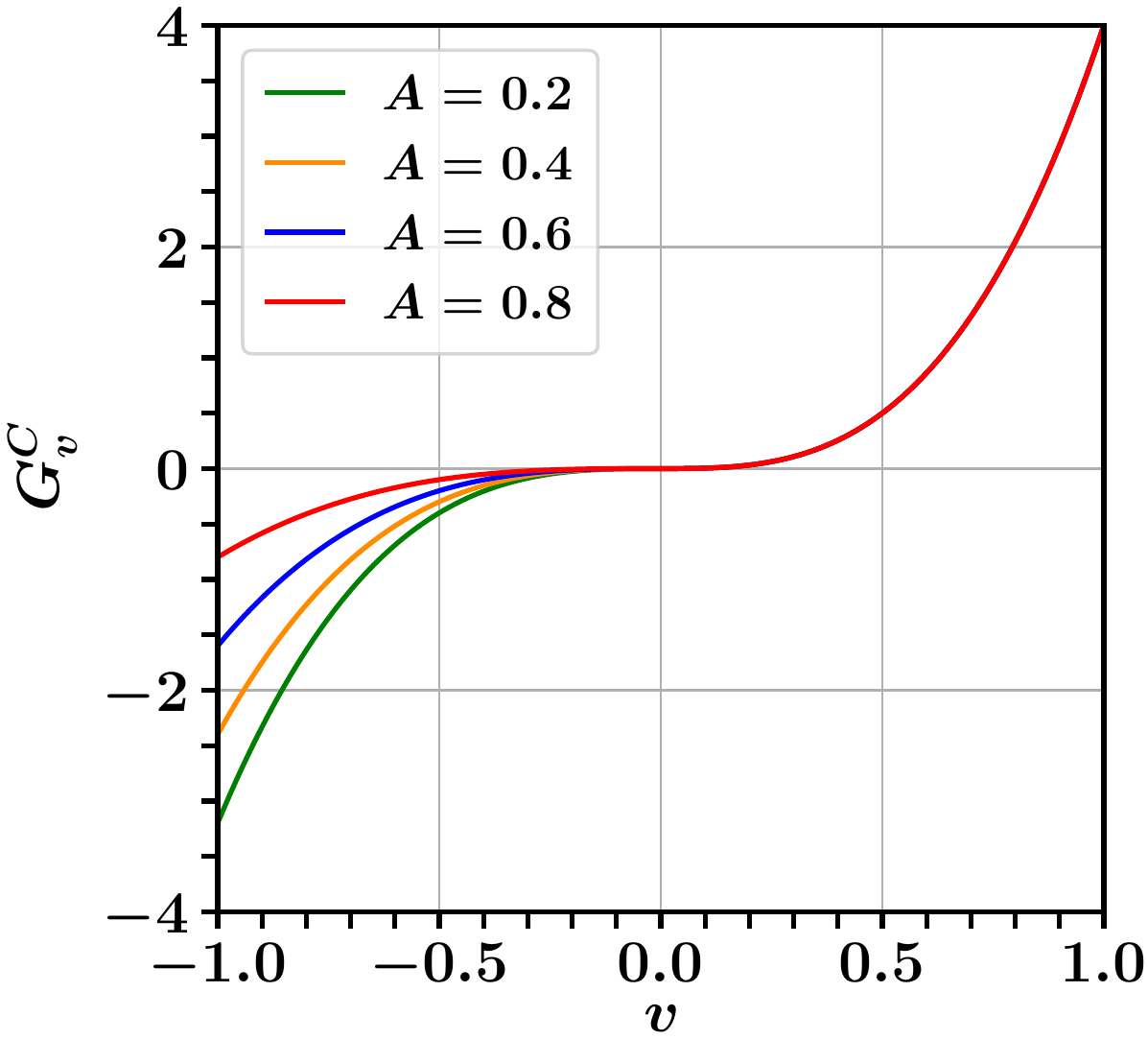}\\[3ex]
	\includegraphics[height=0.65\columnwidth]{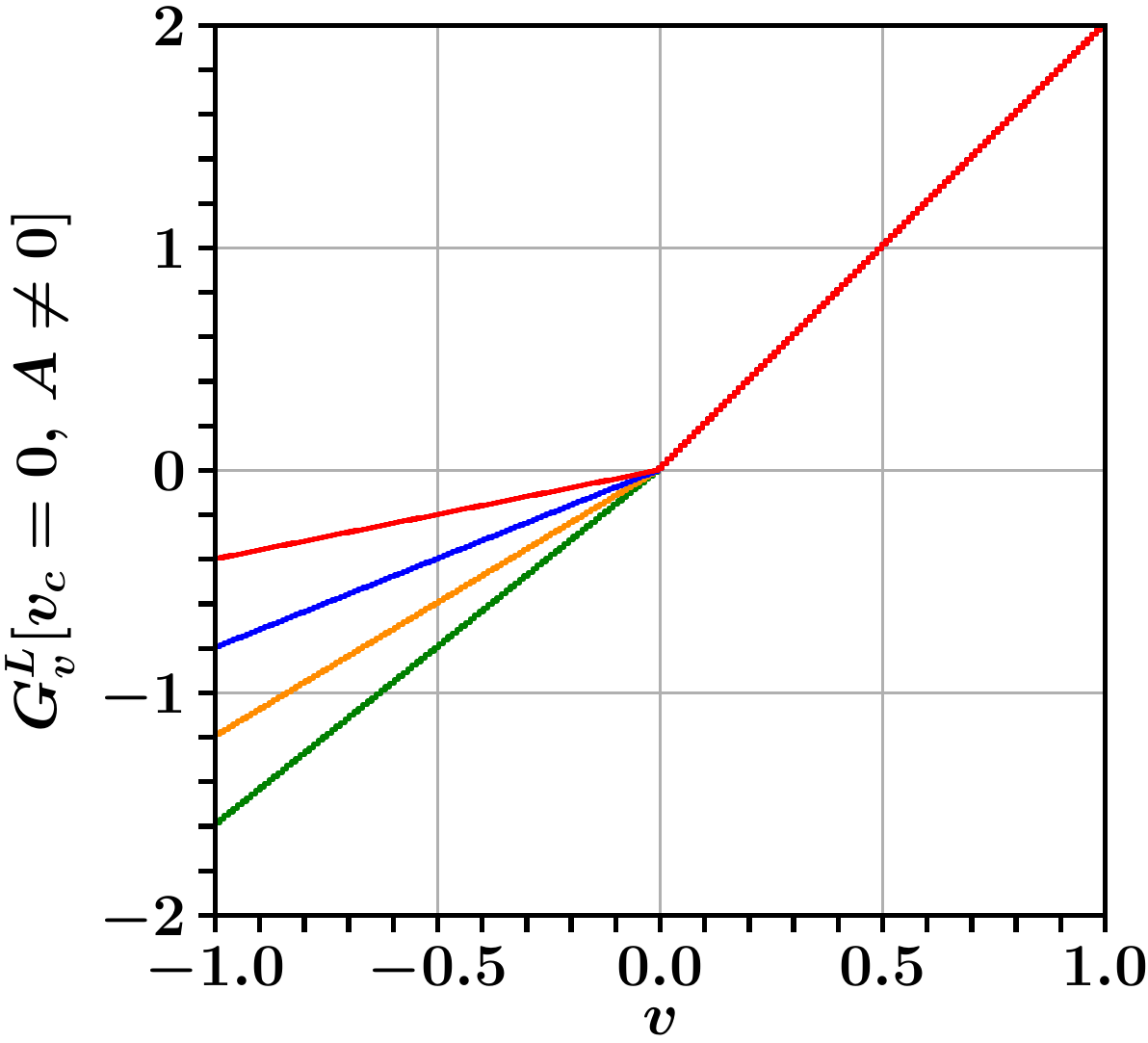}~~\qquad
	\includegraphics[height=0.65\columnwidth]{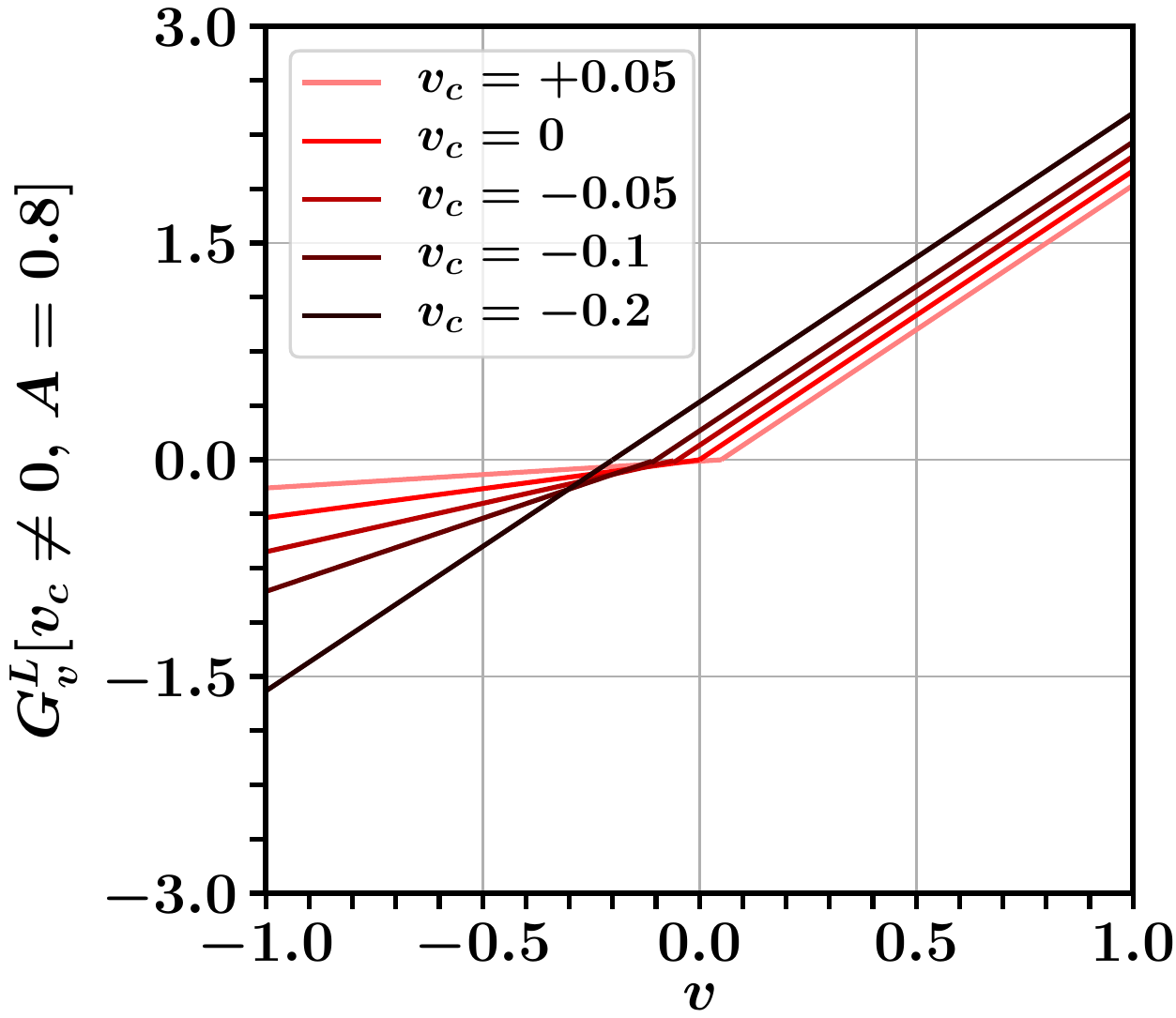}
	\caption{Benchmark ELN models: The four panels show four classes of ELN distributions, $G_v^{B}$, $G_v^{C}$, $G_v^L[v_c =0]$, and $G_v^L[v_c \neq 0]$, as a function of $v$ for different values of the lepton asymmetry $A$ and crossing velocity $v_c$.}
	\label{figELNs}
\end{figure*}
 
Under these assumptions, the flavor content encoded in each $\rho_{v}$ evolves as~\cite{Bhattacharyya:2020dhu,Bhattacharyya:2020jpj}
\begin{align}\label{eom2}
\big(\partial_{t}+v\partial_{z}\big)\mathsf{S}_{v} = \mu_{0}\int_{-1}^{+1} dv{'}G_{v'}\left(1-vv{'}\right) \mathsf{S}_{{v}{'}} \times \mathsf{S}_{v}\,.
\end{align}
\newline
Here $\mathsf{S}_{v}$ is the Bloch vector encoding the flavor state for neutrino modes with velocity $v$, with $|v|<1$. We denote flavor space vectors by sans-serif letters, e.g., $\mathsf S$, and the components parallel {to the ${\hat{\mathsf{e}}}_{3}$ direction by 
$\left(\ldots\right)^{\parallel}$.  The transverse vector confined to the ${\hat{\mathsf{e}}}_{1}-{\hat{\mathsf{e}}}_{2}$ plane for any flavor space vector, for e.g., $\mathsf{S}$, is defined through the following vector formula  
\begin{align}\label{vecform}
\mathsf{S}^{\perp} = \mathsf{S}-\mathsf{S}^{\parallel} {\hat{\mathsf{e}}}_{3}\,.\end{align}
Magnitudes are shown in the usual font, e.g., $S=|\sf S|$.}

The ELN distribution function $G_v$ is the excess of the phase space distribution of $\nu_e$ over $\nu_\mu$ (and $\bar{\nu}_\mu$ over $\bar\nu_e$), integrated over $E^2dE$ and divided by a typical density, say $n_\nu$.  Only the product of $\mu_{0}$ and $G_v$ appears; though, one defines a rate $\mu_{0}\propto G_{\rm F}n_\nu$ as the collective potential. Hereafter, we set $\mu_0=1$, and express $z$ and $t$ in units of $\mu_0^{-1}$. The ELN becomes dimensionless in these units.
\begin{figure*}
	\includegraphics[height=0.29\textwidth]{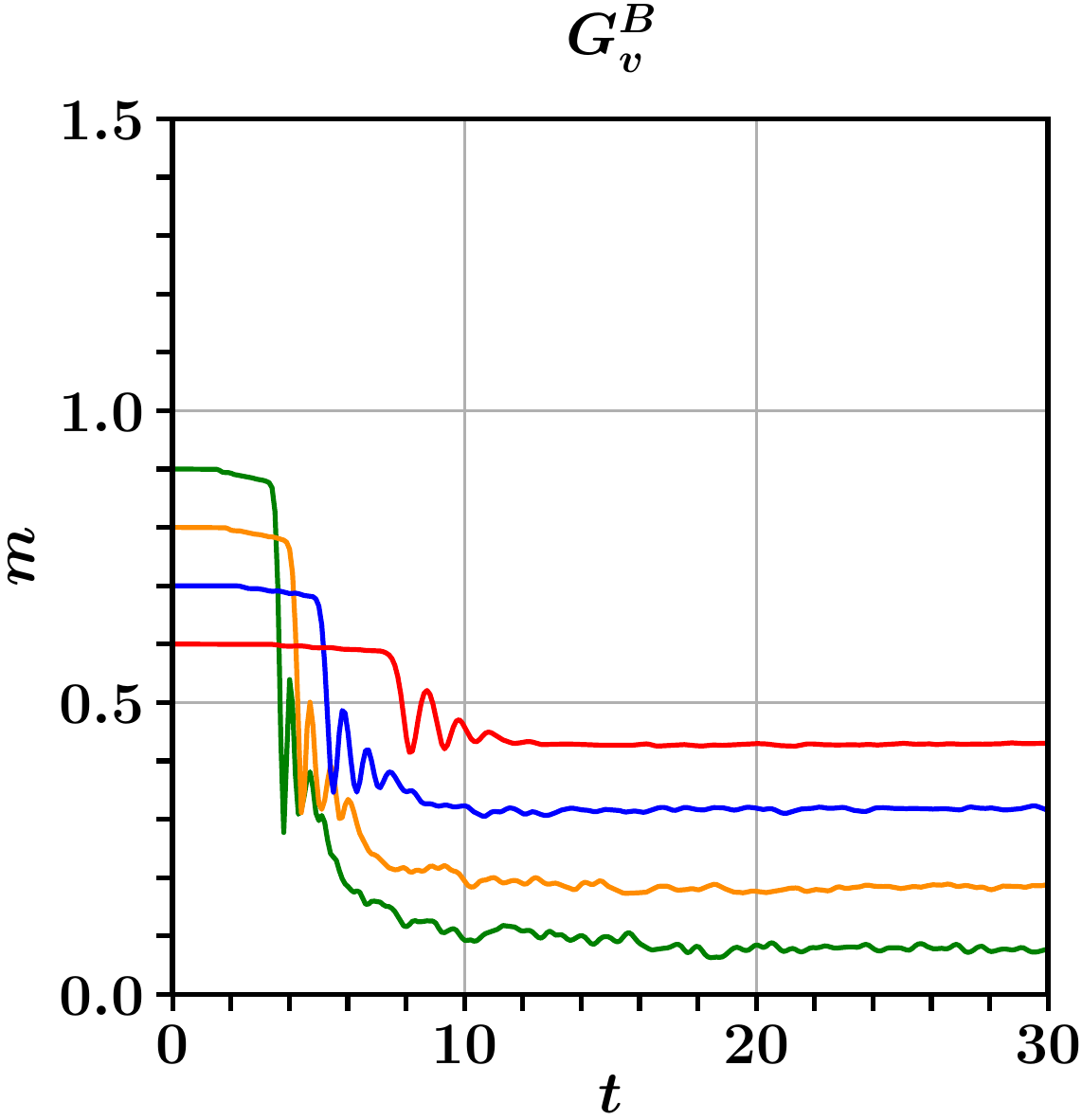}\qquad
	\includegraphics[height=0.29\textwidth]{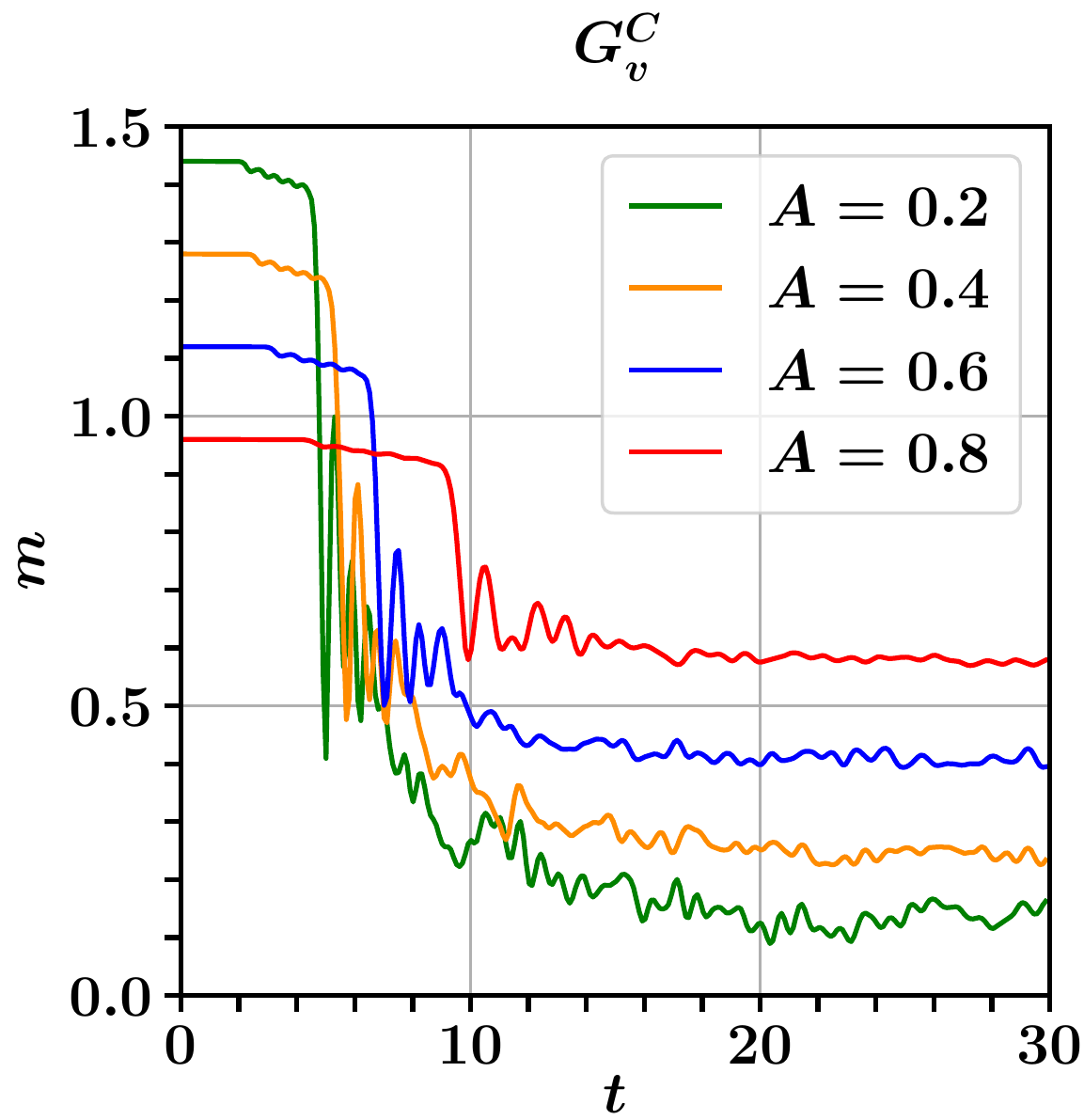}\\[3ex]
	\includegraphics[height=0.29\textwidth]{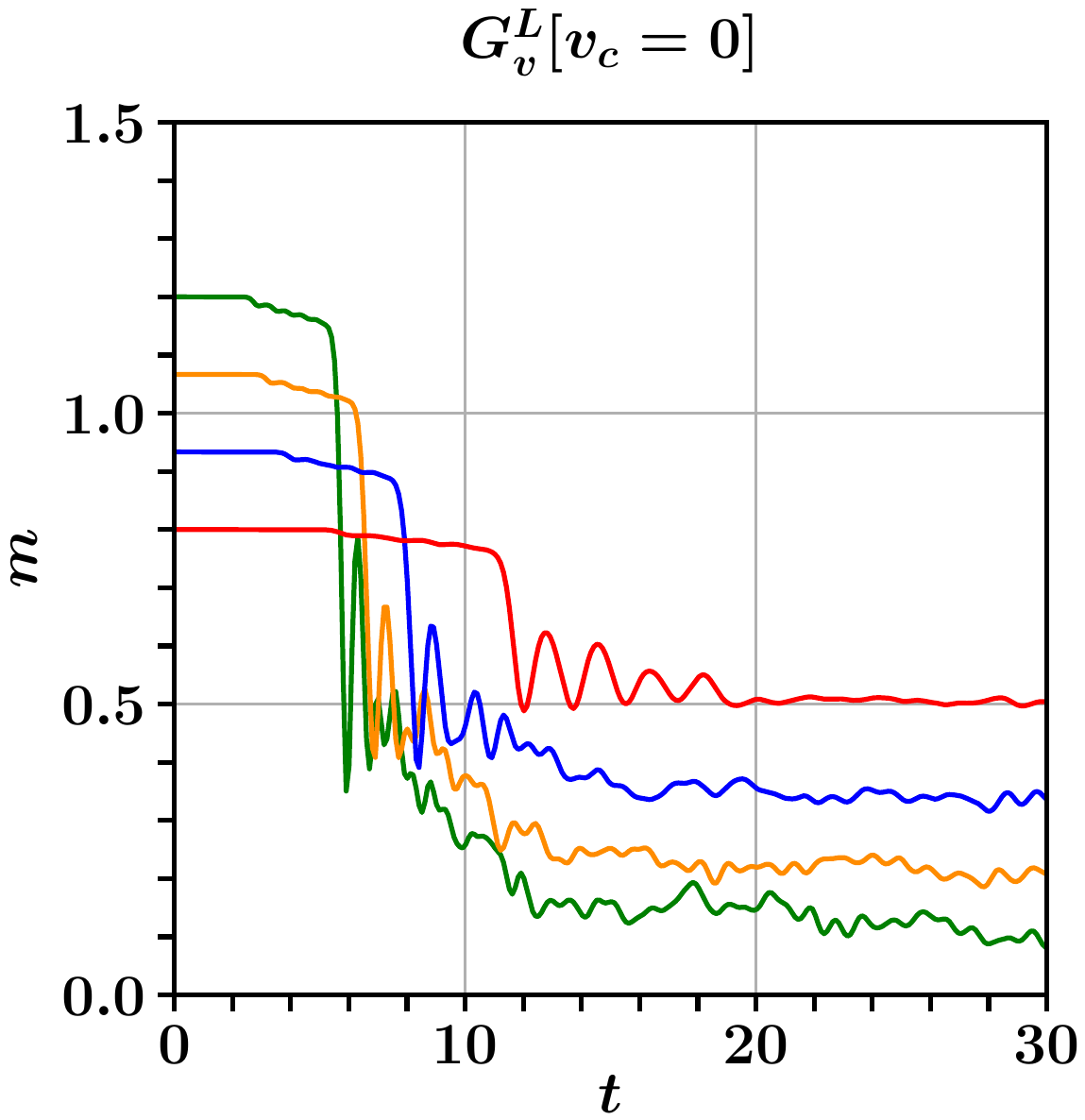}\qquad
	\includegraphics[height=0.29\textwidth]{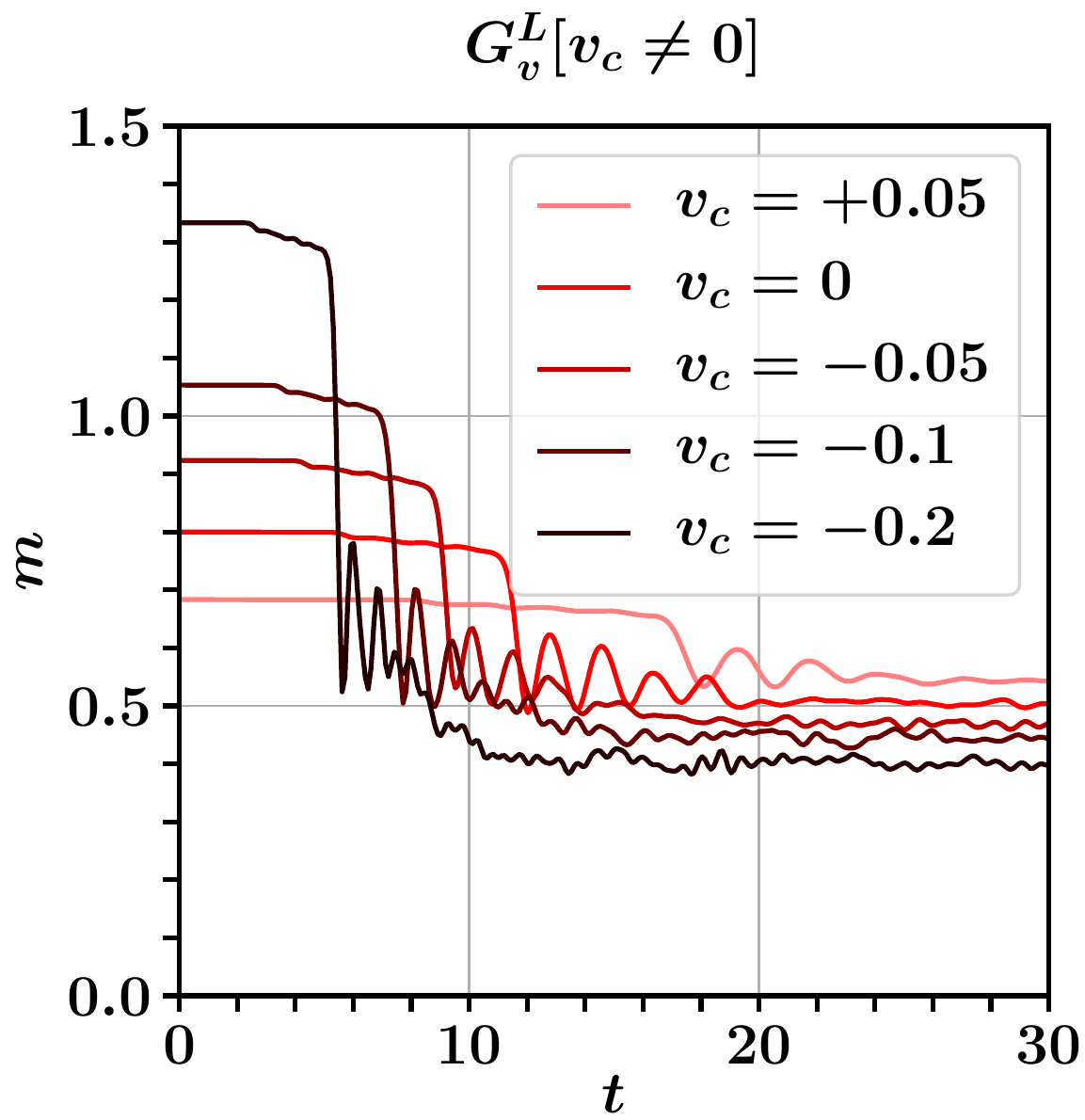}
	\caption{Settling-down of the $\mathsf{M}_1$ pendulum: {We show $m=\tfrac{1}{L}\int_0^L dz\int_{-1}^1 dv \,v\, G_v\, \mathsf{S}_v^{\parallel}[z]$, i.e. the spatially averaged counterpart of the $\mathsf{\hat{e}}_3$ component of $\mathsf{M}_1[z]$}, computed by appropriately averaging the solution to Eq.\eqref{eom2}. It initially oscillates but eventually settles down to a steady value at late times. The four panels are for the ELNs $G_v^B$ , $G_v^C$, and $G_v^L[v_c = 0]$ for various lepton asymmetry $A$, and for $G_v^L[v_c \neq 0]$ with different crossing velocity $v_c$ and $A=0.8$.}
	\label{fig1lep}
\end{figure*}
For this paper, we will mainly consider three families of ELN distributions, shown in Fig.\ref{figELNs}:
\begin{subequations}
\begin{alignat}{1}
 G_v^B &= \begin{cases}
1 &\hphantom{phantomtext12} \text{if $0<v<+1$}\,\\
A-1 &\hphantom{phantomtext12} \text{if $-1<v<0$}\,,
\end{cases}\label{ELNbox}\\[0.4em]	
 G_v^{L}&= \begin{cases}
2 \left(v-v_c\right) &\hphantom{=} \text{if $v_c<v<+1$}\,\\[0.2em]
2 \frac{\left(1-v_c \right)^2-A}{\left(1+v_c\right)^2}\left(v-v_c \right) &\hphantom{=} \text{if $-1<v<v_c$}\,,
\end{cases}\label{ELNvelshift}\\[0.4em]
G_v^{C} &= \begin{cases}
4 \,v^3 \quad &\phantom{phantom.} \text{if $0<v<+1$}\,\\
4 \, (1-A) \, v^3 &\hphantom{phantom.} \text{if $-1<v<0$}\,.
\end{cases}\label{ELNcubic}
\end{alignat}
\end{subequations}
$G_v^B$ is a `\underline{B}ox' spectrum, piecewise constant in $v$ on either side of the crossing at $v=0$. $G_v^L$ is piecewise `\underline{L}inear', with an adjustable crossing at $v_c$. Similarly, $G_v^C$  is `\underline{C}ubic'. 
In addition, we will also study the ELNs $G^{3a}_v$ and $G^{4b}_v$, as defined in Refs.~\cite{Martin:2019gxb, Martin:2021xyl}, as well as the ELN in Ref.~\cite{Wu:2021uvt}, to compare our results. For all the ELNs, the lepton asymmetry is denoted by $A=\int dv\, G_v$. All these ELNs are inspired by SN ELNs with a single crossing, where $\nu_e$ dominates over $\bar\nu_e$ in the forward direction $v>0$ and vice versa, and we restrict our study to $A>0$. We study the dependence of the flavor state on lepton asymmetry $A$ and on the crossing velocity $v_c$.

In all our numerical computations, we solve Eq.\eqref{eom2} with initial conditions that all neutrinos -- with any velocity $v$ and at all locations in the periodic one dimensional box of length $L$ -- are emitted in the electron flavor, i.e., ${\sf S}_v[t=0,z]=+\hat{\sf e}_3$. The numerical set up, i.e., discretizations, dimensionality, tolerances, etc., have been kept exactly the same as in B20b\,\cite{Bhattacharyya:2020jpj}.

To start the flavor evolution, we supply tiny initial perturbations to the transverse components of the Bloch vectors. These are referred to as \emph{seeds}, and are a numerically efficient means of initiating the flavor evolution. In reality, neutrino mass terms would provide the initial misalignment from a pure flavor state, but as we have set them to zero in the  fast oscillation limit we resort to this numerical alternative. See the Supplemental Material of B20b\,\cite{Bhattacharyya:2020jpj} for details, including a discussion of the dependence on seeds. Unless stated otherwise, we will assume a spatially extended seed with transverse perturbations of amplitude $10^{-6}$ and random relative phases.

\section{Flavor Pendulum and Relaxation}
\label{sec:flavor}

Defining the vector $\mathsf{M}_{n}[z, t] = \int_{-1}^{+1}dv\,G_{v}L_{n} \mathsf{S}_{v}[z, t]$ as the ${n}^\textrm{th}$ moment of the Bloch vector $\mathsf{S}_{{v}}$, with $L_{n}[v]$ being the $n^\textrm{th}$ Legendre polynomial in $v$, we can rewrite Eq.\eqref{eom2} in multipole space as \cite{Raffelt:2007yz, EstebanPretel:2009zz},
\begin{align}\label{eom3}
\partial_{t}\mathsf{M}_{n}-\mathsf{M}_{0} \times \mathsf{M}_{n}= \partial_{z} \mathsf{T}_{n}-\mathsf{M}_{1} \times \mathsf{T}_{n}\,,
\end{align}
where 
\begin{align}\label{eom4}
\mathsf{T}_{n} = \frac{n}{2n+1} \mathsf{M}_{n-1}+ \frac{n+1}{2n+1} \mathsf{M}_{n+1}\,.
\end{align}
Using periodic boundary conditions, and the approximation that spatial averaging factorizes over the dot and cross products of vectors, one can write the spatially averaged {or coarse-grained} version of Eq.\eqref{eom3} as
\begin{align}\label{eom5}
\partial_{t} \langle \mathsf{M}_{n} \rangle - \langle \mathsf{M}_{0} \rangle \times \langle \mathsf{M}_{n} \rangle = - \langle \mathsf{M}_{1} \rangle \times \langle \mathsf{T}_{n} \rangle \,.
\end{align}
{\emph{For brevity, hereon we will mostly omit writing $\langle ... \rangle$ for the spatially averaged quantities. Instead, when we occasionally need to refer to quantities which are not spatially averaged, we will explicitly show the $z$-dependence, e.g., $\mathsf{S}_v[z]$, as opposed to the averaged version $\mathsf{S}_v$}}. Hopefully, the distinction will also be clear from the context.

Eq.\eqref{eom5} represents an infinite tower of equations. We will truncate this tower beyond $n = 3$, effectively assuming that the $n \geq 4$ multipoles are negligible. This gives a set of four coupled ODEs:
\begin{subequations}
	\begin{align}
	\partial_{t}{\mathsf{M}}_0 &= 0  \,, \label{10a}\\
	\partial_{t}{\mathsf{M}}_1 &= \mathsf{D} \times  \mathsf{M}_1\,, \label{10b}\\
	\partial_{t}{\mathsf{D}} &= \mathsf{B} \times  \mathsf{M}_1 \,, \label{10c}\\
	\partial_{t}{\mathsf{B}} &= \mathsf{K} \times  \mathsf{M}_1\,, \label{10d}
	\end{align}
\end{subequations}
where 
\begin{subequations}
	\begin{align}
	{\mathsf{D}} &= \frac{\mathsf{M}_0}{3}+\frac{2 \mathsf{M}_2}{3} \,, \label{11a}\\
	{\mathsf{B}} &= \frac{2\mathsf{M}_3}{5}-\frac{9 \mathsf{M}_1}{35} \,, \label{11b}\\
	{\mathsf{K}} &= -\frac{3\mathsf{M}_0}{35}\,. \label{11c}
	\end{align}
\end{subequations}
Note that Eq.\eqref{10a}-\eqref{10d} are written in a frame rotating around $\mathsf{M}_0$ with a frequency $\sqrt{\mathsf{M}_0 \cdot \mathsf{M}_0}$, so that the common rotation of all $\mathsf{M}_n$ around the axis $\mathsf{M}_0$, encapsulated in the second term on the left side of Eq.\eqref{eom5}, is undone.

Eq.\eqref{10a} is the usual lepton number conservation which gives $\mathsf{M}_0=\rm{constant}$, whose 3rd component is $\int_{-1}^{1} G_v dv = A$, which is the lepton asymmetry. Eq.\eqref{10b} and Eq.\eqref{10c} are similar to Eq.(7) in Ref.~\cite{Hannestad:2006nj}, and can be combined to get
\begin{align}\label{eom6}
\mathsf{M}_1 \times \partial_{t}^2{\mathsf{M}}_1 + \left(\mathsf{D}. \mathsf{M}_1 \right)\partial_{t}{\mathsf{M}}_1  = {M}_1^2 \mathsf{B} \times \mathsf{M}_1\,,
\end{align}
which is the familiar pendulum equation for $\mathsf{M}_1$ with length $\sqrt{\mathsf{M}_1 \cdot \mathsf{M}_1}$. However, the vector $\mathsf{B}$ that acts akin to gravity is not a constant, and instead obeys Eq.\eqref{10d}. We remind that $\mathsf{M}_1$ is spatially averaged.

Fig.\ref{fig1lep} shows the numerical solution of Eq.\eqref{eom2} for the parallel component of $\mathsf{M}_1$. The first thing to note is that it does not continue to oscillate forever. Rather, it comes to rest after a few cycles of oscillations. The late-time resting point depends on the lepton asymmetry ($A$), zero crossing position ($v_c$), and the nature of ELN. Note also that relaxation leads to a lower final resting point than the lower turning point of the first few oscillations, especially for the smaller values of $A$.

\subsection{Resting point for ${\sf M}_1$}
 
We now compute the resting point of the ${\sf M}_1$ pendulum starting from the equations of motion. We will not assume that the moment vectors have constant lengths, and instead assume that certain phases randomize. In doing so, our approach departs from J20~\cite{Johns:2019izj}, or the more recent PG21\,\cite{Padilla-Gay:2021haz},  where spatial dependence and relaxation are absent. Thus, rather than deriving the lower turning point of the periodic $\mathsf{M}_1$ pendulum, we focus on deriving the lower \emph{resting point} of the \emph{relaxed} $\mathsf{M}_1$ pendulum.

According to Eqs.\eqref{10a}-\eqref{10d} the energy and spin of $\mathsf{M}_1$ pendulum are
\begin{subequations}
	\begin{align}
	E = \frac{\mathsf{D} \cdot \mathsf{D}}{2}+\mathsf{M}_1 \cdot \mathsf{B} = \rm{const.}\,, \label{15a}\\
	\sigma = \mathsf{M}_1 \cdot \mathsf{D} = \rm{const.}  \,, \label{15b} 
	\end{align}
\end{subequations}
which are conserved quantities in time $t$. The motion of $\mathsf{B}$ in Eq.\eqref{10d} allows us to write two more conserved quantities in time, i.e.,
\begin{subequations} 
	\begin{align}
	\frac{\mathsf{B} \cdot \mathsf{B}}{2} +\mathsf{K} \cdot \mathsf{D} = \rm{const.}\,, \label{18a} \\
	\mathsf{B} \cdot \mathsf{K} = \rm{const.} \,. \label{18b} 
	\end{align}
\end{subequations}

We give a name to the parallel component of the pendulum vector $\mathsf{M}_1$:
	\begin{align}
	{\mathsf{M}^{\parallel}_1 = \frac{1}{L}\int_0^L dz\int_{-1}^1 dv \,v\, G_v\, \mathsf{S}_v^{\parallel}[z]\equiv m \,.} \label{19a}
\end{align}
We also use some temporary shorthand notation to eliminate excess clutter in the derivation to follow, until Eq.\eqref{26d}:
\begin{subequations}
	\begin{align}
	\mathsf{D}^{\parallel} \equiv {u} \,, \label{19c}\\
	\mathsf{B}^{\parallel} \equiv b \,, \label{19d}\\
	\mathsf{K}^{\parallel} \equiv k \,. \label{19e}
	\end{align}
\end{subequations}
The quantities in Eqs.\eqref{19a}-\eqref{19e} will be denoted with subscripts, $i$ at the initial time $t = 0$, and $f$ at the late time when the system becomes steady. We do not use subscripts for quantities that are constant in time, e.g., for $b$ and $k$.

Our aim is to derive $m_f$, i.e., the $\hat{\sf e}_3$ component of the steady-state relaxed ${\sf M}_1$ pendulum. The key idea is to use the steady state condition and to eliminate any unknown late-time perpendicular components in terms of conserved quantities.

\begin{figure*}[t!]
	\includegraphics[width=0.57\columnwidth]{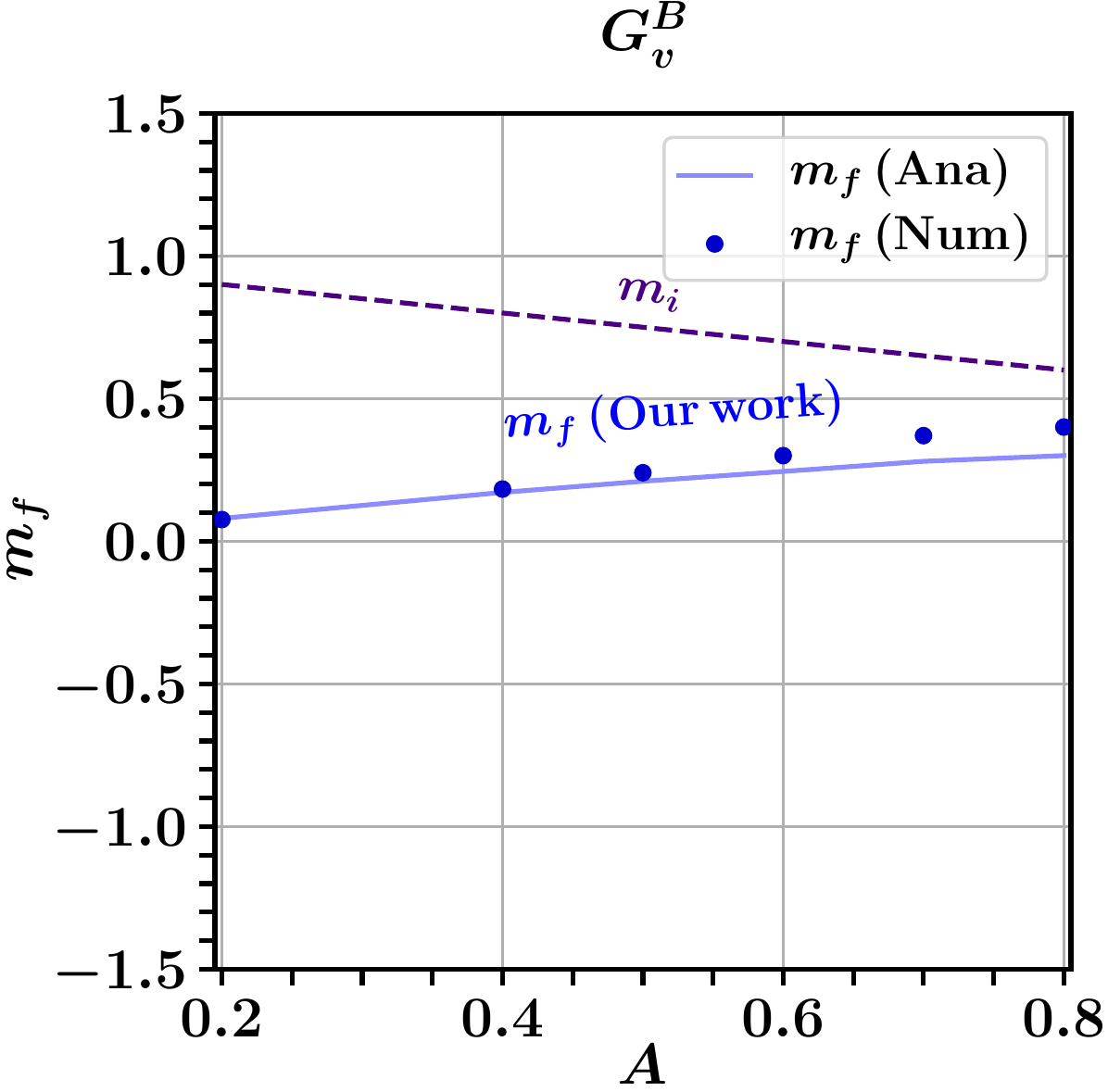}~\qquad
	\includegraphics[width=0.57\columnwidth]{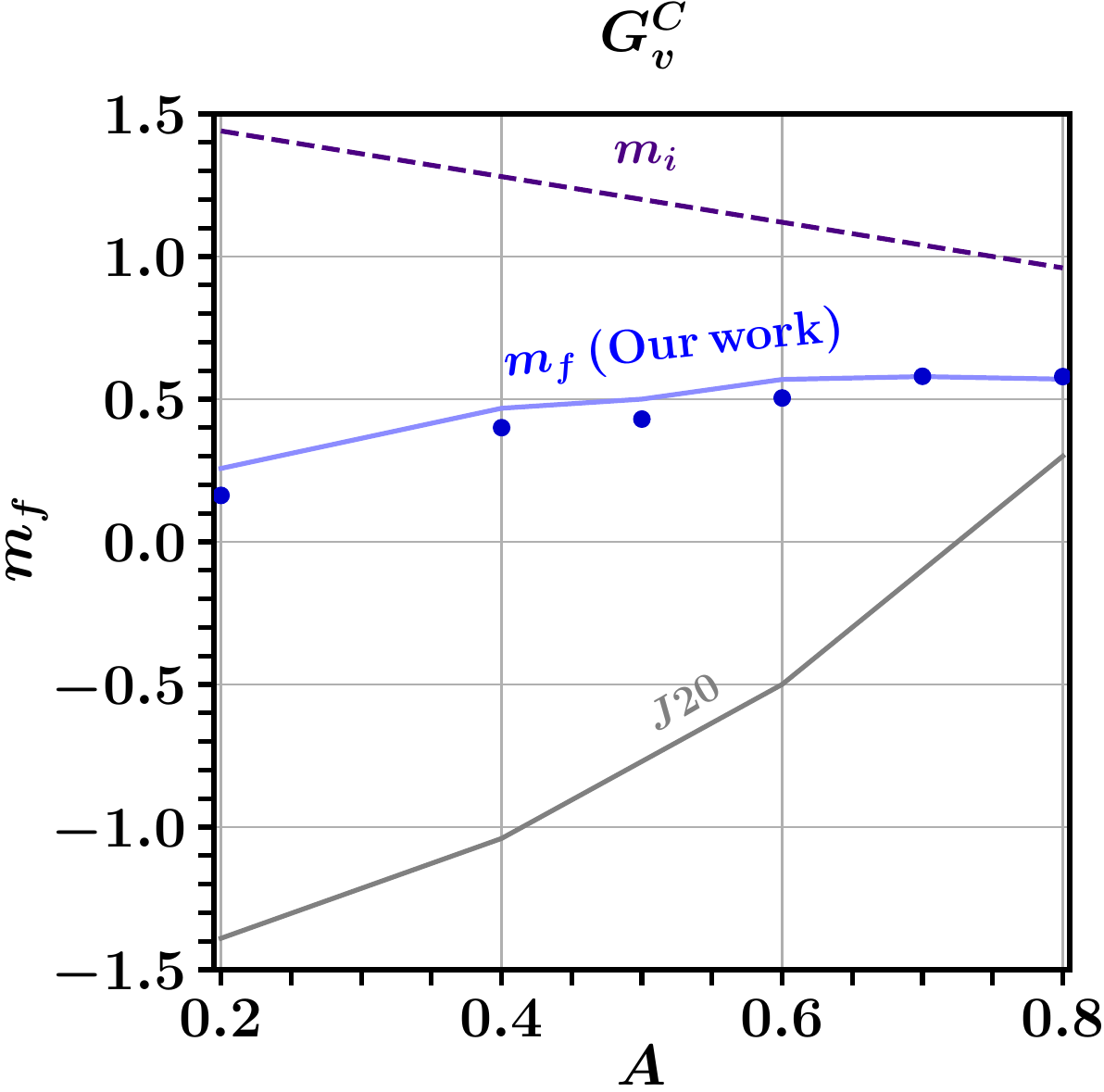}\\[2ex]
	\includegraphics[width=0.57\columnwidth]{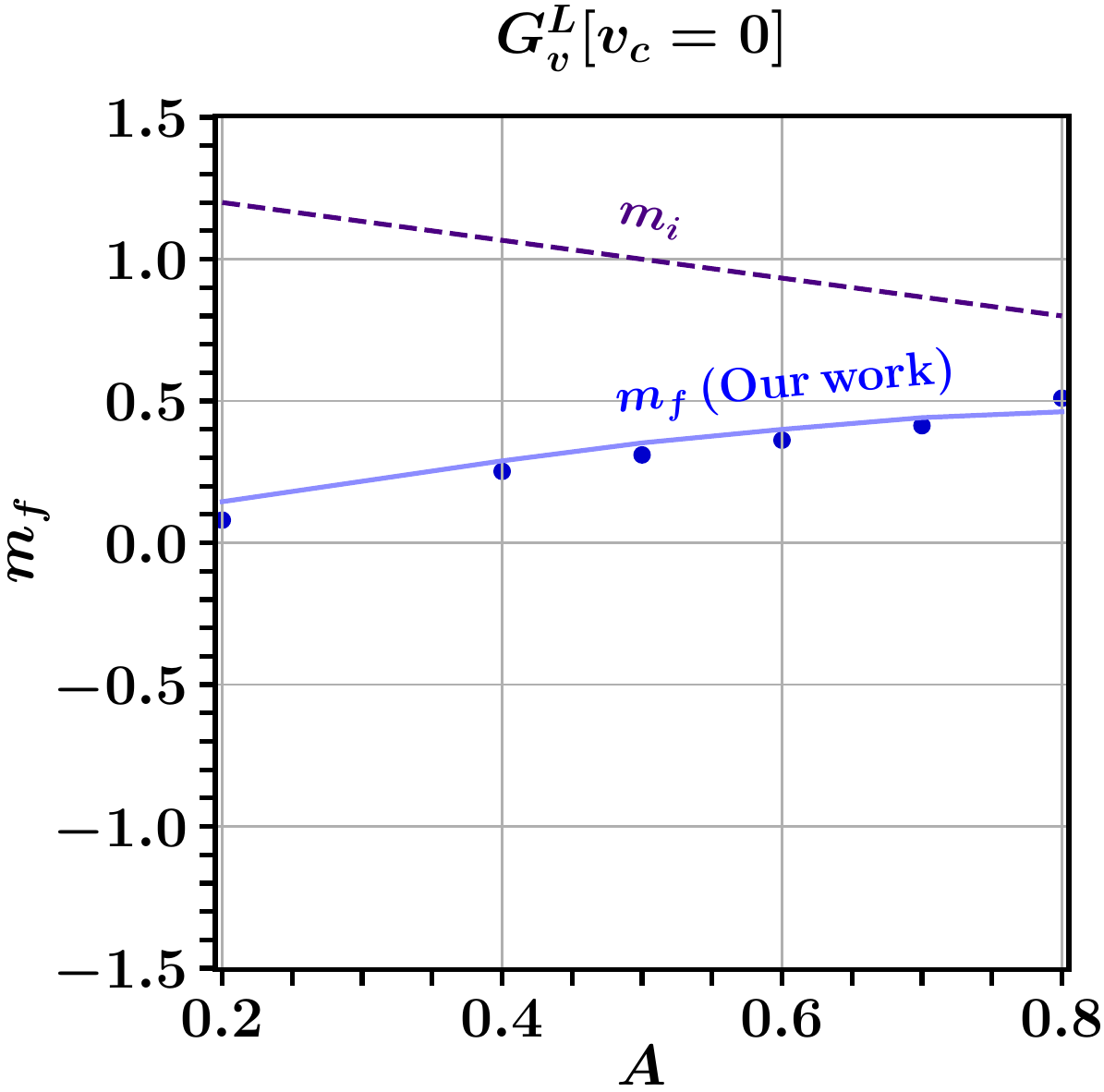}~\qquad
	\includegraphics[width=0.57\columnwidth]{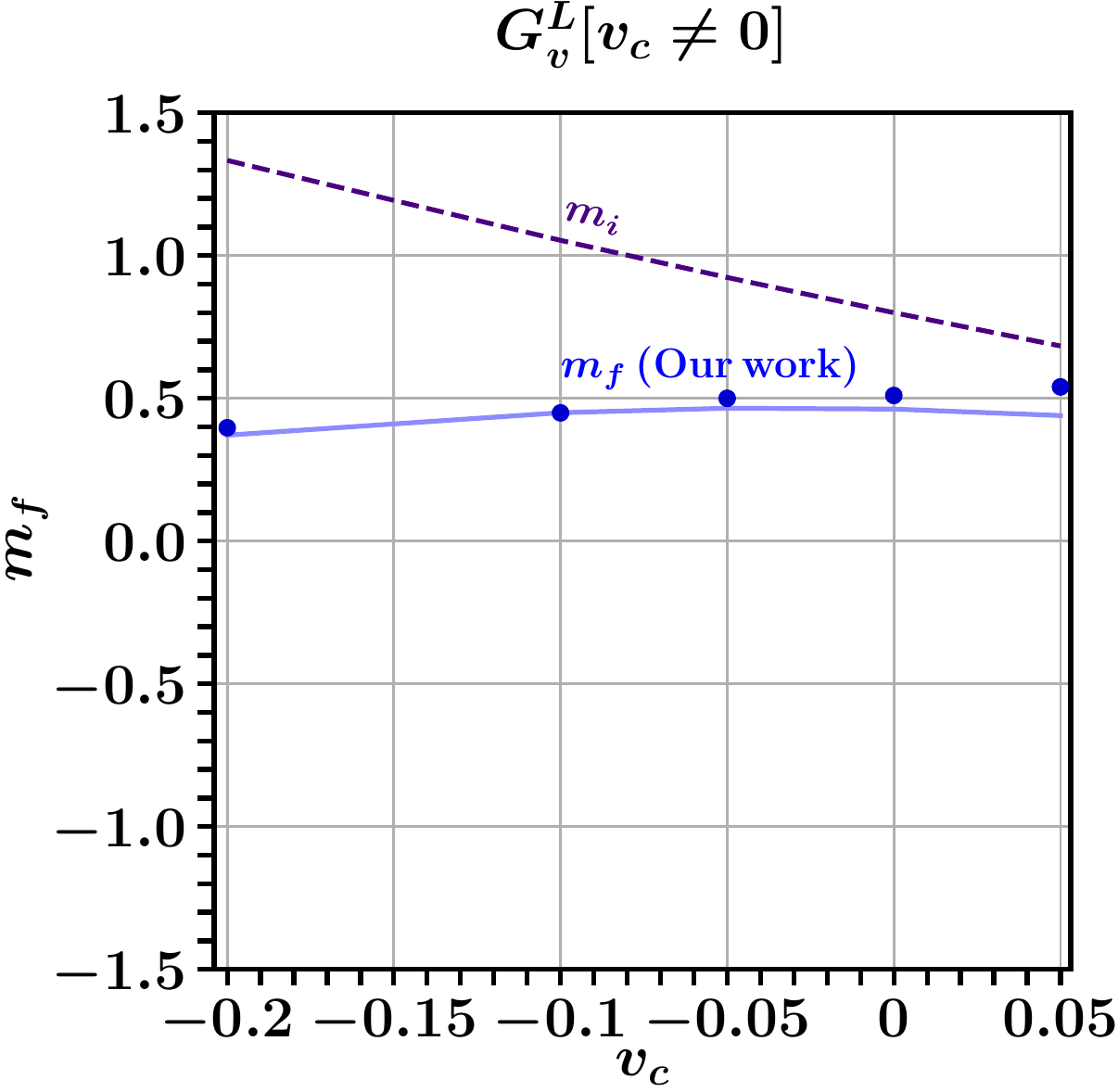}
	\caption{Lower resting point of ${\sf M}_1$: The dots in the four panels show $m_f$ as a function of $A$ for $G_v^{B}$, $G_v^{C}$, $G_v^L[v_c =0]$, and as a function of $v_c$ for $G_v^L[v_c \neq 0]$ with $A = 0.8$, computed by appropriately averaging the solution to Eq.\eqref{eom2}. We have also shown the comparison with the analytical prediction via Eq.\eqref{26c} (blue curve, labeled ``our work'') and the lower \emph{turning} point estimated by J20~\cite{Johns:2019izj} (gray curve, for the cubic ELN only). The purple dashed curve shows the initial value $m_i$.}
	\label{fig3mf}
\end{figure*}

At the resting point of $\mathsf{M}_1$ one must have 
\begin{align}\label{20v}
\partial_{t}{{m}}|_f = 0\,.
\end{align}
{Note that this \emph{resting point}, defined above, allows for relaxed solutions arising due to assumption of dephasing in deriving our approximate Eq.\eqref{eom5} from the exact Eq.\eqref{eom3}. In contrast, the \emph{turning point} of $\bf D_1$ (as given in Eq.14 of J20 or Eq.13 of PG21) explicitly excludes the spatial dependence of $\bf D_1$}.
Equations\,\eqref{20v} and \eqref{10b} imply $\mathsf{D}_{f} = \left(0, 0, u_{f} \right)$. This along with the  conservation of $\sigma$ between the initial and final positions of $\mathsf{M}_1$ pendulum implies
\begin{align}\label{21v}
m_{i} u_{i} = m_{f} u_{f}\,.
\end{align}
Equation\,\eqref{21v} and the conservation of $E$ and $\frac{\mathsf{B}^2}{2} +\mathsf{K} \cdot \mathsf{D}$ predict that
\begin{subequations}
	\begin{align}
	\frac{u_{f}^2-u_{i}^2}{2} = b \left(m_{i}-m_{f}\right)-{M}_{1,f}^{\perp}B_f^{\perp} \, \cos\theta^{MB}_f\,, \label{23a}\\
	\left(B_f^{\perp}\right)^2 = 2 k \left(u_{i}-u_{f}\right)\,. \label{23b}
	\end{align}
\end{subequations}
Note that $\theta_{f}^{MB}$ is the angle between ${\sf M}_{1,f}^\perp$ and $\mathsf{B}^{\perp}$ at $t = t_f$, and the lengths ${M}_{1,f}^\perp$ and $B_f^{\perp}$ have to be calculated by first taking the magnitudes of ${\sf M}_{1,f}^\perp[z]$ and  $\mathsf{B}^{\perp}_f[z]$ at each spatial location, followed by spatial averaging. 

At the resting position of the $\mathsf{M}_1$ pendulum, $\partial_t m|_f \approx 0$, and from Eq.\eqref{10b} one has ${\sf D}^\perp_f\approx 0$. Then, using the conservation of $\sigma$ per Eq.\eqref{15b} implies 
\begin{equation}
\partial_t \mathsf{D}^{\parallel} |_{f} = -\frac{m_i u_i}{m_f^2} \partial_t m |_{f} \approx 0\,,
\end{equation}
and gives $\mathsf{B} \parallel \mathsf{M}_1$ at $t = t_f$, i.e.,  $\theta_f^{MB} \approx 0$, resulting in
\begin{align}\label{24}
\left(\frac{b}{{B}_{f}^{\perp}} \right)^2 = \left(\frac{m_{f}}{{M}_{1,f}^\perp} \right)^2\,.
\end{align}
Using Eqs.\eqref{23a} and \eqref{23b}, and ignoring the trivial solution $m_f = m_i$, help to simplify Eq.\eqref{24} in terms of the desired variable $m_{f}$:
\begin{equation}\label{mfeq}
 \left(2b^2m_{f}^2+(m_{f}+m_{i})bu_{i}^2\right)^2-16k^2u_{i}^2m_{f}^4 = 0\,.
\end{equation}
The solutions of Eq.\eqref{mfeq} are
\begin{subequations}
	\begin{align}
	m_{f}^{++} = \frac{-bu^2_{i} + \sqrt{b^2u^4_{i}-4 \left(2b^2 + 4ku_{i}\right) bu^2_{i} m_{i}}}{4b^2 + 8ku_{i} } \,, \label{26a}\\
	m_{f}^{-+} =  \frac{-bu^2_{i} - \sqrt{b^2u^4_{i}-4 \left(2b^2 + 4ku_{i}\right) bu^2_{i} m_{i}}}{4b^2 + 8ku_{i} } \,, \label{26b} \\
    m_{f}^{+-} = \frac{-bu^2_{i} + \sqrt{b^2u^4_{i}-4 \left(2b^2 - 4ku_{i}\right) bu^2_{i} m_{i}}}{4b^2 - 8ku_{i} }\,, \label{26c}\\
    m_{f}^{--}  =  \frac{-bu^2_{i} - \sqrt{b^2u^4_{i}-4 \left(2b^2 - 4ku_{i}\right) bu^2_{i} m_{i}}}{4b^2 - 8ku_{i} } \,. \label{26d}
	\end{align}
\end{subequations}
Only $m_f^{+-}$ in Eq.\eqref{26c} has the correct qualitative behavior with $A$ and $v_c$ to qualify as a solution. Note that $m_i$, $u_i$, $b$, and $k$, and thus $m_f^{+-}$, are known from the ELN. 

In Fig.\,\ref{fig3mf} we plot $m_f^{+-}$, as obtained from Eq.\eqref{26c}, in case of $G_v^B,\, G_v^C,\, G_v^L$. In all these plots, the blue disks show the numerical solution of $m$ at $t \sim 30$. We find excellent agreement, with correct dependence on $A$ and $v_c$. The other solutions $m_f^{++}, m_f^{-+}, m_f^{--}$ are spurious and do not have the correct scaling with $A$ and $v_c$.

The qualitative dependence of the resting point on $A$ or $v_c$ can be understood as follows: The kinetic energy of the $\mathsf{M}_1$ pendulum is
\begin{equation}\label{Eq20}
E_k = \frac{\mathsf{D} \cdot \mathsf{D}}{2} \sim \frac{(\partial_{t}{\mathsf{M}}_1)^2}{2 \mathsf{M}_1 \cdot \mathsf{M}_1}+ \frac{\sigma^2}{2 \mathsf{M}_1 \cdot \mathsf{M}_1}\,.
\end{equation}
Clearly $\sigma$ is a constant and can be determined from the initial conditions. For example if we consider $G_v^L[v_c]$, 
\begin{equation}\label{Eq21}
\sigma \sim \frac{A}{2}+\left(\frac{2}{3}-\frac{A}{3}\right)v_c\,.
\end{equation}
For the case with $A \neq 0$ and $ v_c = 0$, one has $\sigma \sim {A}/{2}$, whereas for $A = 0.8, v_c \neq 0$, one has $\sigma \sim 0.4 \, \left(1+v_c \right)$. As a result, in the limit $A \to 0$ or $v_c\to -1$, we have $\sigma \to 0$ so that the  impact of internal spin $\sigma$ is small. Thus, the pendulum swings like an ordinary pendulum resting at a smaller $m_f$. In the other limit when $\sigma $ is large one can approximately neglect the $\mathsf{M}_1 \times \partial_{t}^2 \mathsf{M}_1$ term in Eq.\eqref{eom6} so that the $\mathsf{M}_1$ pendulum equation becomes a simple spin-precession equation indicating $m_f \approx m_i$. This is roughly the case with large $A$ or $v_c \to +1$.

\begin{figure*}[t]		
				\includegraphics[width=0.59\columnwidth]{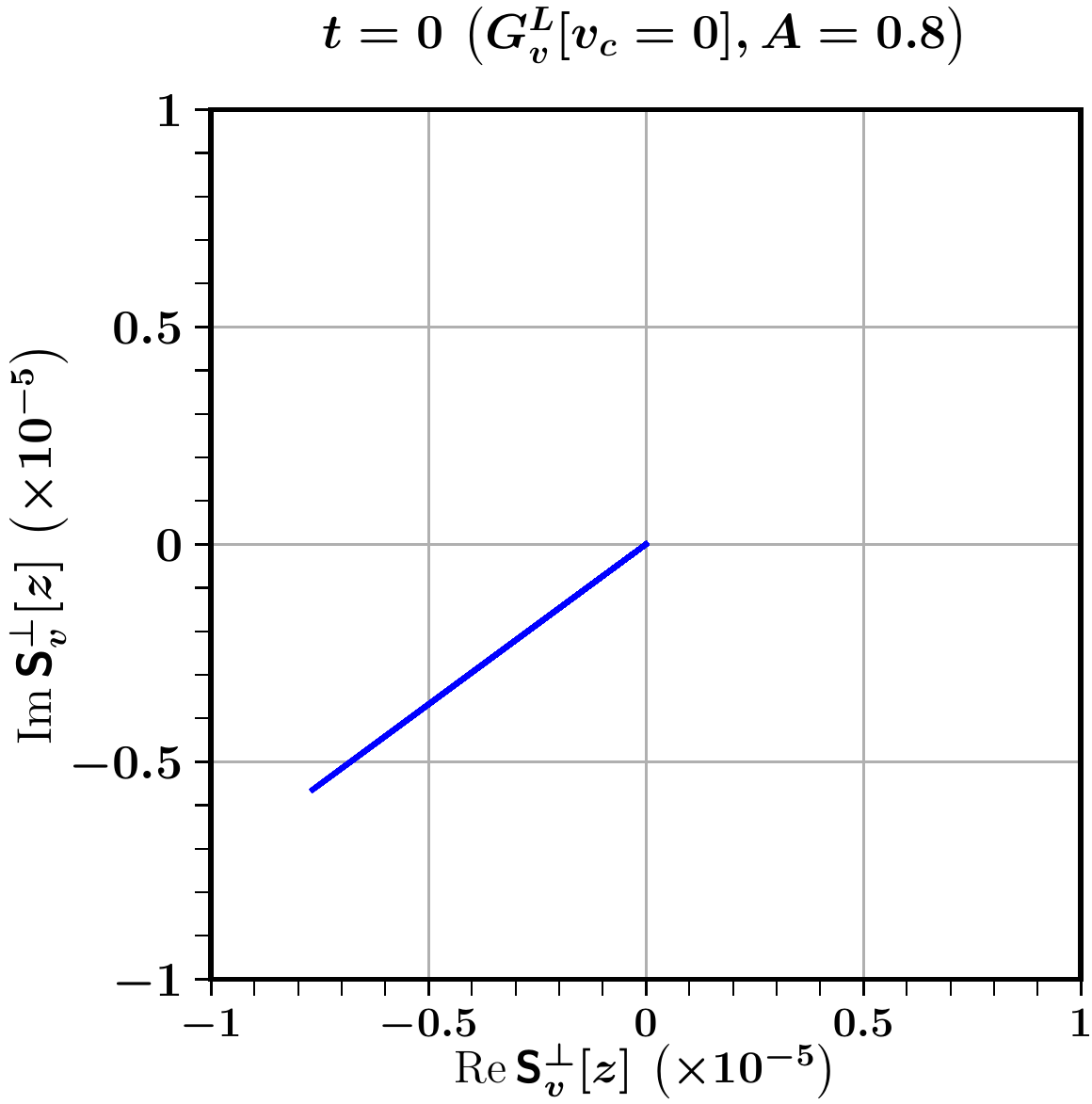}
	\hspace{0.3cm}	\includegraphics[width=0.59\columnwidth]{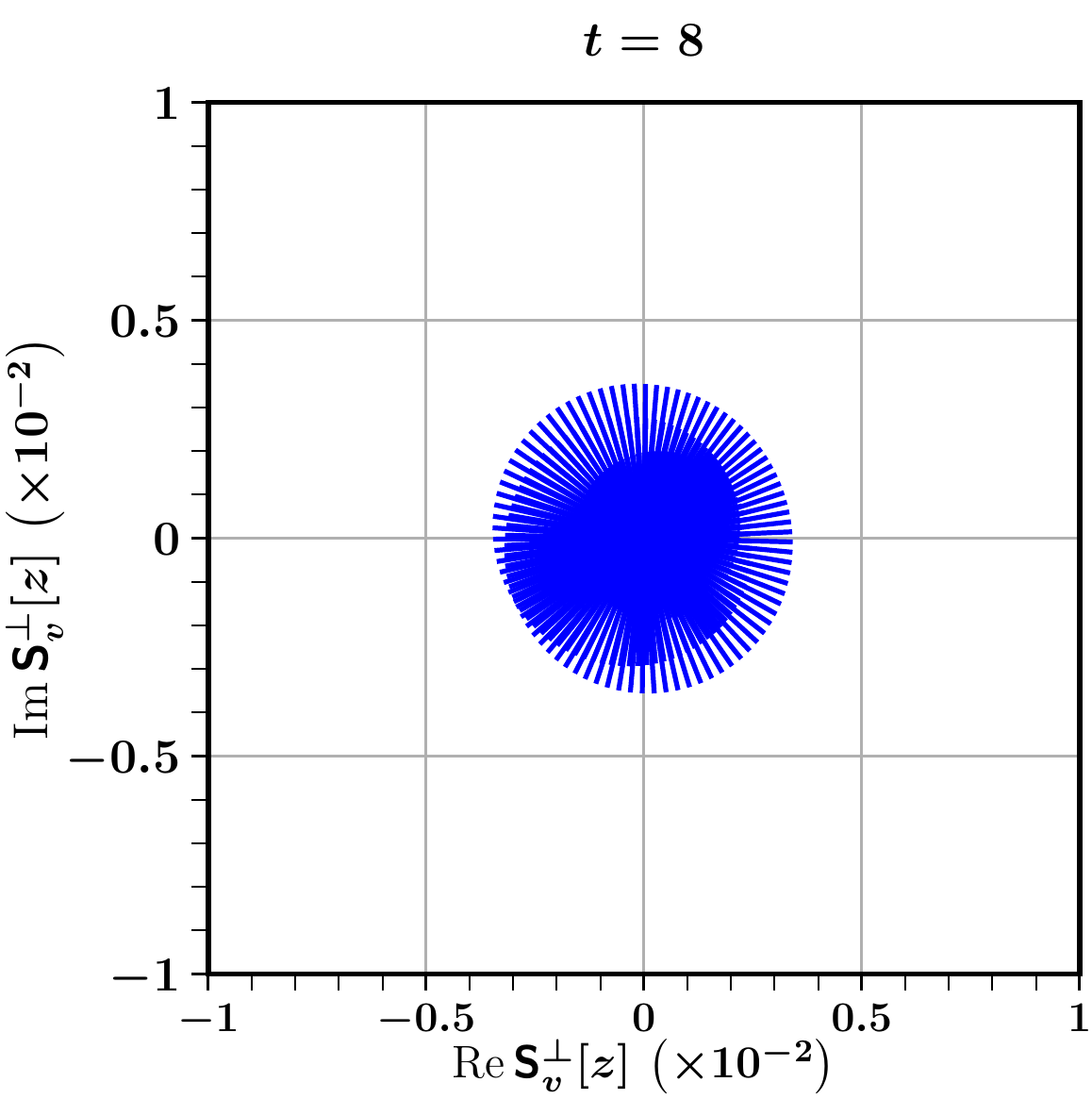}
	\hspace{0.3cm}	\includegraphics[width=0.59\columnwidth]{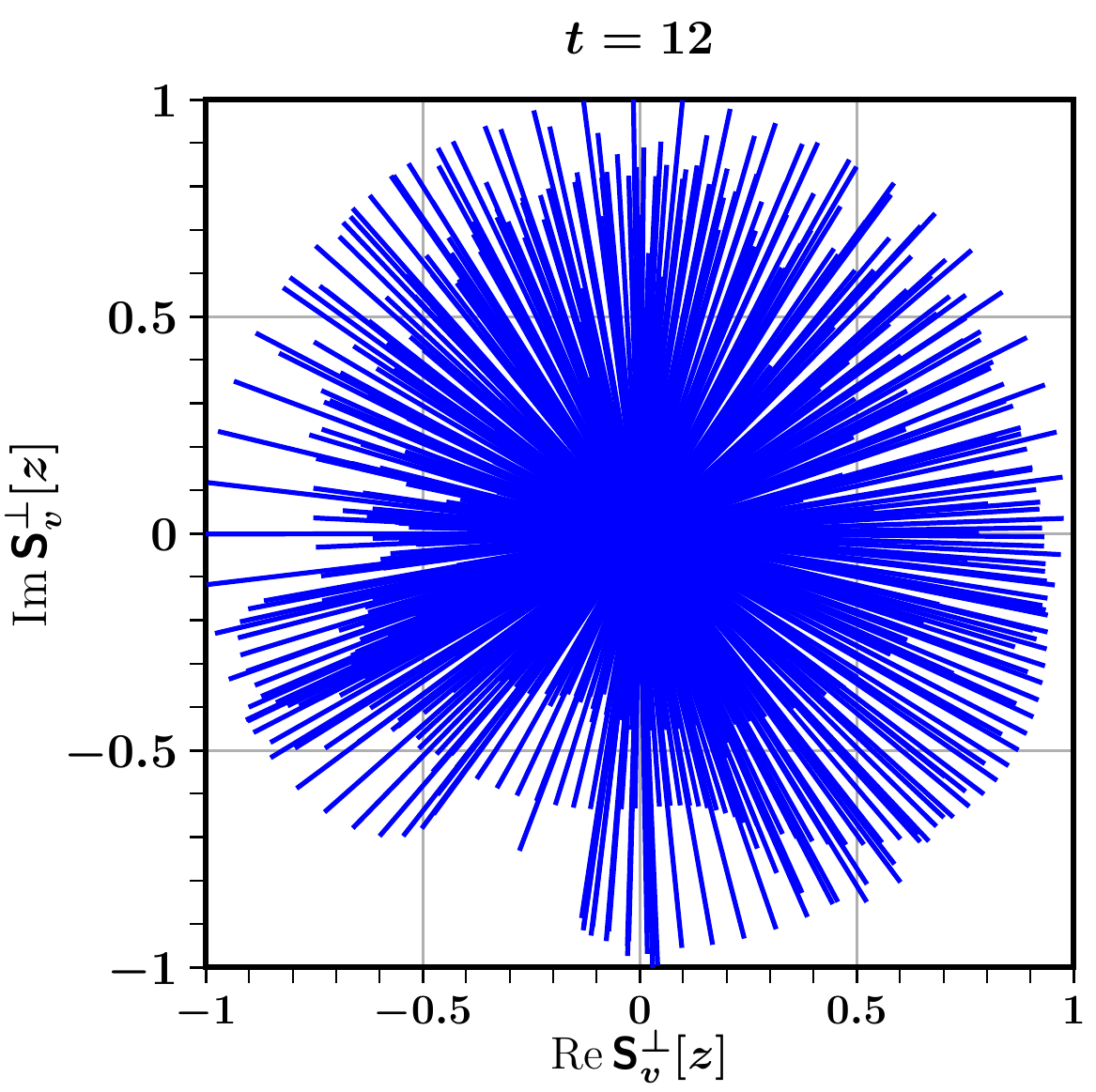}	
	\caption{Relative dephasing of Bloch vectors: Each blue line starting from the origin indicates the size and orientation of the transverse Bloch vector, ${\rm Im} \, {\mathsf S}^{\perp}_{v}[z]$ vs. ${\rm Re} \, {\mathsf S}^{\perp}_{v} [z]$, at $4096$ different spatial locations at three different instants in time $t = 0$, $t = 8$ and $t = 12$. For this calculation, we used a non-random seed (see text). All panels show the data for $v =-0.5$ for $G_v^L[v_c = 0]$ and $A = 0.8$. One sees that at late times, here after $t\approx12$, the transverse Bloch vectors become large and random.}
	\label{figT2relax}
\end{figure*}

\begin{figure*}[t]
    ~~\includegraphics[height=0.51\columnwidth]{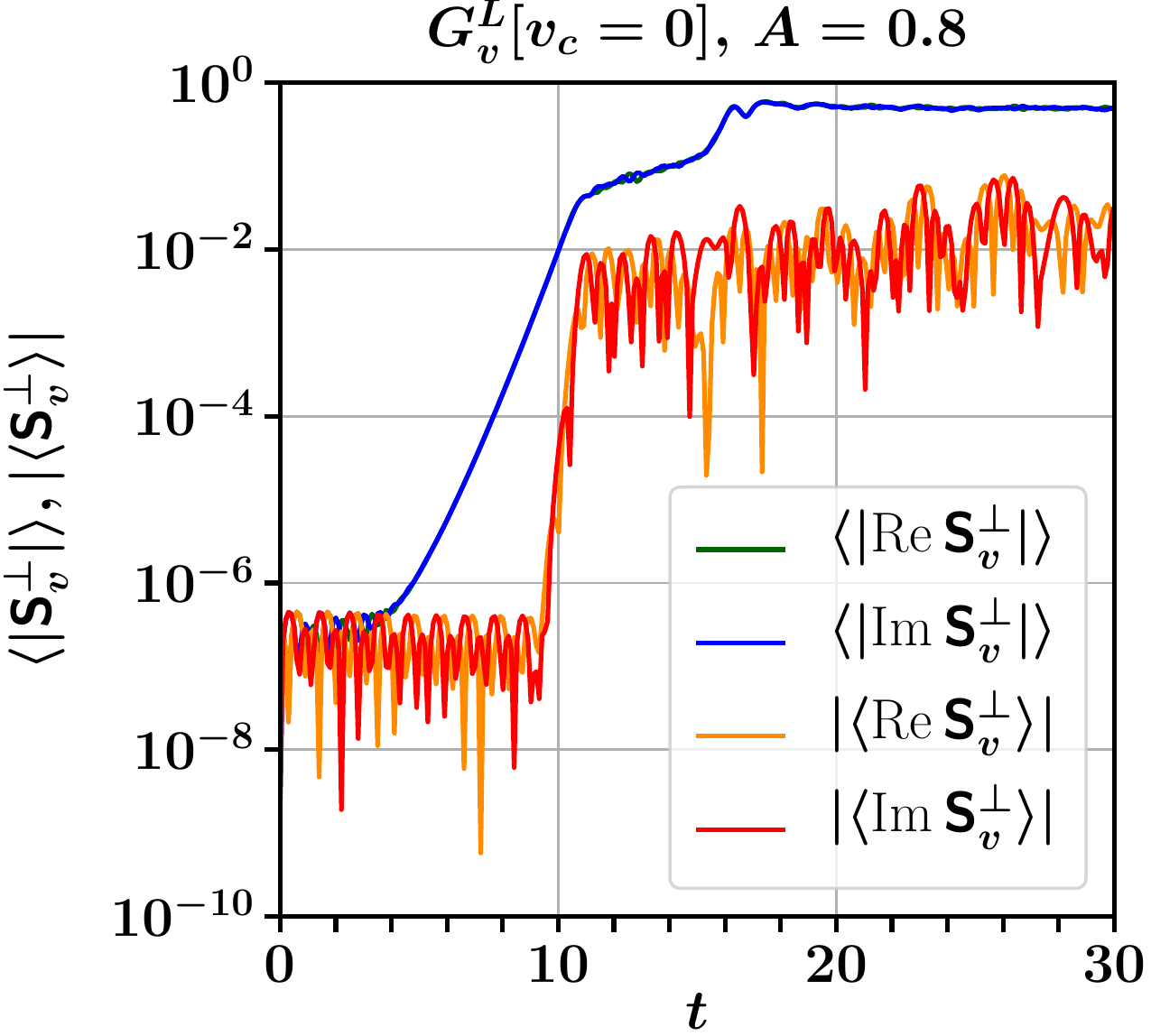}
   \qquad \includegraphics[height=0.52\columnwidth]{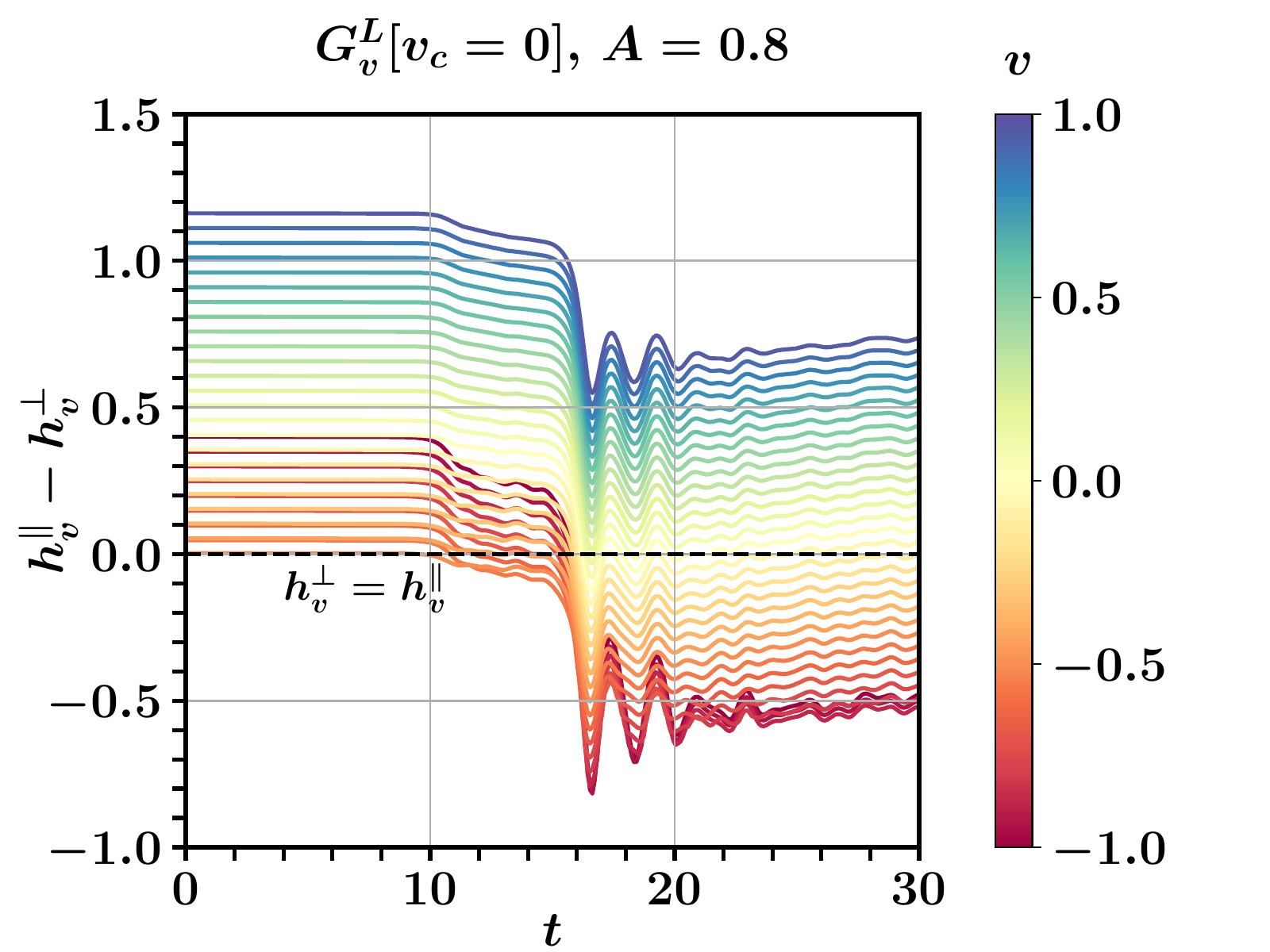}
    \quad \includegraphics[height=0.52\columnwidth]{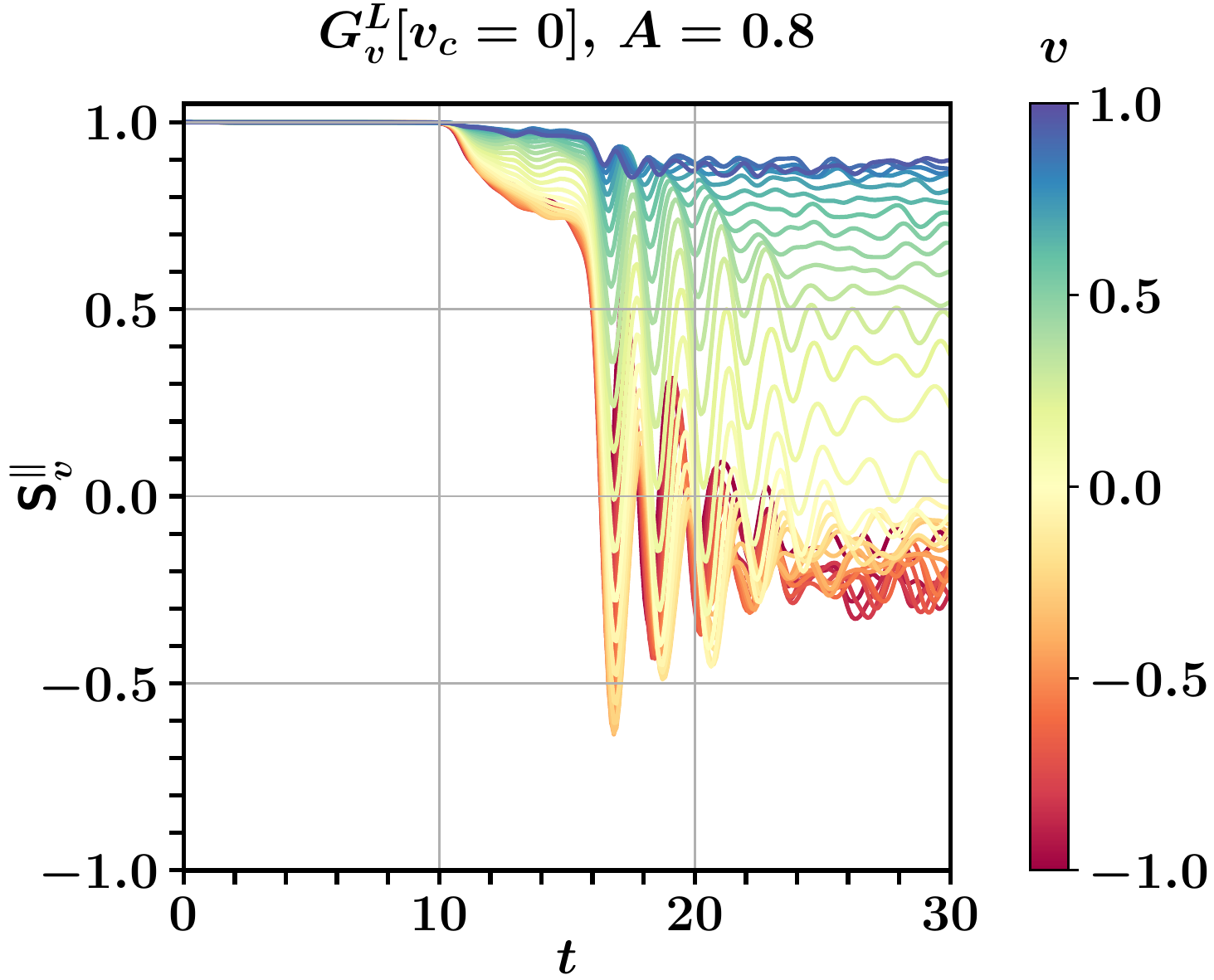}
 \caption{Relaxation and Hamiltonians: Leftmost plot shows the growth of transverse vectors for $v=-0.5$. The other two plots show that velocity modes that experience longitudinal depolarization are those that feel a large transverse Hamiltonian, as seen from the relative sizes of $h_v^{\parallel}$ and $ h_v^{\perp}$ vs. $t$ in the middle panel and behavior of $\mathsf{S}_v^{\parallel}$ vs. $t$ in the rightmost panel. The ELN is $G^L_v[v_c=0]$ with $A=0.8$. Note that the ${\sf M}_1$ pendulum has its first dip at $t\approx12$ in this case (see Fig.\,\ref{fig1lep}), coinciding with the epoch of transverse relaxation.}
    	\label{Srelax}
\end{figure*}

\begin{figure*}[t]		
				\includegraphics[width=0.57\columnwidth]{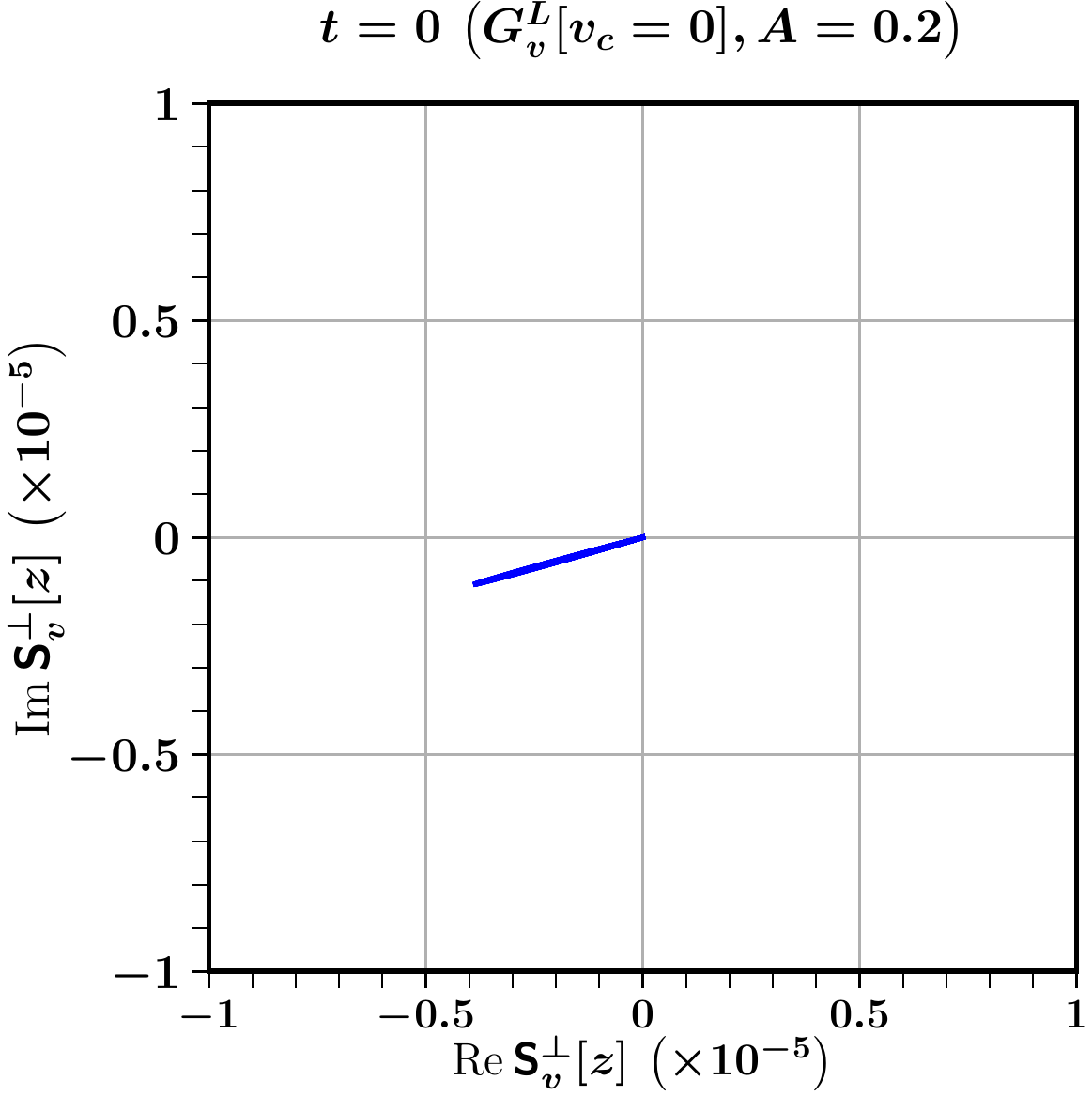}
	\hspace{0.3cm}	\includegraphics[width=0.57\columnwidth]{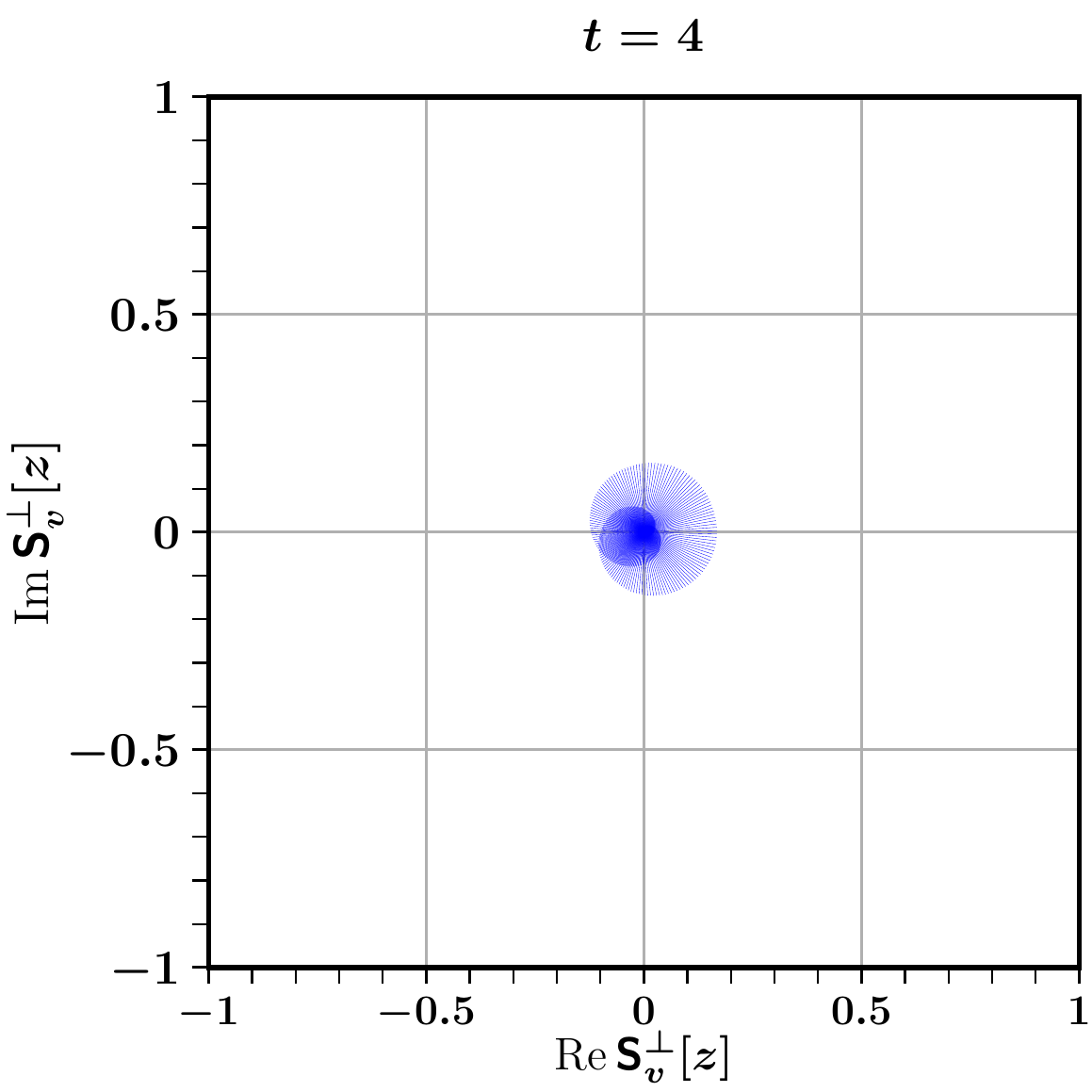}
	\hspace{0.3cm}	\includegraphics[width=0.57\columnwidth]{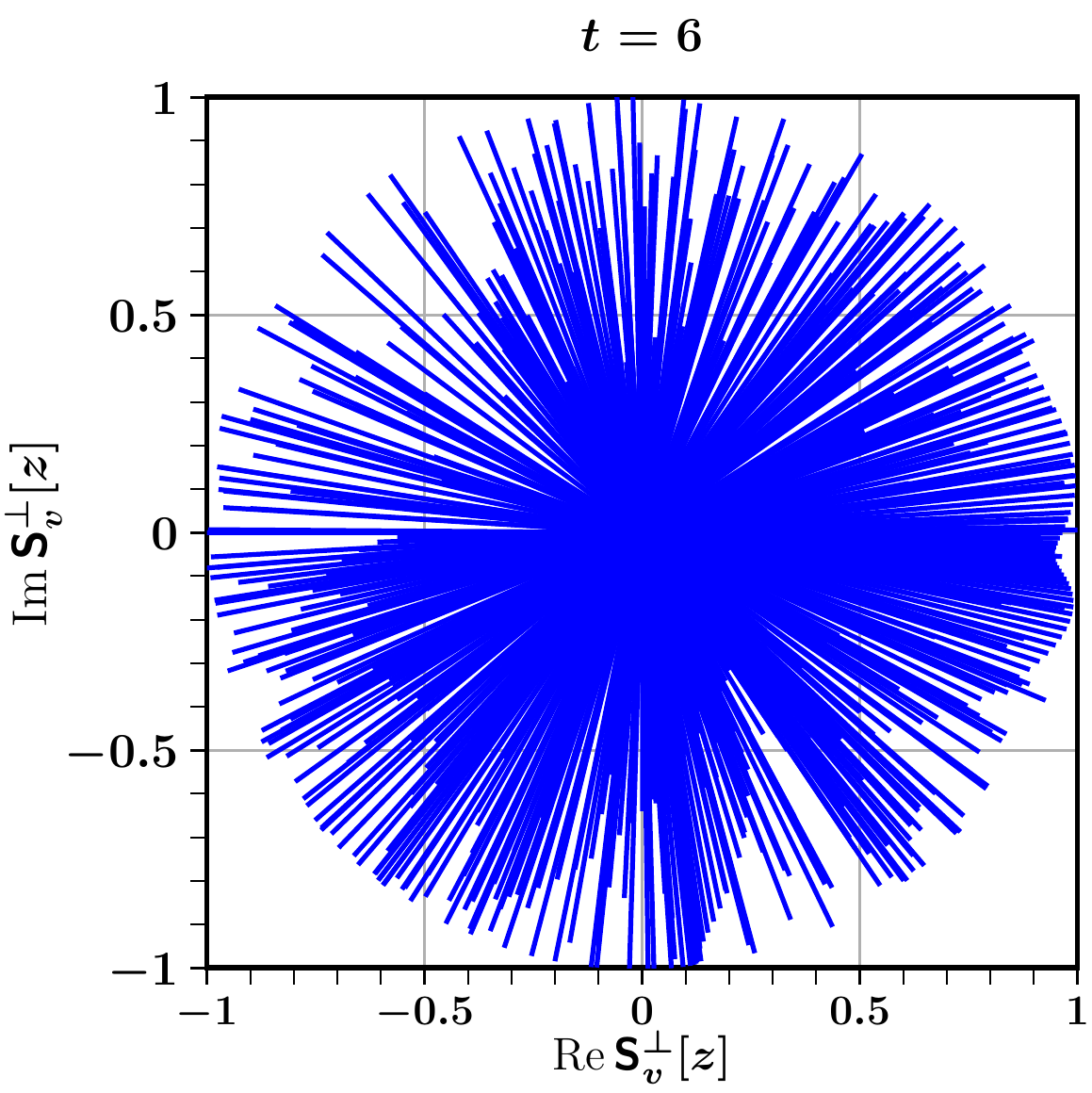}	
	\caption{Relative dephasing of Bloch vectors: Each blue line starting from the origin indicates the size and orientation of the transverse Bloch vector, ${\rm Im} \, {\mathsf S}^{\perp}_{v}[z]$ vs. ${\rm Re} \, {\mathsf S}^{\perp}_{v} [z]$, at $4096$ different spatial locations at three different instants in time $t = 0$, $t = 4$ and $t = 6$. For this calculation, we used a non-random seed (see text). All panels show the data for $v = -0.5$ for $G_v^L[v_c = 0]$ and $A = 0.2$. At late times, here between $t=4$ and 6, the transverse Bloch vectors get large and random.}
	\label{figT2relax2}
\end{figure*}

\begin{figure*}[t]
    ~~\includegraphics[height=0.5\columnwidth]{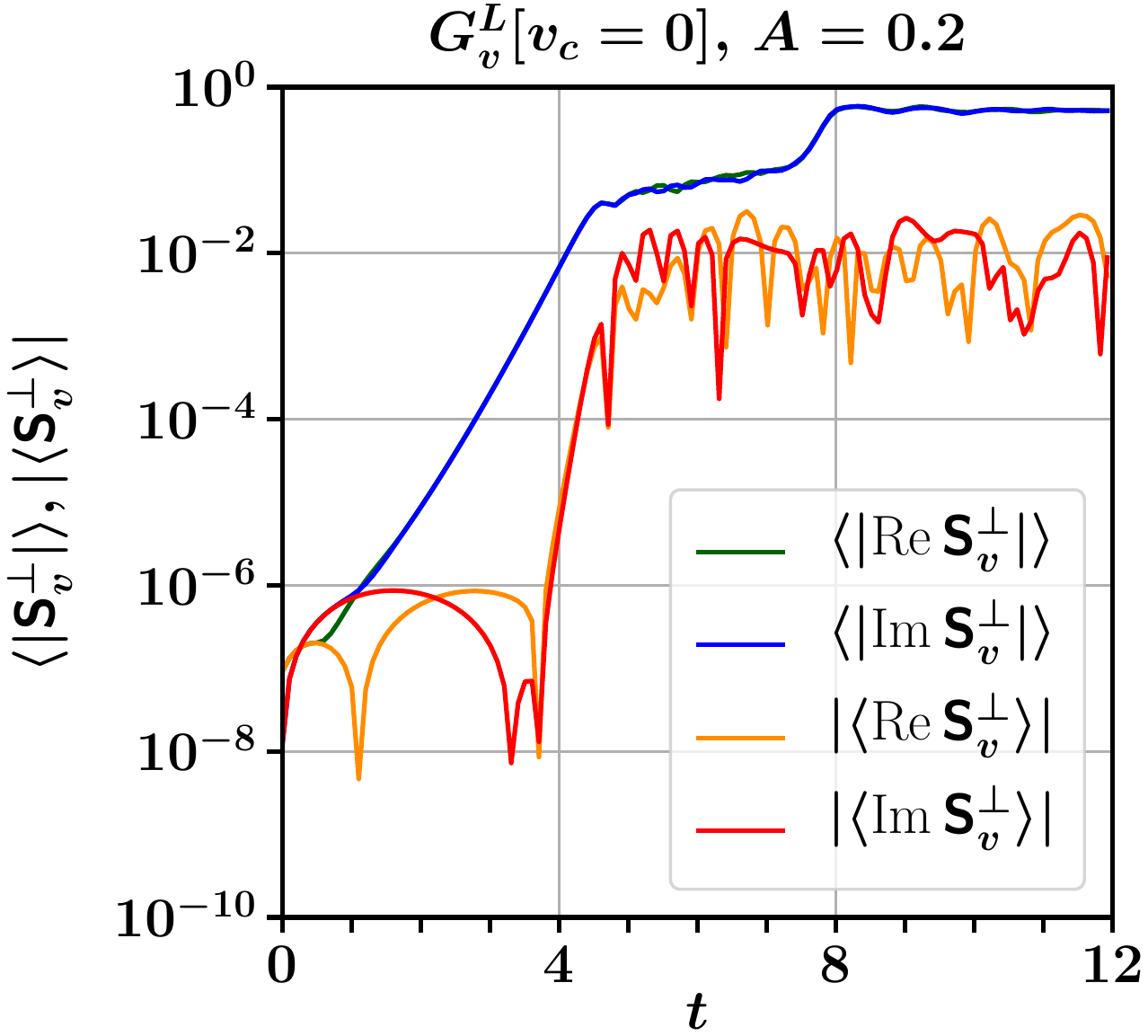}
   \hspace{0.1 cm}\quad \includegraphics[height=0.51\columnwidth]{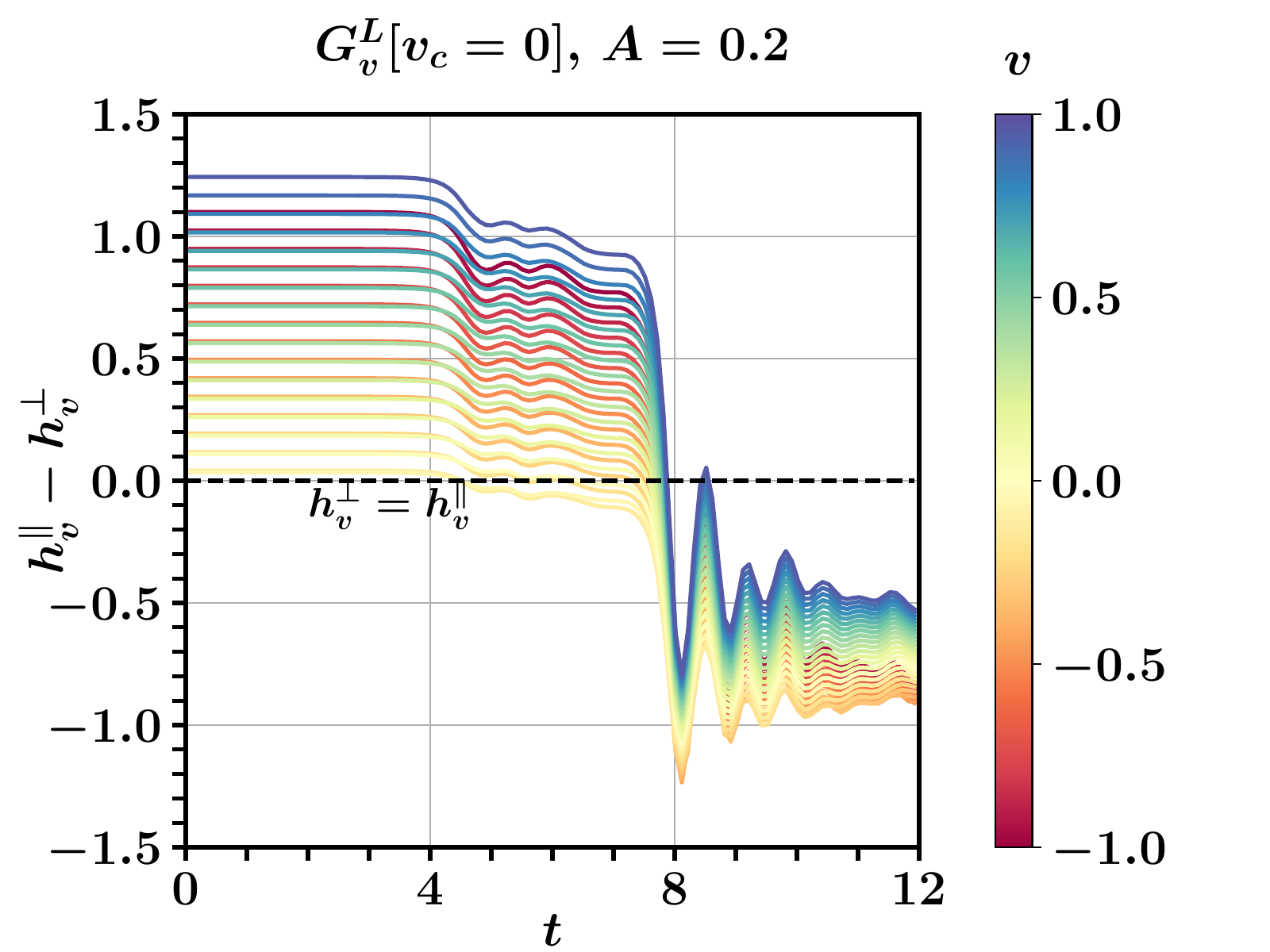}
    \hspace{-0.1 cm}\quad \includegraphics[height=0.52\columnwidth]{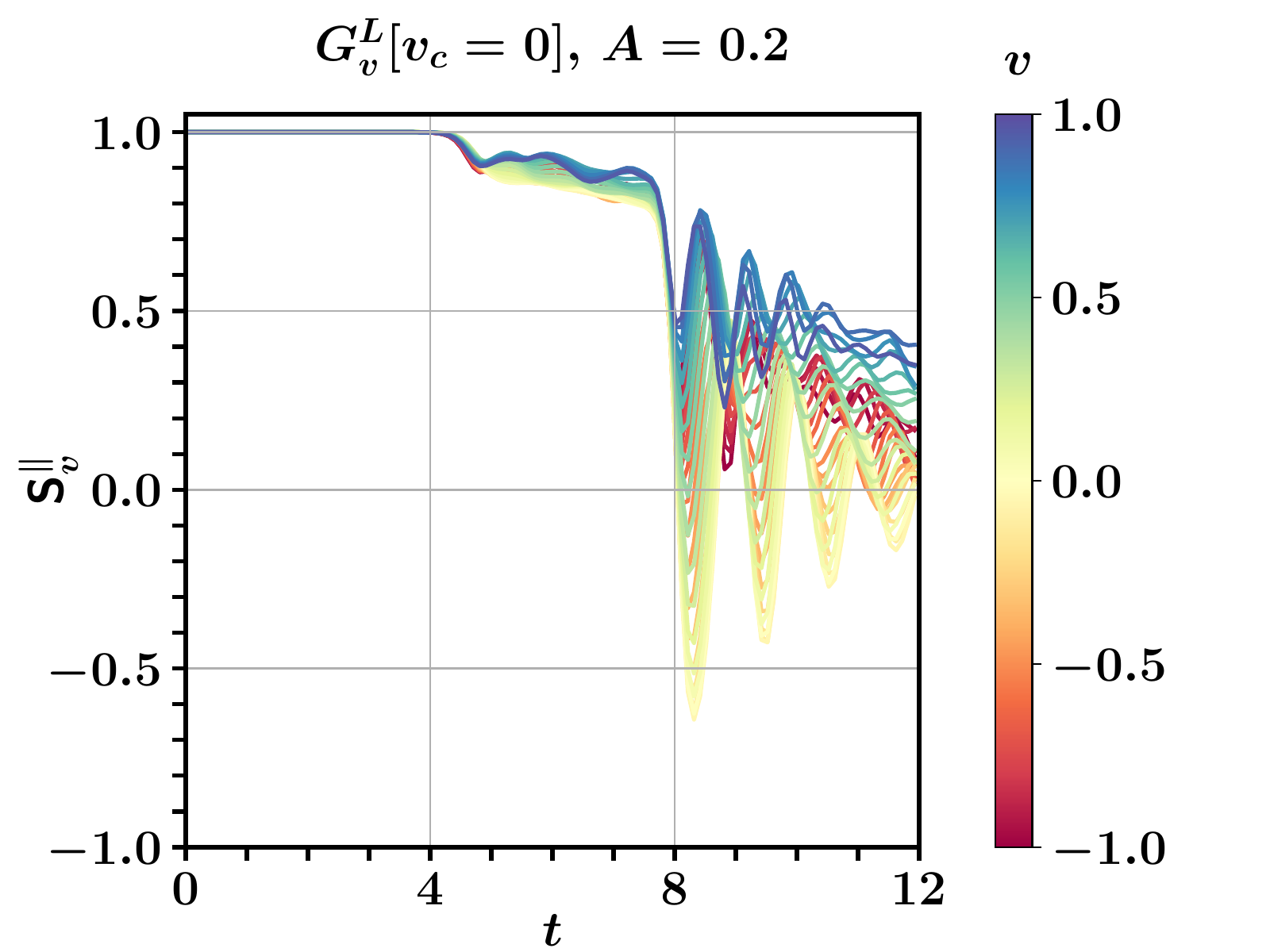} \caption{Relaxation and Hamiltonians: Leftmost plot shows the growth of transverse vectors for $v=-0.5$. The other two plots show that velocity modes that experience longitudinal depolarization are those that feel a large transverse Hamiltonian, as seen from the relative sizes of $h_v^{\parallel}$ and $ h_v^{\perp}$ vs. $t$ in the middle panel and behavior of $\mathsf{S}_v^{\parallel}$ vs. $t$ in the rightmost panel. The ELN is $G^L_v[v_c=0]$ with $A=0.2$. Note that the ${\sf M}_1$ pendulum has its first dip at $t\approx6$ in this case (see Fig.\,\ref{fig1lep}), coinciding with the epoch of transverse relaxation.
    }
    	\label{Srelax2}
\end{figure*}

The resting point need not coincide with the lower turning point given in Eq.(14) of J20~\cite{Johns:2019izj} (the factor of 9 therein should be 5/4 -- which has been corrected here). For the ELNs $G_v^B$ and $G_v^L$, which do not have a cubic term, one does not expect a sensible estimate from Eq.(14) of J20 (since it formally diverges for $D_3 = 0$). However, even for the ELNs where a cubic term is present, and one ought to get a sensible estimate, we see that the turning point is not the resting point (gray line marked as J20 in the top right plot of Fig.\,\ref{fig3mf}). PG21~\cite{Padilla-Gay:2021haz} solves the fast flavor pendulum, assuming homogeneity. Eq.(13) therein is an accurate description of the strictly homogeneous evolution, but it cannot be applied in more general inhomogeneous settings, e.g., to our Eq.\eqref{eom2}. In fact, the homogeneous mode is typically stable for our ELNs. They also note that the truncated multipole approach is not accurate for homogeneous evolution. However, note that in our case inhomogeneity and dephasing are present. For all ELNs we have checked, the $m_f$ in Eq.\eqref{26c}, derived by assuming dephasing and a truncated multipole tower, agrees well with the spatial average of ${\sf M}_1[z,t\to\infty]$ computed using Eq.\eqref{eom2}.

\subsection{Transverse relaxation}
\label{sec:TR}

The \emph{spatially averaged} version of the flavor evolution given by Eq.\eqref{eom2} can be derived using approximations similar to those used in deriving Eq.\eqref{eom5} from Eq.\eqref{eom3}, to get
\begin{equation}\label{corot}
\partial_{t} \mathsf{S}_v  = \mathsf{h}_v \times \mathsf{S}_v = \Bigg(-\bigg(\frac{1}{3}A+\frac{2}{3} \mathsf{M}_2^{\parallel}\bigg) \hat{\mathsf{M}}_0 - v \mathsf{M}_1 \Bigg) \times \mathsf{S}_v \,,
\end{equation}
in a \emph{special corotating frame} where we ensure that Hamiltonian for $\mathsf{M}_1$ is purely transverse. {Note $\hat{\mathsf{M}}_0 = {{\mathsf{M}}_0}/{\sqrt{{\mathsf{M}}_0 \cdot {\mathsf{M}}_0}}$ denotes the unit vector along ${\mathsf{M}}_0$}. While this frame has a complicated motion in general, if we neglect changes in length of $\mathsf{M}_2^{\parallel}$ it simply rotates about $\mathsf{M}_0$ with an extra frequency $\frac{4}{3}A+\frac{2}{3} \mathsf{M}_2^{\parallel}$ relative to the frame in which Eqs.\eqref{10a}-\eqref{10d} are written. See Sec.\,IIC.2 of B20a~\cite{Bhattacharyya:2020dhu} for a derivation.  In the remainder of this paper, corotating frame will refer to this special frame.

The length of each coarse-grained Bloch spin ${\sf S}_v$ is predicted to remain constant according to Eq.\eqref{corot}. This isn't borne out by numerical calculations. The reason is simply that the spatially averaged equations are approximate in the first place. To understand this analytically, one needs to study the pre-coarse-grained partial differential equation in Eq.\eqref{eom3}. Here we draw an analogy to the nuclear magnetic resonance of macroscopic samples, to obtain a semi-quantitative understanding.

In Eq.\eqref{corot}, the Bloch vector $\mathsf{S}_v$ can be interpreted as the net spin of a macroscopic sample volume being acted on by a magnetic field equivalent to the corotating Hamiltonian $\mathsf{h}_v$. In reality, the macroscopic spin is composed of several microscopic spins bunched together, similar to how we have defined the coarse-grained $\mathsf{S}_v$ from the pre-coarse-grained $\mathsf{S}_v[z]$. Initially, $\mathsf{h}_v$ is along the $\hat{\mathsf{e}}^{(3)}$ direction and the Bloch spins remain aligned with $\mathsf{h}_v$. However, as $\mathsf{M}_1$ tips over, $h_v^{\parallel}$ decreases and concomitantly $h_v^{\perp}$ increases. As a result, for some velocity modes, the transverse component of $\mathsf{h}_v$ can become of similar size as its parallel component. {We remind, the spatially averaged lengths of the parallel and transverse components of $\mathsf{h}_v$, i.e., $h_v^{\parallel}$ and $h_v^{\perp}$, respectively, are defined as follows:
\begin{subequations}
	\begin{eqnarray}
	h_v^{\parallel} &=& \frac{1}{L} \int_{0}^{L}\,dz \,\Big|-\frac{A}{3}-\frac{2}{3} \mathsf{M}_2^{\parallel}-v \mathsf{M}_1^{\parallel}\Big| \,, \label{25a}\\
	h_v^{\perp} &=& \frac{1}{L} \int_{0}^{L}\,dz \,\Big|\mathsf{M}_0^{\perp}[z]-v \mathsf{M}_1^{\perp}[z]\Big|\,. \label{25b}
	\end{eqnarray}
\end{subequations}
In Eq.\eqref{25b}, we have used the fact that the length of the transverse vector remains invariant under the rotation about $\hat{\sf e}_3$.} The Bloch spins for those velocity modes develop a large precession angle to reach the transverse plane. At this juncture, the dispersion of the magnetic field $\mathsf{h}_v[z]$, within the coarse-graining volume, can lead to the constituent microscopic spins to precess at different rates at different locations within the coarse-graining volume. This causes the transverse component of the macroscopic spins to become smaller over a timescale T2 -- a process known as T2 relaxation. The transverse components of ${\sf M}_1$ relax in the same way. As ${\sf M}_1$ oscillates, for some velocity modes the relaxation turns on and off repeatedly. 

The above analogy predicts that relaxation is strongest when the co-rotating Hamiltonian develops a large transverse component, i.e., when $h_{v}^{\perp} \sim h_{v}^{\parallel}$. Roughly, this must coincide with ${\sf M}_1$ developing a large transverse component. Further, one expects that transverse relaxation is prominent for those velocity modes for which $h_{v}^{\perp}$ becomes comparable to $h_{v}^{\parallel}$. Conversely, for velocity modes whose transverse corotating Hamiltonians never grow too large, relaxation should be less efficient.

We will demonstrate the development of transverse relaxation using our numerical results for $G^L_v[v_c = 0]$, for two values of lepton asymmetry $A=0.8$ (Figs.\,\ref{figT2relax} and \ref{Srelax}) and $A=0.2$ (Figs.\,\ref{figT2relax2} and \ref{Srelax2}). The former case shows a slower rate and lesser degree of relaxation, while the latter shows faster and more extensive relaxation.  For these four plots, we use a localized non-random seed: ${\sf S}^{\perp}_v[z] = 10^{-6} \exp[-(z-L/2)^2/5] (\hat{\sf e}_1+i\hat{\sf e}_2)$, i.e., the initial transverse components are taken to be $\approx10^{-6}$, with fixed relative phase, localized around the centre of the box. This choice of seed (i.e., not random, unlike elsewhere in this paper) is to emphasize that even if we do not put random relative phases by hand, the system generates effective random phases on its own. The long-term results will be at best mildly sensitive to the seeds.

We begin with $G^L_v[v_c = 0]$ with $A=0.8$.  In Fig.\,\ref{figT2relax}, at $t=0$ the transverse components of the Bloch vectors at all locations start out in phase  (as set by the initial seeds in this case). By $t\approx8$ they begin to get dephased relative to each other, though the transverse vector is still very small at most locations. By $t\approx12$, this dephasing is essentially complete. In other words, the transverse components of  $\mathsf{S}_v^{\perp}[z]$ become large and randomized across different locations $z$, as was shown in the bottom panel results of Fig.\,6 in Ref.\cite{Bhattacharyya:2020dhu}. One thing to note is that $S_v^{\perp}[z] \neq 0$ without coarse-graining in $z$, but  vanishes upon coarse-graining in $z$. Obviously, the transverse components of the coarse-grained multipole moments of $\mathsf{S}_v^{\perp}$ also decay due to this relaxation. In Fig.\,\ref{Srelax} the left panel shows the growth of the transverse components for $v=-0.5$. One sees that up to $t\approx8$ the evolution in linear. Yet, relative dephasing causes $|\langle {\sf S}^\perp_v \rangle|$ to become smaller than $\langle |{\sf S}^\perp_v |\rangle$. Around $t\approx10$, close to the time of the first dip of the ${\sf M}_1$ pendulum, nonlinearity sets in. The transverse components quickly grow to ${\cal O}(10^{-2})$ and saturate. In the middle panel one sees  that $h_{v}^{\parallel}-h_{v}^{\perp}$ starts decreasing around $t\approx10$, owing to the growth of $h_{v}^{\perp}$. For the $v<0$ modes, after $t\approx15$ one has $h_{v}^{\parallel}-h_{v}^{\perp} < 0$ intermittently or  permanently, leading to the relaxation of corresponding modes (as seen in the right panel). For $v>0$ one has $h_{v}^{\parallel} > h_{v}^{\perp}$ so that $\mathsf{S}_v^{\parallel} \approx 1$ always. 

For $G^L_v[v_c = 0]$ with $A=0.2$ the relaxation is quicker, stronger, and more ubiquitous.  In Fig.\,\ref{figT2relax2}, one can see, the transverse components of the Bloch vectors at all locations start out in phase  (as set by the initial seeds in this case). By $t\approx4$ they start to get dephased relative to each other, though the transverse vector is still very small at most locations. By $t\approx6$, this dephasing is essentially complete. In Fig.\,\ref{Srelax2} the left panel shows the growth of the transverse components for $v=-0.5$. One sees that up to $t\approx4$ the evolution in linear. Yet, relative dephasing causes $|\langle {\sf S}^\perp_v \rangle|$ to become smaller than $\langle |{\sf S}^\perp_v |\rangle$. Around $t\approx4$, close to the time of the first dip of the ${\sf M}_1$ pendulum, nonlinearity sets in. The transverse components quickly grow to ${\cal O}(10^{-2})$ and saturate. In the middle panel one sees  that $h_{v}^{\parallel}-h_{v}^{\perp}$ starts decreasing around $t\approx4$, owing to the growth of $h_{v}^{\perp}$. After $t\approx8$ one has $h_{v}^{\parallel}-h_{v}^{\perp} < 0$ for all the velocity modes, leading to the relaxation of corresponding modes (as seen in the right panel). In other words, one finds that the depolarization of $\mathsf{S}_v^{\parallel}[z]$ occurs if and when one has $h_{v}^{\perp} \gtrsim h_{v}^{\parallel}$. We have found this expected correlation for all the ELNs we have considered in this paper.

\begin{figure}[t]
\includegraphics[height=0.63\columnwidth]{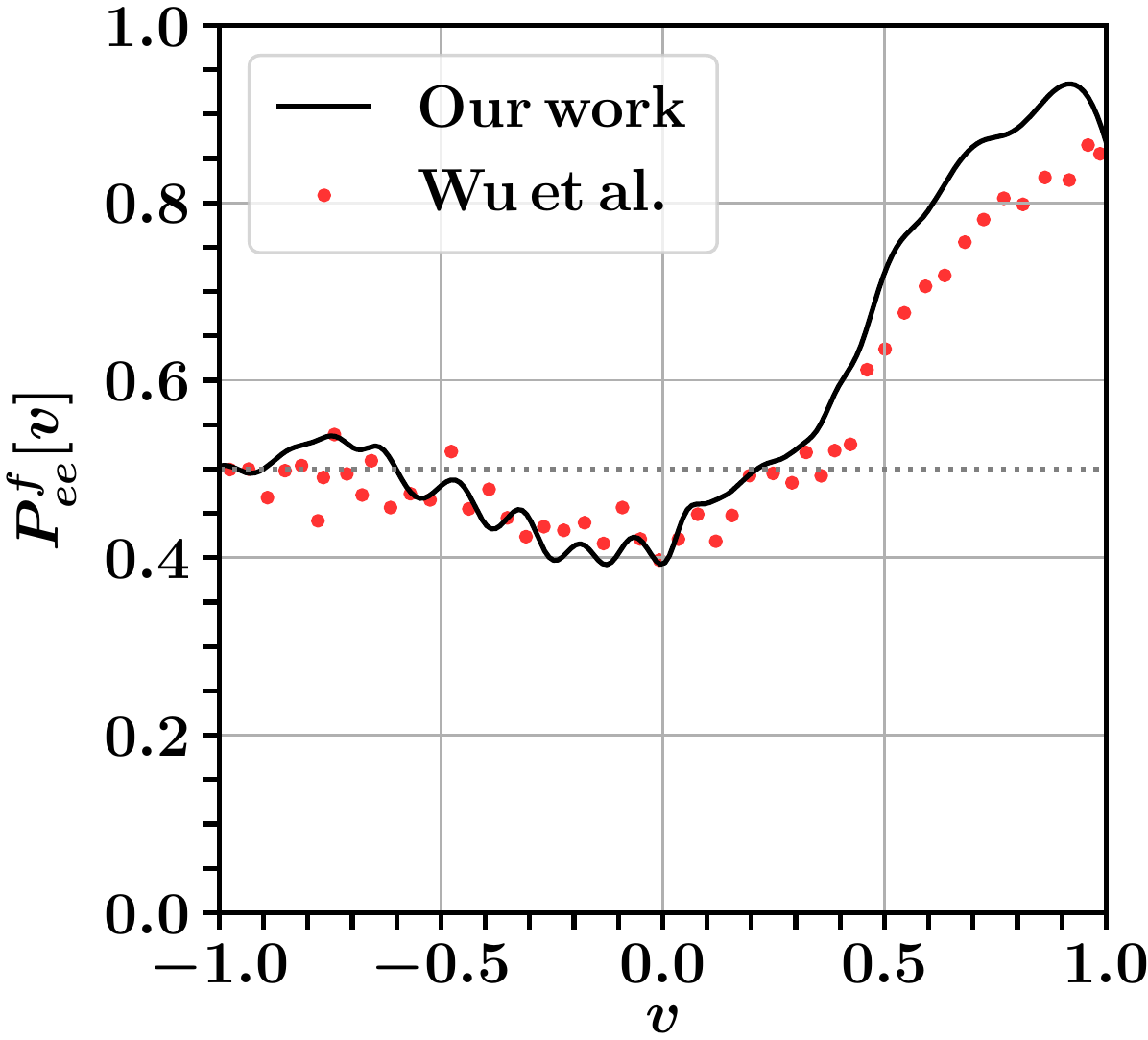}
	\caption{Comparison with Wu21\,\cite{Wu:2021uvt}: Survival probability for different velocity modes at late time. The red dots are obtained by digitizing the red curve in top panel of Fig.6 in Wu21\,\cite{Wu:2021uvt}. The black curve is our result for $g_{\nu}[v]-1.1g_{\bar{\nu}}[v]$, where $g_{\nu/\bar{\nu}}[v]$ is the same as given in Eq.(5) of Wu21\,\cite{Wu:2021uvt}.}
	\label{fig10}
\end{figure}
\begin{figure}[t]
\includegraphics[height=0.21\textwidth]{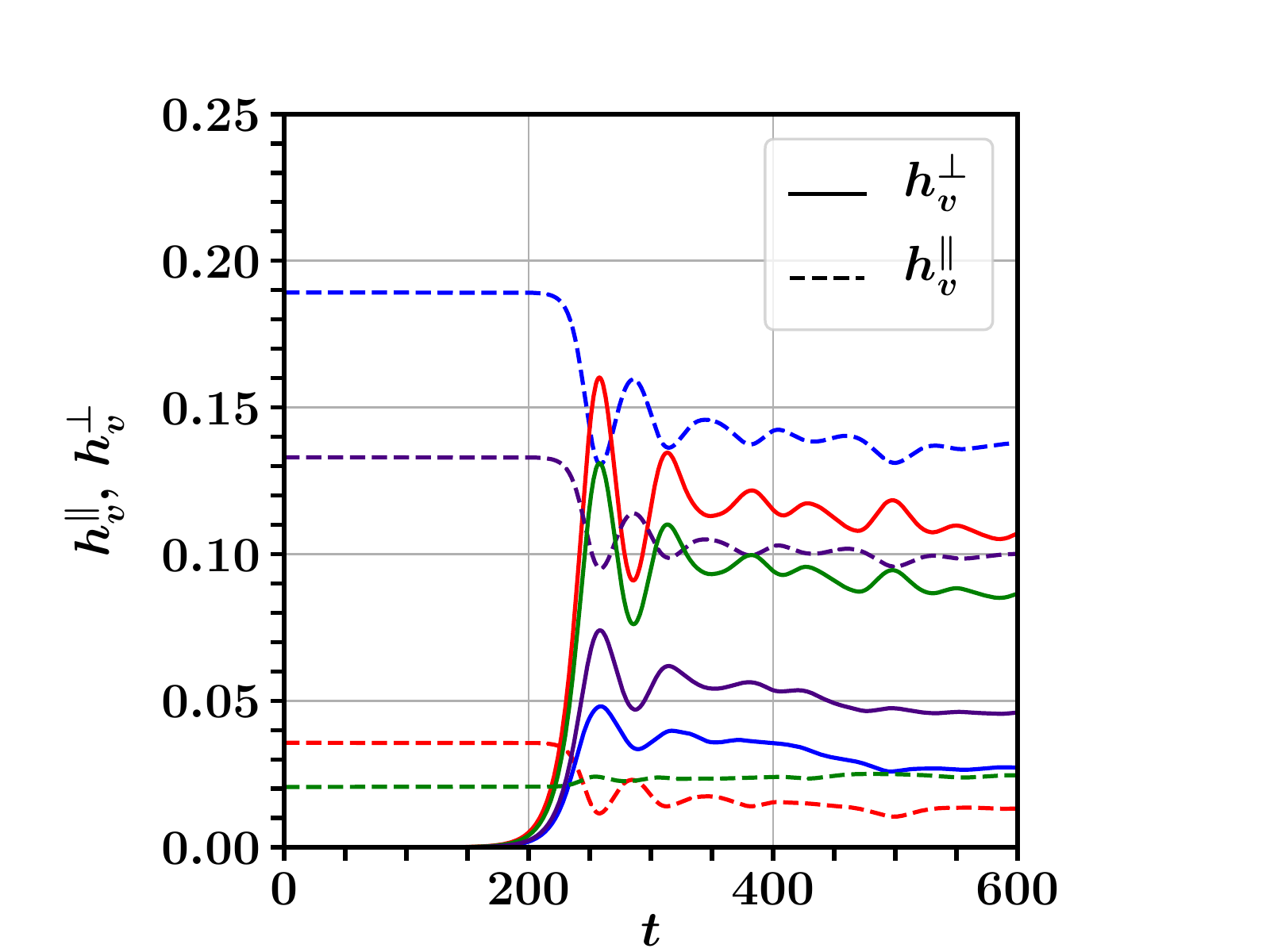}~
	\includegraphics[height=0.19\textwidth]{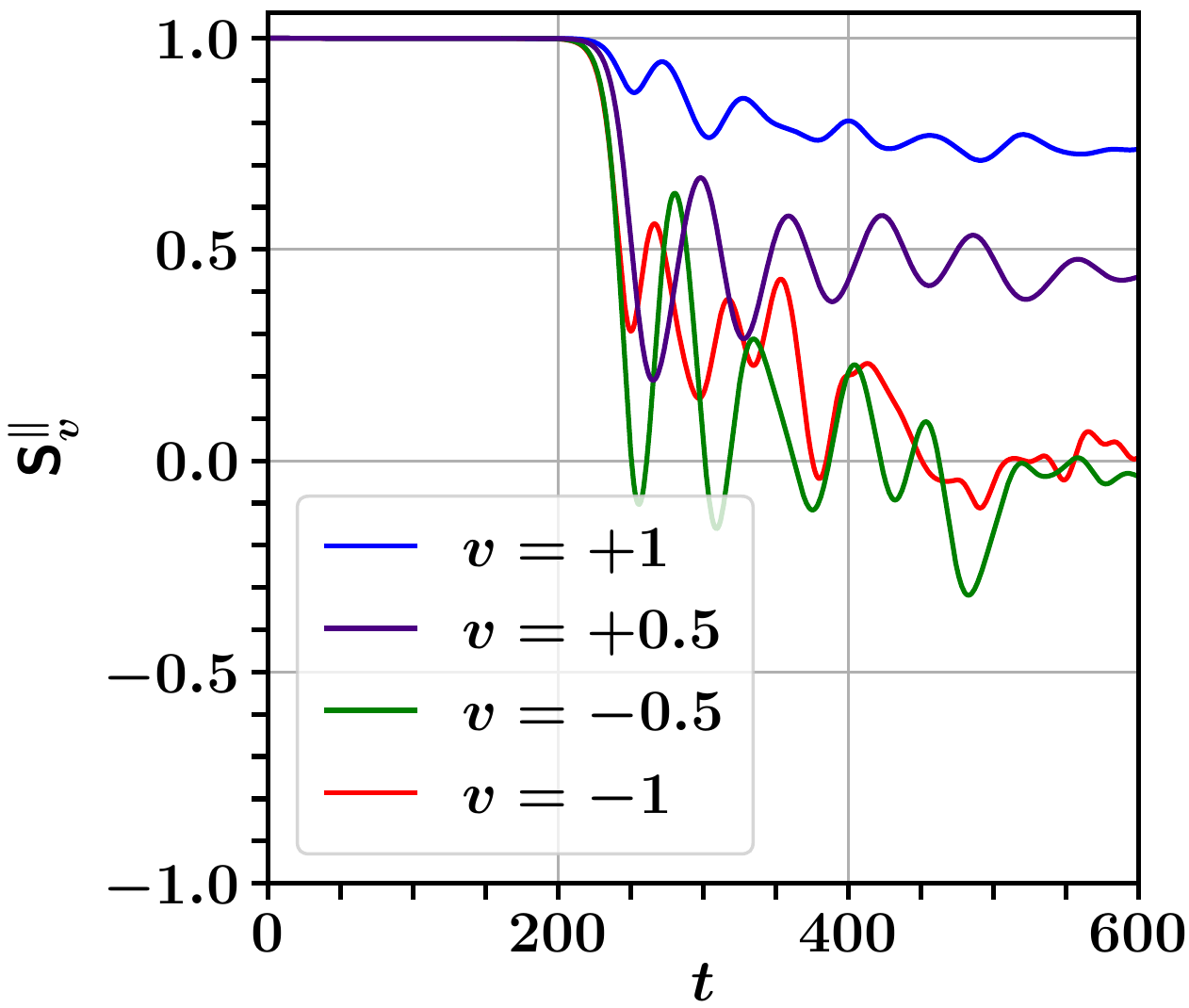}
	\caption{T2 relaxation in the ELN models considered in Wu21\,\cite{Wu:2021uvt}: $h_v^{\parallel}, h_v^{\perp}$ vs. $t$ in left panel and $\mathsf{S}_v^{\parallel}$ vs. $t$ in right panel. The ELN is same as in Fig.\,\ref{fig10}.}
	\label{fig11}
\end{figure}

In Wu21~\cite{Wu:2021uvt} doubts were raised whether the transverse relaxation mechanism holds in general, as they failed to find the correlation noted above. We investigated by repeating the computations of Wu21. Our results for the survival probability $P_{ee}$, as shown in Fig.\,\ref{fig10}, are in quite good agreement for the survival probabilities, showing partial depolarization.  However, unlike Wu21, we clearly see the correlation expected from T2 relaxation. As shown in Fig.\,\ref{fig11}, for $v<0$ modes one has $h_v^{\perp} \sim h_v^{\parallel}$ at around $t \sim 250$, and around the same time $\mathsf{S}_v^{\parallel} \to 0$. For $v = +1$ mode $h_v^{\perp} \ll h_v^{\parallel}$ but for $v = +0.5$ mode $h_v^{\perp} < h_v^{\parallel}$ but the difference is less compared to $v = +1$ mode and thus one finds almost no depolarization for $v = +1$ mode whereas partial depolarization for $v = + 0.5$ mode. In Wu21 this comparsion was not made in the corotating frame and $h_v^{\perp}$ was computed as the magnitude of the spatial average of the \emph{vectors} ${\sf h}_v^{\perp}[z]$ (which is always close to zero due to dephasing), as opposed to average of the \emph{magnitudes}, leading to their conflicting observation. Correcting for these misunderstandings, we find that the computations in B20a, B20b as well as in Wu21 are consistent with each other. The final depolarized state is almost entirely identical in both computations, and more importantly, the mechanism of T2 relaxation seems to work as we predicted when their example is analyzed as we recommended. There are minor differences because our code uses Fast Fourier Transform for differentiation in a way that creates ring-down effects around features that are sharp on the scale of the discretization scale\footnote{We thank Meng-Ru Wu and Zewei Xiong for helpful and collegial discussions that pinpointed this to us.}. Smoother initial conditions do not get affected by this. Despite this difference, our predictions for the final survival probability agree to better than 5\% r.m.s. error for the tested example.

\subsection{Multipole cascade}
\label{sec:MC}
The discussion in the previous subsection was limited to the first four multipole moments. In this subsection we review the nonlinear behavior of the higher multipole moments, as given in B20b\,\cite{Bhattacharyya:2020jpj}. Spatially averaging over Eq.\eqref{eom3} assuming periodic boundary conditions on $z$, and taking $n\gg1$, gives
\begin{align}\label{19}
\partial_{t}{M_{n}} = \frac{{M}_{1}}{2} \left(\partial^{2}_{n}{M}_{n}  + \frac{1}{n}\partial_{n} {M}_{n} \right)\,.
\end{align}
Note in our approximation $|\langle \mathsf{A} \cdot \mathsf{B} \rangle| \sim |\langle \mathsf{A} \times \mathsf{B} \rangle| \sim \langle A \rangle \langle B \rangle$ and $\langle \partial_{n, t} \mathsf{A}\rangle \sim  \partial_{n, t}\langle \mathsf{A}\rangle$. See Supplementary Material of Ref.\cite{Bhattacharyya:2020jpj} for the detailed derivation. Eq.\eqref{19} is a diffusion-advection equation where $n$ plays the role of space and $ M_1 $ is the diffusion coefficient. Note $M_n[t] = \frac{1}{L}\int M_n[z, t] \, dz $ is the net power present in each multipole $n$. Using the $n \to an$ and $t \to a^2t$ scaling invariance one can derive the solution for Eq.\eqref{19} as,
\begin{align}\label{20}
{M_{n}}[t] = c_{1}\,\text{Ei}\big[-{n^{2}}/\left({2 \, M_{1} t}\right)\big]+c_{2}\,,
\end{align}
where $c_1, c_2$ are integration constants with $\text{Ei}[x]=\int_{-\infty}^{x}dy\,e^{y}/y$. Eq.\eqref{19} and Eq.\eqref{20} together indeed indicate that there is a diffusion of the quantity ${M_{n}}[t]$ from {low to high $n$ multipoles} as time passes causing irreversibility in the system. Due to such leakage of power from smaller moments, $ {M_{n}}[t]$  for large $n$ starting from some initial value grows exponentially to peak roughly around $t^{\rm peak}_n \approx n^2/(2 \, M_{1})$ and then asymptotes to some steady final value at late times. Note $t^{\rm peak}_n$ increases with $n$. In B20b\,\cite{Bhattacharyya:2020jpj}, we had shown this for the box-type ELNs denoted here by $G_v^B$. We have now verified it hold for all the ELNs considered in this paper. An example is shown in the top panel plot of Fig.\ref{figdepol1n} with $G_v^L$ for $v_c = 0$ and $A = 0.2$. The takeaway is that the flavor difference increasingly gets moved to high multipoles. If a physical process does not distinguish closely spaced momentum modes, it no longer sees the flavor difference stored in high-multipoles.

\subsection{Mixing of flavor waves}
\label{sec:BFW}
Flavor waves also cascade to smaller distance scales, similar to the cascading to smaller momentum scales we just discussed.  This was shown very clearly in R21a~\cite{Richers:2021nbx} and R21b~\cite{Richers:2021xtf}. To understand this we take the Fourier transform of Eq.\eqref{eom2} using $\mathsf{S}_v[k, t] = \int_{0}^{L} e^{ikx} \mathsf{S}_v[x, t] dx$ to rewrite the following equation: 
\begin{equation}\label{21n}
\begin{split}
\left({\partial_{t}}+ivk\right)\mathsf{S}^{\perp}_{v}[k, t] = i\mu_{0}\int_{-\infty}^{+\infty} dk^{'} \int_{-1}^{+1} dv{'}G_{v'}\left(1-vv{'}\right) \\  \left(-\mathsf{S}_{{v}}^{\perp}[k^{'}, t]\mathsf{S}_{{v}^{'}}^{\parallel}[k-k^{'}, t] + \mathsf{S}_{{v}}^{\parallel}[k^{'}, t]\mathsf{S}_{{v}^{'}}^{\perp}[k-k^{'}, t]\right)\,.
\end{split}
\end{equation}
Initially, in the linear regime,  one has $\mathsf{S}_v^{\parallel}[k^{'}, t] \approx 1$ for all $k'$. Thus, in Eq.\eqref{21n}, $\mathsf{S}_v^{\perp}[k, t]$ for different $k$-modes evolve independently. When the system reaches nonlinearity, $\mathsf{S}_v^{\parallel}[k^{'}, t]$ and $  \mathsf{S}_{v^{'}}^{\parallel}[k-k^{'}, t]$ start deviating from unity, and the different $k$-modes get coupled.

\begin{figure}[t]
	\hspace{1cm}\includegraphics[width=0.85\columnwidth]{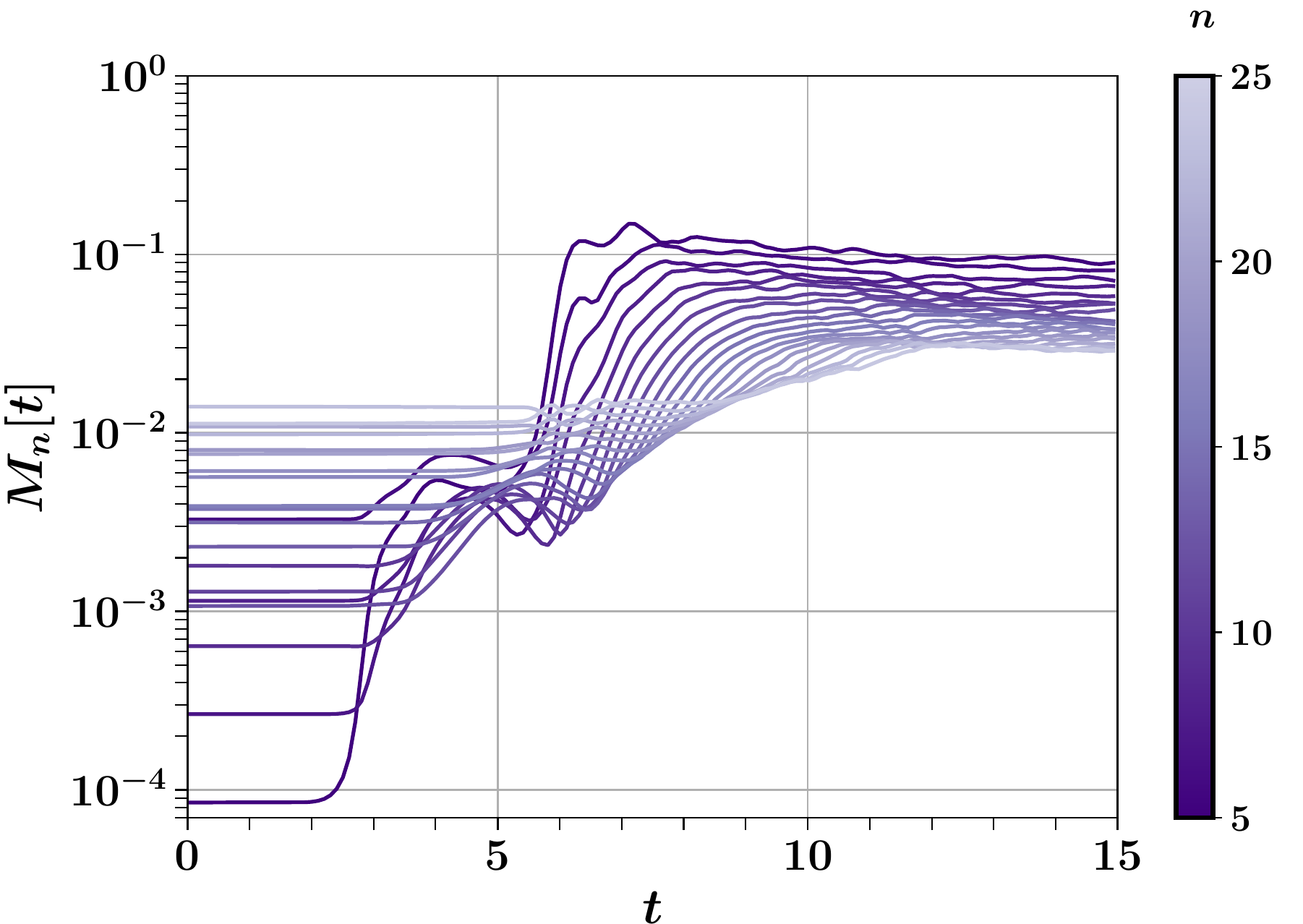}\\[2ex]
	\includegraphics[width=0.79\columnwidth]{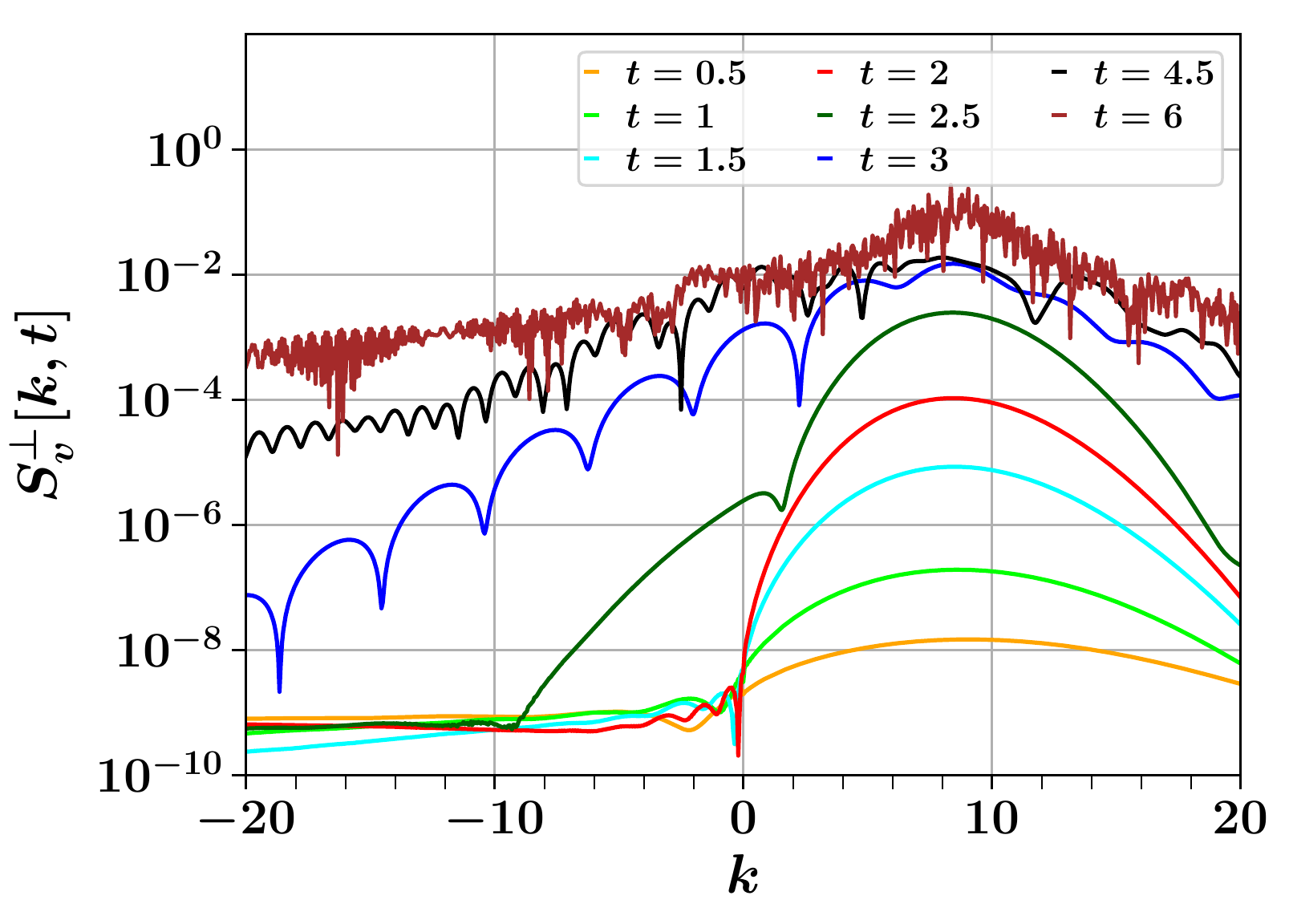}
	\caption{Mixing in phase space: Top panel shows $M_n[t]$ vs. $t$ and bottom panel shows $S_v[k, t]$ vs. $k$ at six different times for $v=-0.5$, for $G_v^L$ with $v_c = 0$ and $A = 0.2$. One sees multipole cascades in velocity space (in top panel) and $k$-mode mixing indicated by development of wiggles (in bottom panel). All solutions are computed using Eq.\eqref{eom2}.}
	\label{figdepol1n}
\end{figure}

We check this behavior by plotting ${S}_{v}[k, t]$ as a function of  various $k$ modes at different time $t$, as shown in the bottom panel of Fig.\,\ref{figdepol1n}. We see that for $t \leq 2$ the system is in the linear regime and the power, defined to be $S_v^{\perp}[k, t]$, for a specific $k$ mode does not cascade to other $k$ modes. Until about $t=2$, each curve grows with time exponentially for each $k$ but with its characteristic $k$-dependent linear growth rate ${{\rm Im} \, \Omega[k]}$. In the linear regime one can clearly see that the footprint of instability is limited to the $k$-modes between $\sim(5-15)$ for our chosen example. By $t\approx 2.5$ the modes close to $k\approx8$ have become large and they start affecting the growth of  modes close to $k \sim 0$, enhancing them considerably. This sudden distortion is a signature of mode-coupling in Eq.\eqref{21n}. Further, mode-coupling also allows the large-$|k|$ modes with smaller amplitude to grow in a cascade, at the expense of the modes that start with higher amplitudes ,and thus spread the flavor instability to almost all $k$ modes. This moves the flavor differences to smaller and smaller distance scales. If a physical process does not distinguish closely spaced locations, it does not see the flavor difference that is now stored in very high-$|k|$ modes. Multipole diffusion and mode-coupling, together create extremely fine structures in the phase space, which upon coarse graining present themselves as effective depolarization.

\begin{figure}[t]
	\includegraphics[width=0.48\columnwidth]{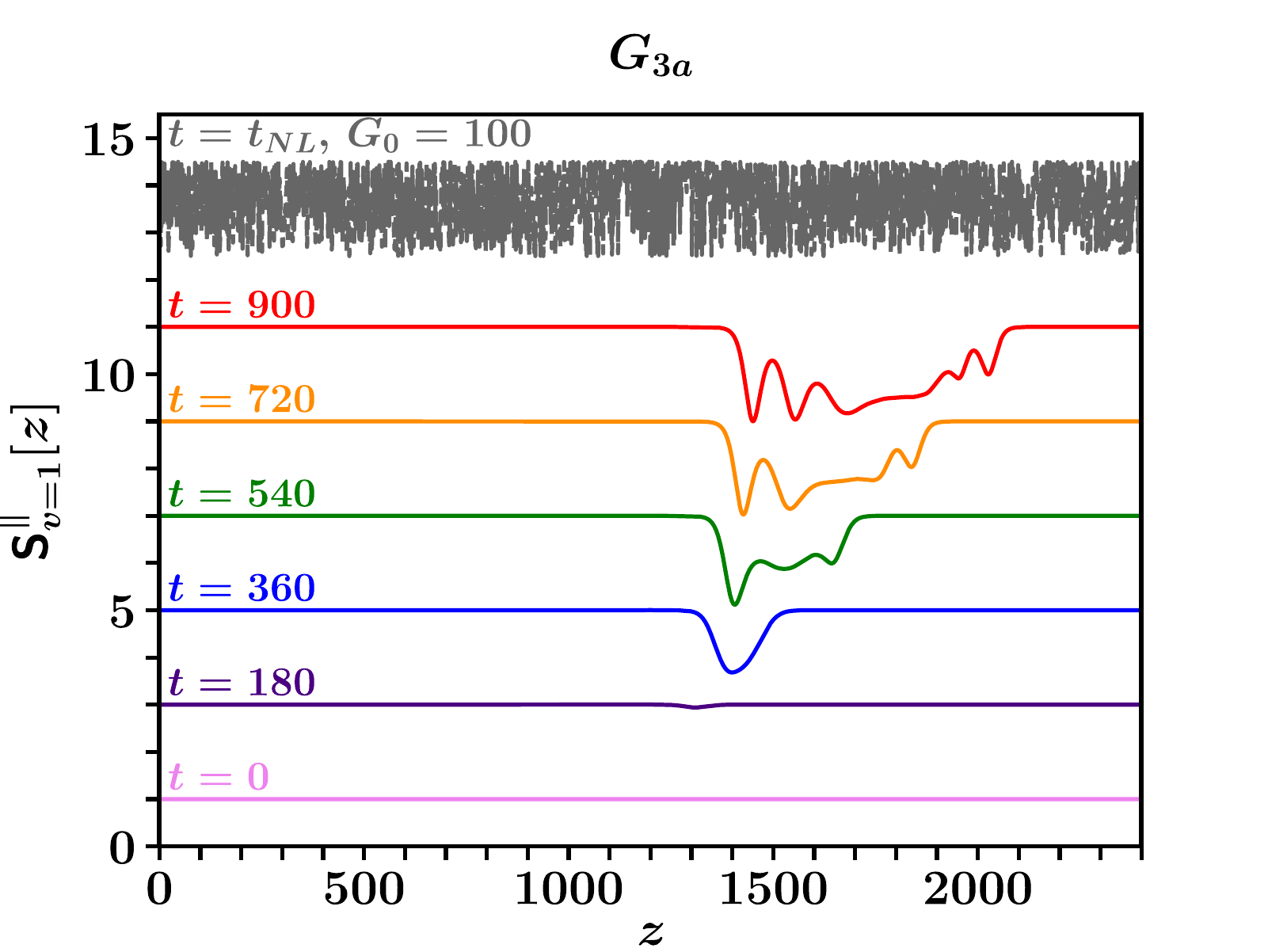}
	\includegraphics[width=0.48\columnwidth]{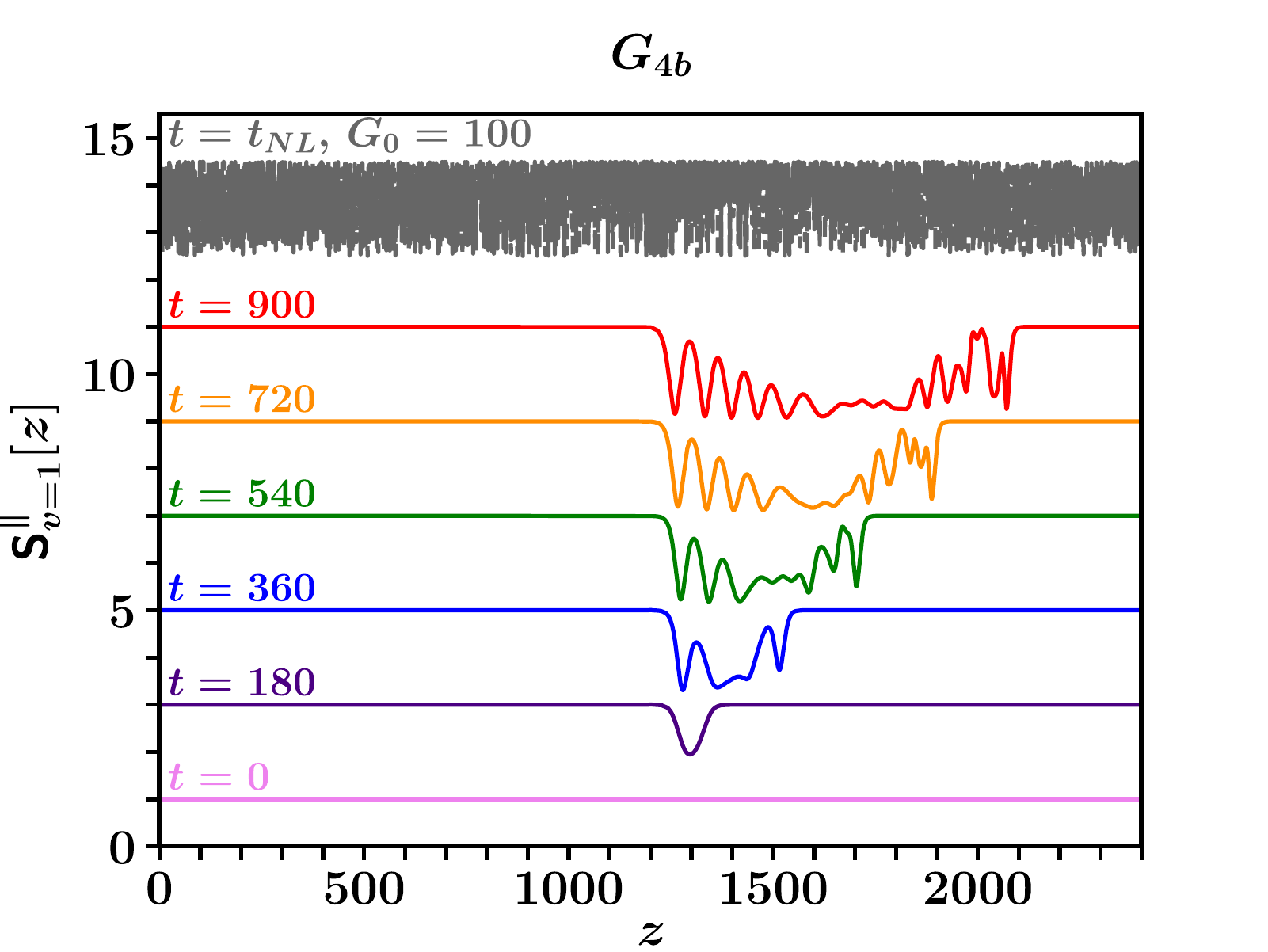}\\[2ex]
	\includegraphics[width=0.48\columnwidth]{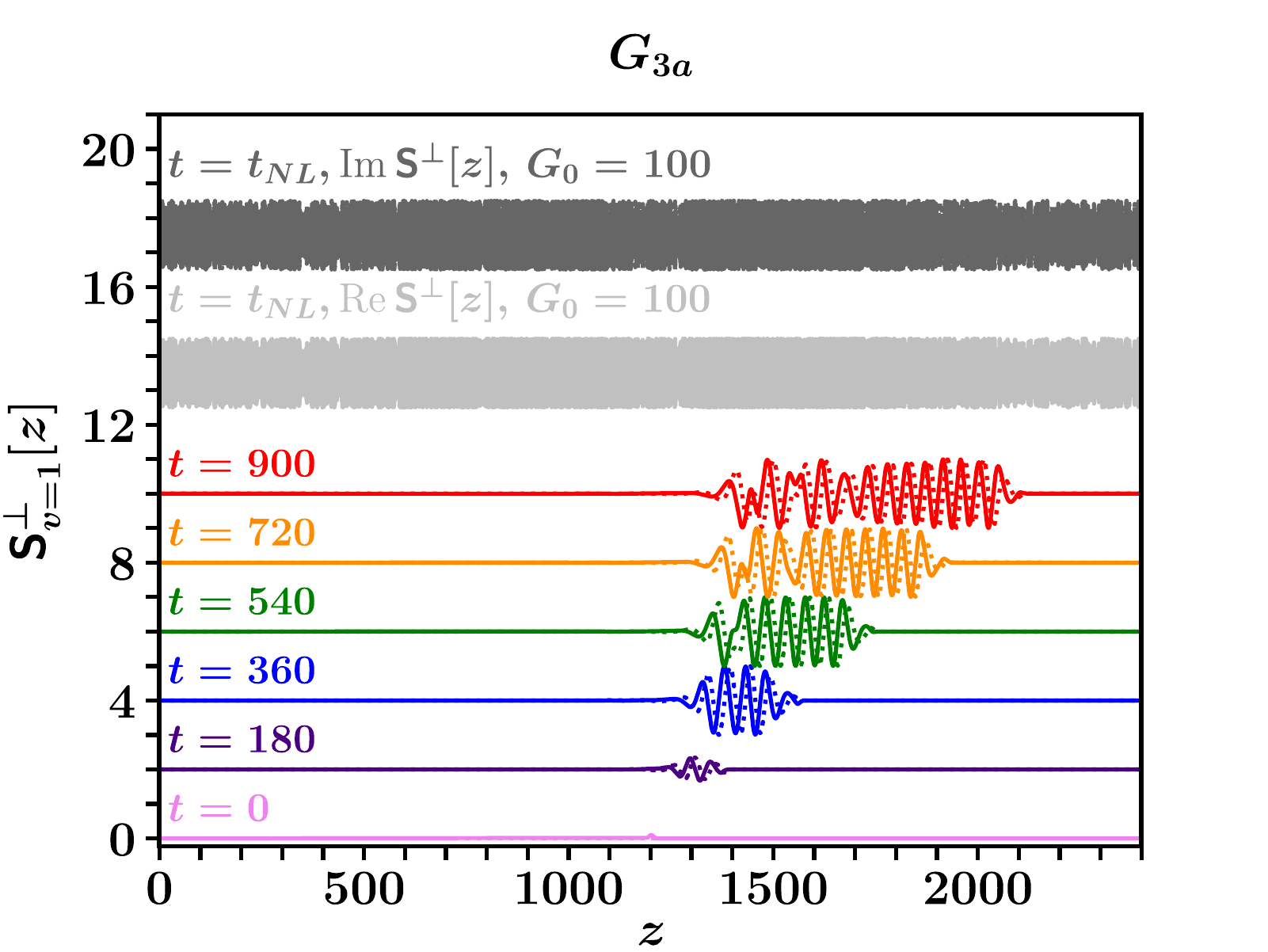}
	\includegraphics[width=0.48\columnwidth]{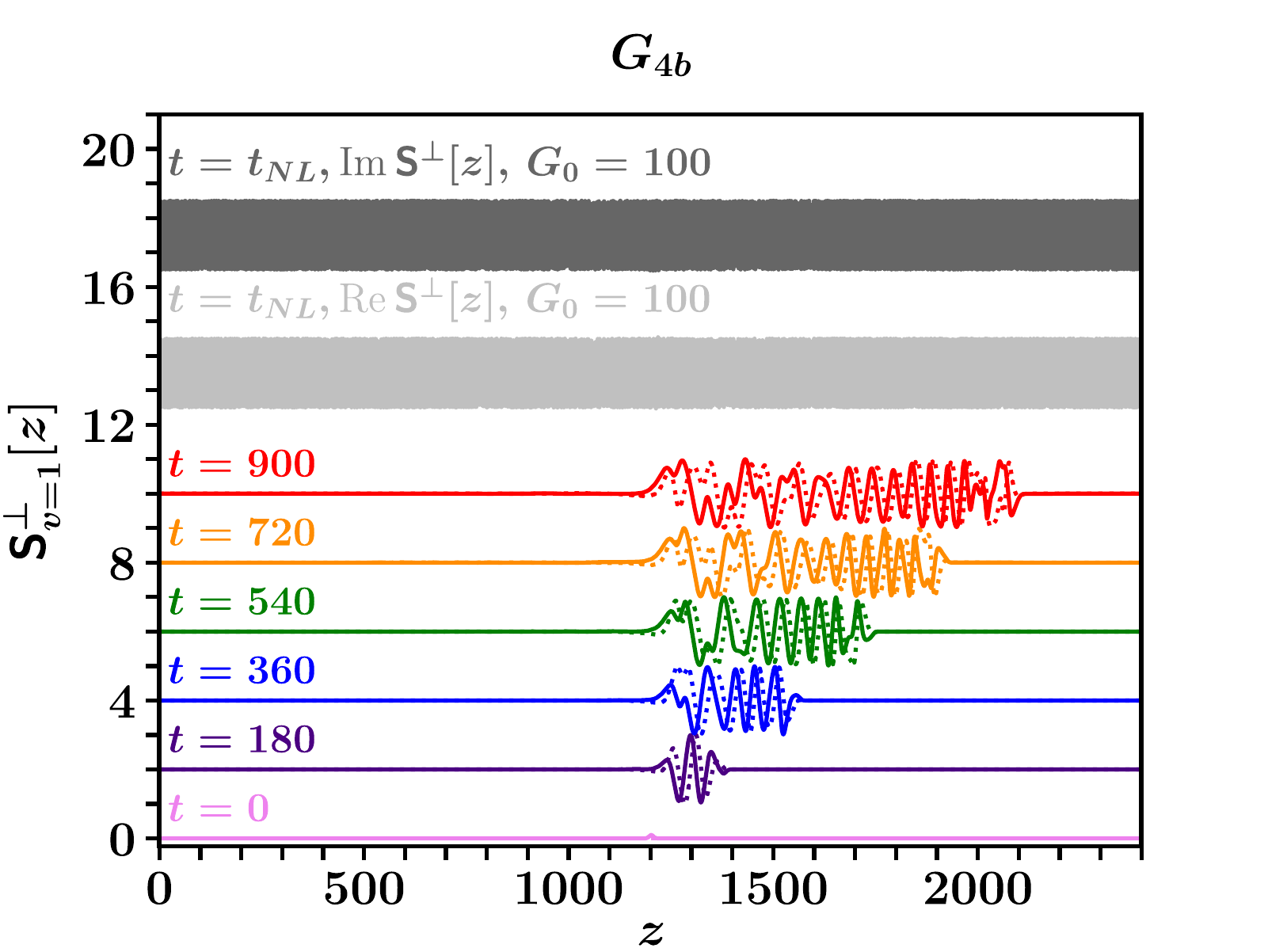}\\[2ex]
	\includegraphics[width=0.48\columnwidth]{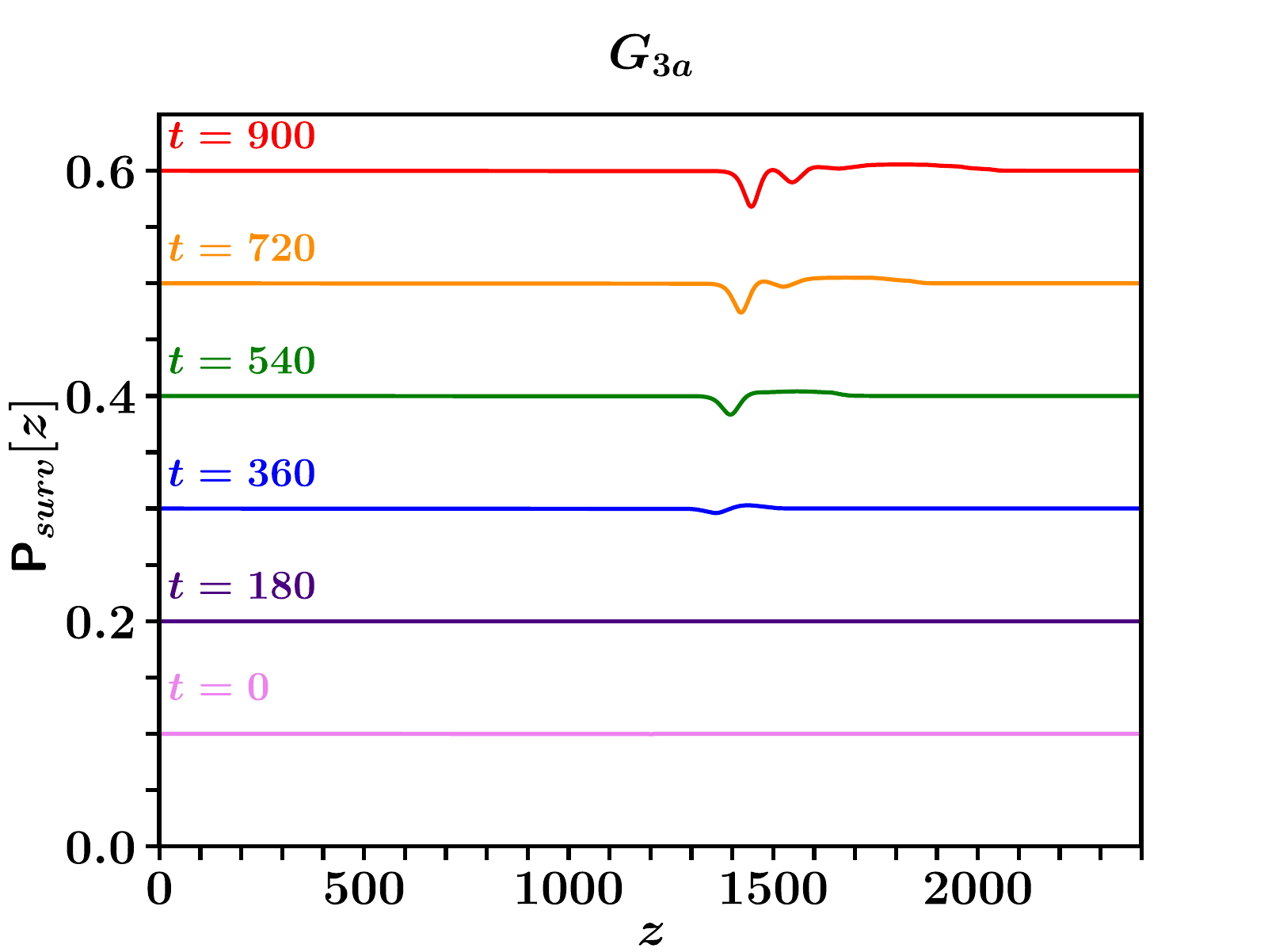}
	\includegraphics[width=0.48\columnwidth]{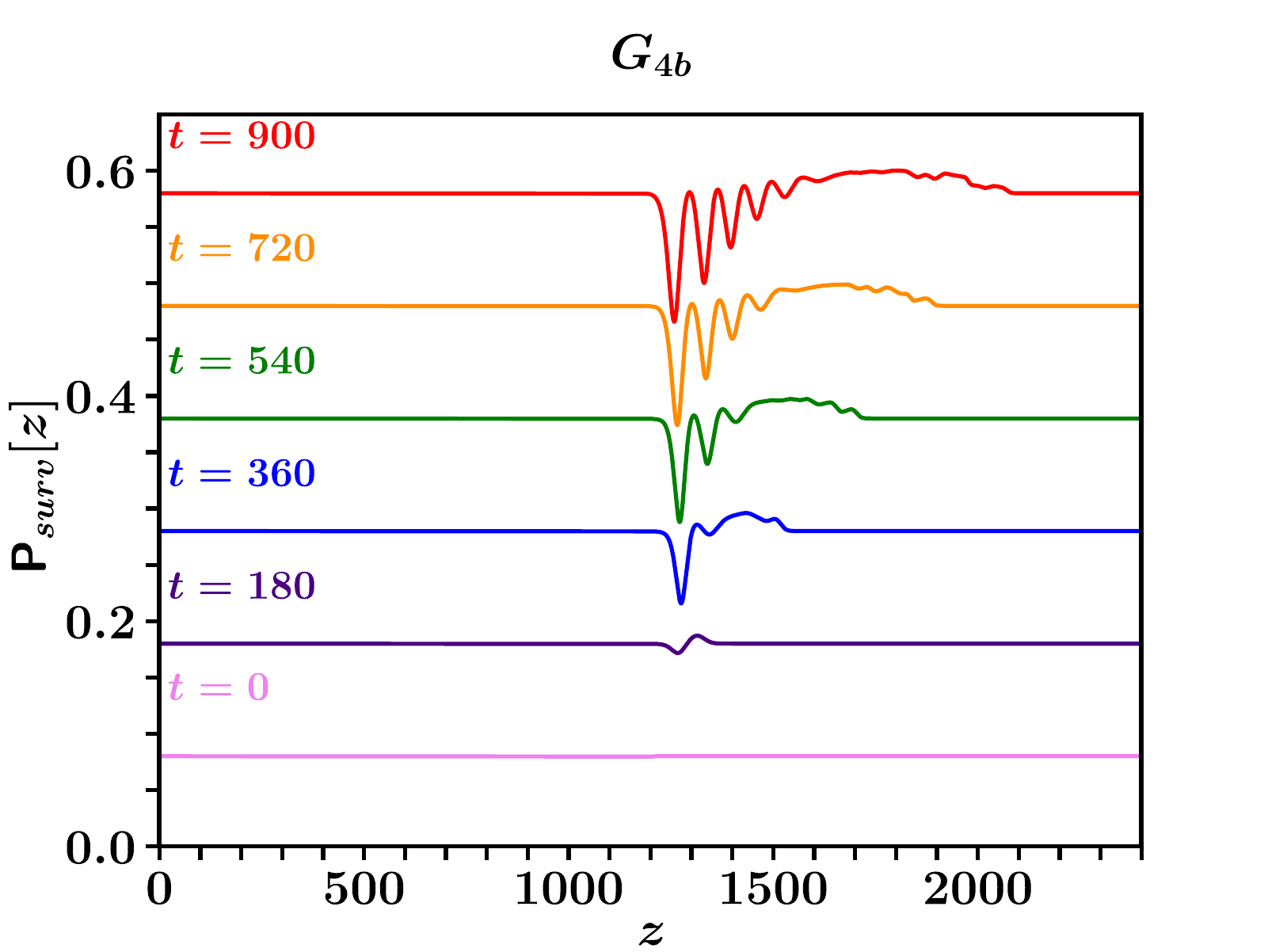}
	\caption{Comparison with Fig.3 of M20\,\cite{Martin:2019gxb}$: \mathsf{S}_{v=1}^{\parallel}[z]$ as a function of $z$ in top panel and $\mathsf{S}_{v=1}^{\perp}[z]$ in the middle panel (with real part as solid and imaginary part as dotted), and $\mathsf{P}_{surv}[z]$ in the bottom panel. For legibility, the curves are vertically offset from each other; by 2 in the top and middle panels and 0.1 in the bottom panel. The left and right panels are for $G_{3a}$ and $G_{4b}$, respectively. The curves above $t = 900$ shown in the top (dark gray line) and middle panel are for the same ELN scaled by a factor $G_0 = 100$ at $t_{NL}=12$, showing the state at late times.}
	\label{fig9}
\end{figure}

\begin{figure}[t]
	\includegraphics[width=0.75\columnwidth]{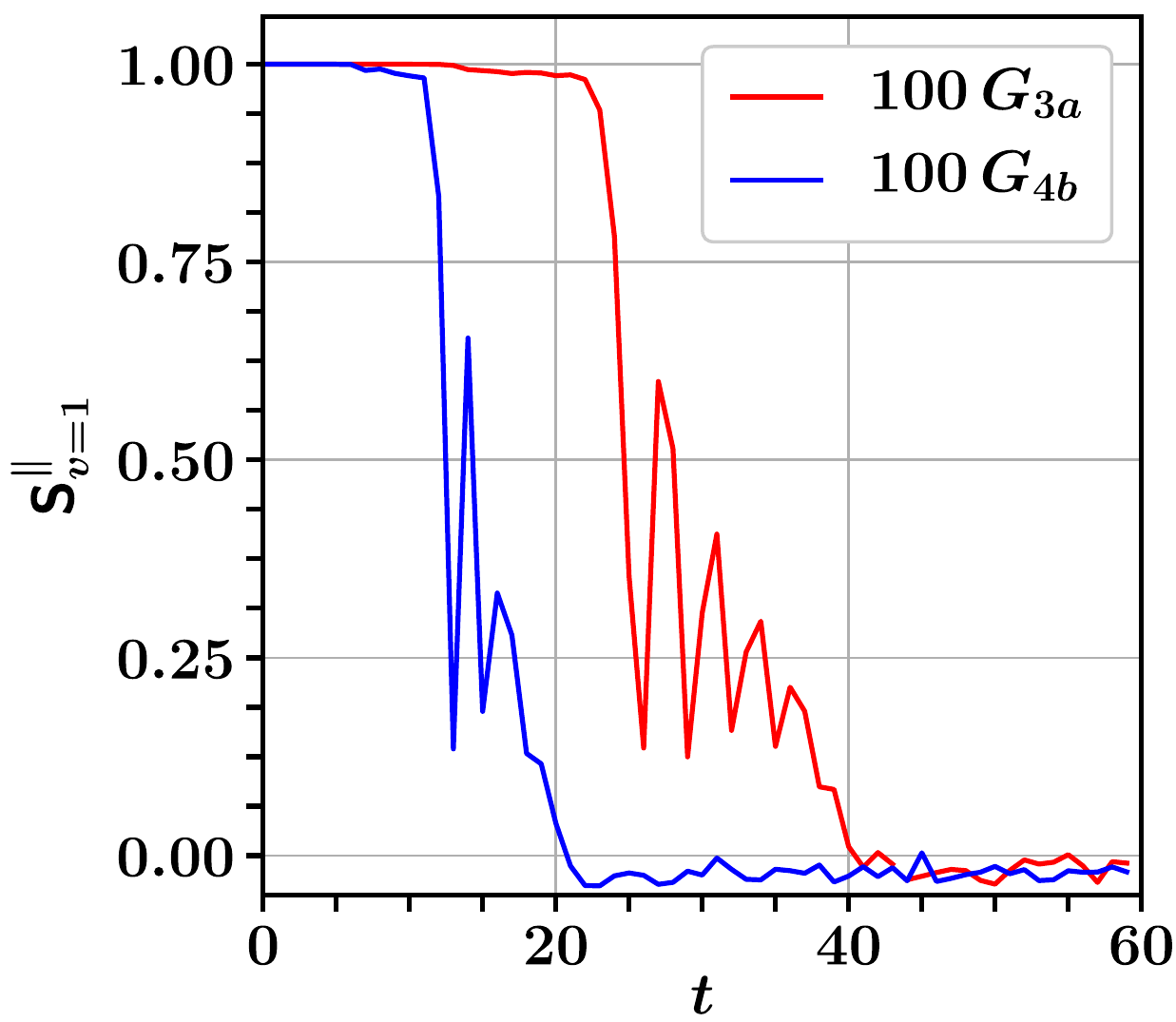}
	\caption{Depolarization for the ELNs inspired by M20\,\cite{Martin:2019gxb}: $\mathsf{S}_{v = +1}^{\parallel}$ vs. $t$ for the ELNs $100\, G_{3a}$ and $100\, G_{4b}$.}
	\label{fig10a}
\end{figure}

\subsection{Flavor waves vs. depolarization}
\label{sec:3E}
In M21~\cite{Martin:2021xyl}, the authors speculated that the simulation tools used in our previous work in B20a\,\cite{Bhattacharyya:2020dhu} and B20b\,\cite{Bhattacharyya:2020jpj} may have failed to maintain causality. This speculation stemmed from the persistence of wavelike numerical solutions found in M20\,\cite{Martin:2019gxb} and M21\,\cite{Martin:2021xyl}, as opposed to a depolarized state. In the mean time, other groups have found results that are broadly consistent with depolarization seen in our previous works (see e.g., \cite{Wu:2021uvt, Richers:2021xtf, Richers:2021nbx}). Here, we reproduce the key results of M20~\cite{Martin:2019gxb}, to show our code produces results identical to theirs, if restricted to the regime they have explored. If extended to longer times, one finds depolarization.

To benchmark our code against the calculation in M20\,\cite{Martin:2019gxb}, we focus on their $G_{3a}$ and $G_{4b}$ ELNs.  Our results for $\mathsf{S}^{\perp}_{v = 1}[z]$, $\mathsf{S}_{v = 1}^{\parallel}[z]$ and $\mathsf{P}_{surv}[z] = \int_{-1}^{1} G_v \mathsf{S}_v^{\parallel}[z] \, dv$ as a function of $z$ at various time snapshots up to $t = 900$ are shown in Fig.\,\ref{fig9}. The results agree, with excellent fidelity, with their counterparts in Fig.3 of M20\,\cite{Martin:2019gxb}. One clearly notices flavor waves in space and the region over which they exist extends with time as they propagate. Note the flavor waves show convective and absolute nature for $G_{3a}$ and $G_{4b}$, respectively.

\begin{figure*}[t!]	
	\includegraphics[height=0.6\columnwidth]{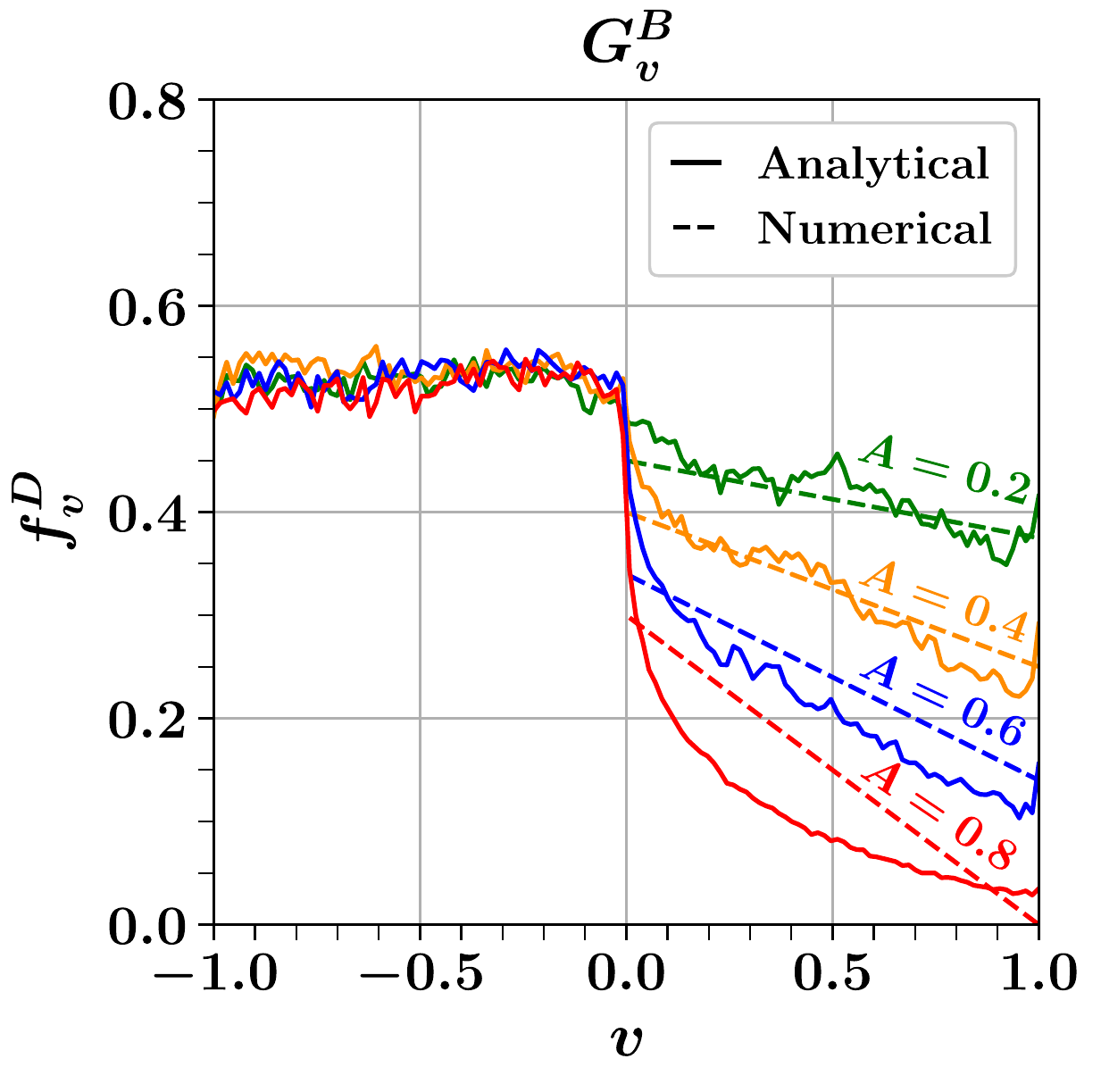}
	\includegraphics[height=0.6\columnwidth]{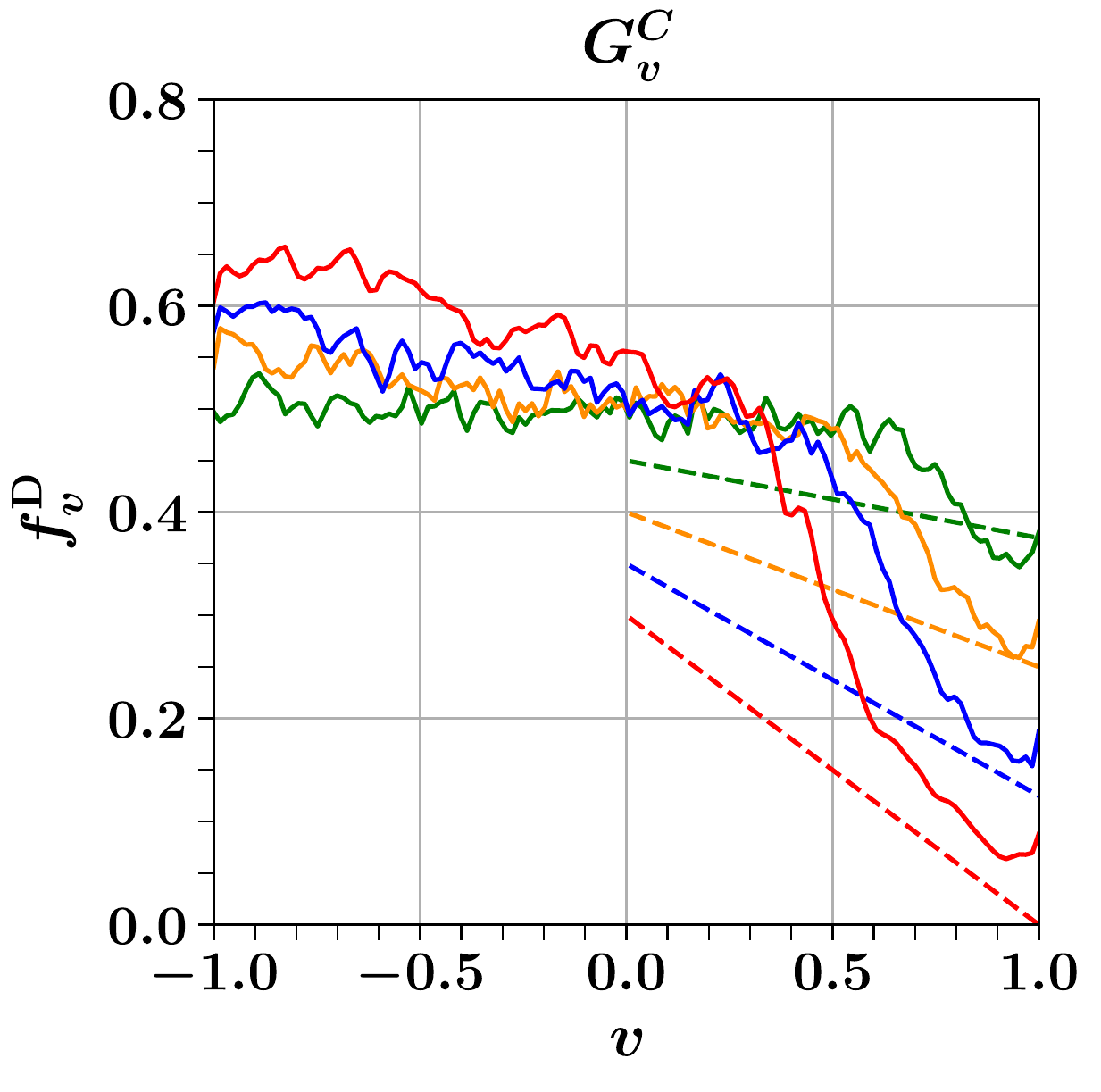}\\
	\includegraphics[height=0.6\columnwidth]{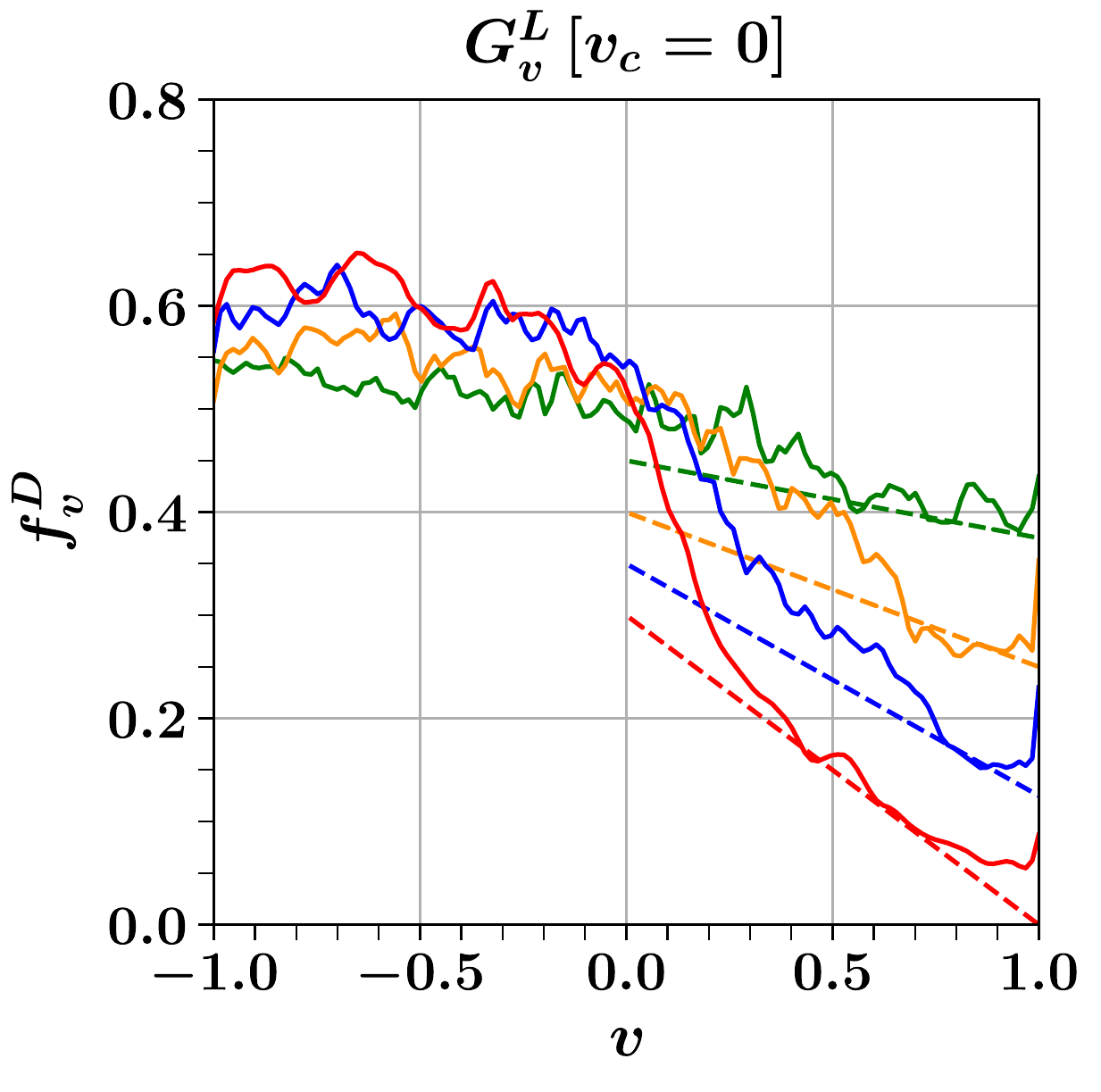}
	\includegraphics[height=0.6\columnwidth]{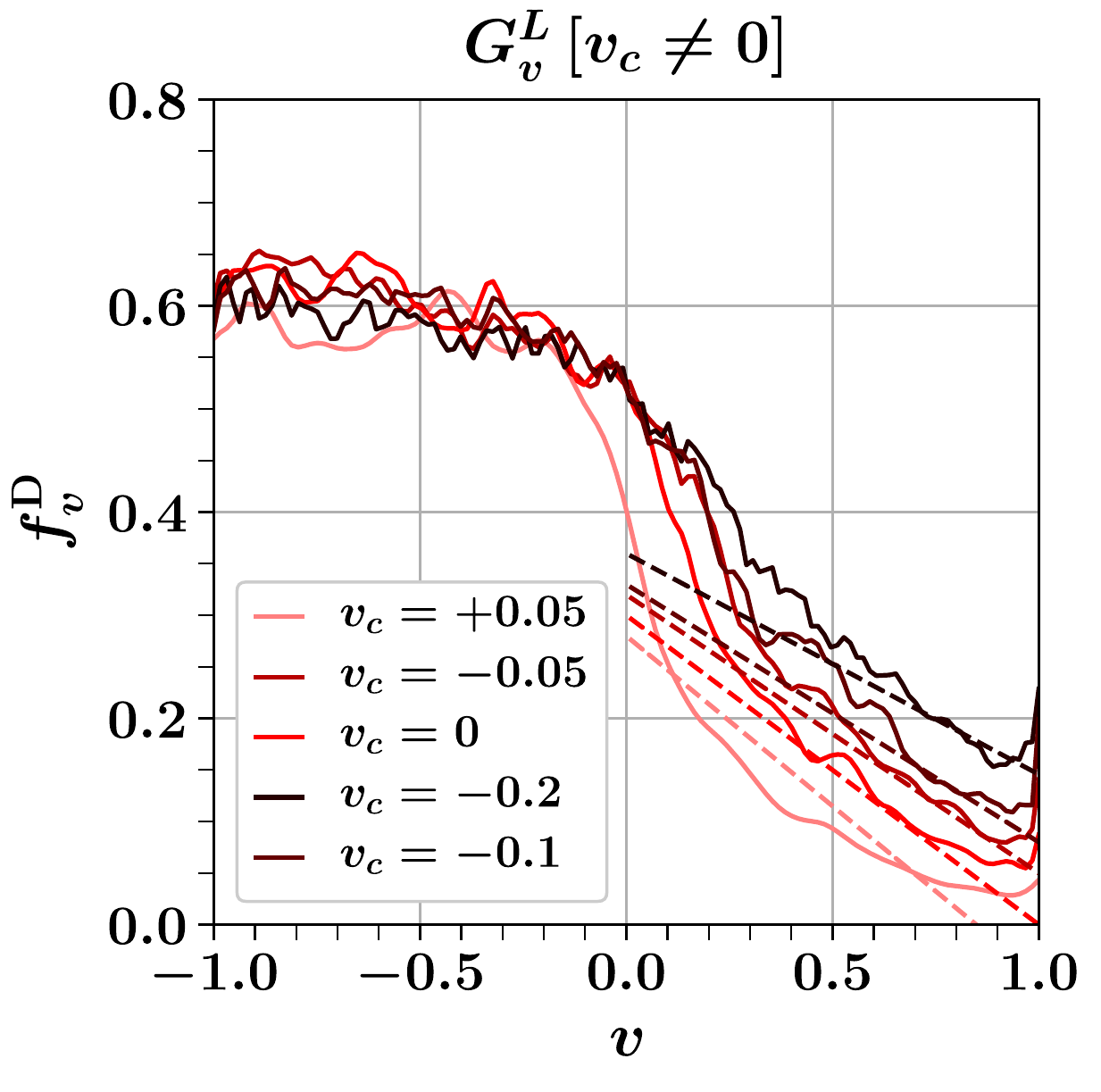}
	\caption{Extent of depolarization: $f_v^{\rm{D}}$ vs. $v$ at $t = t_f \approx 30$ in bottom-left for $G_v^L[v_c =0]$, in top-left for $G_v^B$, in top-right for $G_v^C$ and in bottom-right for $G_v^{L}[v_c \neq 0]$. Various colors indicate the different $A$ value in the first three plots but different $v_c$ values in the bottom-right plot. The continuous and dashed lines represent the numerical and analytical solutions, respectively.}
	\label{figdepol1}
\end{figure*}

However, we believe that two important issues were ignored in in M20\,\cite{Martin:2019gxb} and M21\,\cite{Martin:2021xyl}. Firstly, the numerical results were shown only up to $t = 900$ when the system does not reach sufficient nonlinearity. Secondly, the quantities  were not coarse-grained over a spatial volume. Both of these were important to obtain the irreversible steady state depolarized solution in our previous work. To clarify these two points, we scale up the neutrino ELNs $G_{3a}$ and $G_{4b}$ by a factor $G_0 = 100$ {(i.e., instead of $G_{3a}$ and $G_{4b}$, we consider the ELNs to be $100\times G_{3a}$ and $100\times G_{4b}$, respectively)} and otherwise retain exactly the same specifications, {i.e., same box size, spatial discretization, initial condition, boundary condition, and so on}, as in M20\,\cite{Martin:2019gxb}. Since with $G_0 = 1$ both of the examples show a linear growth rate of ${\cal O}(10^{-2})$, choosing $G_0= 100$ makes the initial flavor evolution 100 times faster. Thus, instead of $t \sim {\cal O}(100)$ the systems now reach nonlinearity roughly around $t_{NL} \sim {\cal O}(1)$. With this scaled ELN, we check for the extreme nonlinear behavior of the solution. In the top most curves for the top and middle panels of Fig.\,\ref{fig9} we show our numerical results for $\mathsf{S}_{v=1}^{\parallel}[z]$ (dark gray) and $\mathsf{S}_{v=1}^{\perp}[z]$ (dark or light gray lines) as a function of $z$ choosing $G_0 = 100$ and $t_{NL} = 12$. One can clearly see that flavor waves break down at after reaching nonlinearity. {Note that this respects $L/2 \gg t_{NL}$, required to avoid boundary effects due to the periodic boundary condition at late times}. We show $\mathsf{S}^{\parallel}_{v = 1}$ (after spatial averaging) vs. $t$ in Fig.\,\ref{fig10a}, which shows that the system indeed reaches a flavor depolarized steady state.

\section{Flavor Depolarization}
\label{sec:range}

To quantify the amount of flavor depolarization we define a depolarization factor in the following way:
\begin{equation}\label{23}
	f_v^{\rm{D}} = \frac{1}{2}\, \left(1-\frac{\mathsf{S}_v^{\parallel}[t_f]}{\mathsf{S}_v^{\parallel}[0]} \right)\,.
\end{equation}
Note that $t_f$ is chosen to be large enough, such that the system has reached steady state. Full flavor depolarization leads to $f^\text{D}_{v} = 0.5$, whereas no depolarization is given by $f^\text{D}_{v} = 0$, and partial depolarization by $f^\text{D}_{v}$ between $0$ to $0.5$. Sometimes one may find $f^\text{D}_{v}>0.5$. This happens because the system first changes flavor almost completely, corresponding to a flavor conversion probability of 1, and then depolarized partially. 

We show our numerical solution for $f^\text{D}_{v}$ as a function of different velocity modes $v$  in Fig.\ref{figdepol1} considering $G_v^B$, $G_v^L$ and $G_v^C$ for various choices of $A>0$. Our numerical analysis suggests that depolarization is velocity-dependent: the negative velocity modes are almost always fully flavor depolarized for $A>0$, but the positive ones are partially flavor depolarized. The extent of partial flavor depolarization depends on lepton asymmetry $A$, zero crossing position of neutrino angular distributions $v_c$.

\subsection{Extent of depolarization}
In this subsection we analytically explain the functional dependence of $f^\text{D}_{v}$ on $A, v, v_c$ and give an explicit linearized formula for $f^\text{D}_{v}$ in terms of quantities determined from initial conditions. To derive this we use the numerical observation that $\mathsf{S}_v^{\parallel}[t_f] \approx 0$, i.e., $f^\text{D}_{v} \approx \frac{1}{2}$, for $v < 0$, in all four cases based on our numerical analysis in Fig.\ref{figdepol1}. This assumption, for  $A>0$,  is motivated by our qualitative understanding of which modes get more depolarized. Using this and enforcing lepton number conservation $\int_{-1}^1 dv \, G_v \mathsf{S}_v^{\parallel} = A$, to zeroth order in $v$ one can write $\mathsf{S}_v^{\parallel}[t_f] \approx \frac{A}{\gamma_0}$ for $v>0$ modes where we define the ``forward'' moments of the ELN as
\begin{equation}
\gamma_n = \int_{0}^1 v^n dv\, G_v\,.
\label{eq:ELNmom}
\end{equation}
To obtain the linear order correction to the above result, we expand $\mathsf{S}_v^{\parallel}[t_f]$ as a function of $v$ as
\begin{align} \label{SecB24}
\mathsf{S}_v^{\parallel}[t_f] \approx \frac{{s}_0}{2} + \frac{3{s}_1}{2} \, v \,,
\end{align}
where ${s}_0$, ${s}_1$ are space-time independent constants but can depend on $A$  and the nature of $G_v$. Note ${s}_0$, ${s}_1$ can be determined from the following formula:
\begin{align}\label{SecB241}
	{s}_n =  \int_{-1}^{1} \mathsf{S}_v^{\parallel}[t_f] L_{n}[v] dv\,.
\end{align}
For our chosen form of $G_v$, with $A>0$ and a forward excess, we use $\mathsf{S}_{v>0}^{\parallel}[t_f] \approx \frac{A}{\gamma_0}$ and $\mathsf{S}_{v<0}^{\parallel}[t_f] \approx 0$ to deduce ${s}_0, {s}_1$ from Eq.\eqref{SecB241} as
\begin{align}\label{31}
 {s}_0 & \approx  \frac{A}{\gamma_0}\,,
\end{align}
and
\begin{align}\label{32}
 {s}_1 & \approx  \frac{A}{2\gamma_0}\,.
\end{align}
Using Eqs.\eqref{SecB24},\eqref{31}, and \eqref{32} we can  write $f_v^{\rm{D}}$ as
\begin{align}\label{29}
f^\textrm{D}_{v}  \approx 
\begin{cases}
\frac{1}{2}-\frac{A }{4\, \gamma_0}-\frac{3 A}{8 \gamma_0}\, v & \text{if ${\phantom-}1\geq v\geq 0$}\,,\\
\frac{1}{2} & \text{if $-1 \leq v \leq 0$}\,.
\end{cases}
\end{align} 
In case of $G_v^B, G_v^C$ we find $\gamma_0 = 1$ but for $G_v^L$, $\gamma_0 = 1-2v_c$. Clearly the functional dependence in Eq.\eqref{29} indicates $f^\textrm{D}_{v}$ for $v>0$ modes decreases with increase in $A$ and decrease in $\gamma_0$ (or in other words $v_c \to 1$). Plugging in the values for $\gamma_0$, $A$ and $v_c$ we get a good agreement between the numerical and analytical solution of $f^\textrm{D}_{v}$ as a function $v$ with $v > 0$ modes for all the cases except $G_v^C$ as shown in Fig.\,\ref{figdepol1}. For cubic ELN, our linear approximations used in the above derivation might be inapppropriate since $G_v^C$ itself contains only terms higher than linear order. Also, even for $v<0$ modes, $\mathsf{S}_v^{\parallel}[t_f] $ to $0$ is not entirely correct as we see. However, for reasonable values of asymmetry $A\approx0.2$, our prescriptions seems to work quite reasonably because the naive equilibration hypothesis with $f^\textrm{D}_{v}=0.5$ for all modes is already a good approximation, and one only needs to ``fix'' the lepton number conservation constraint that is violated by naive equilibration. A small linear correction, as provided by our approach, provides such an improved estimate.

\subsection{Three flavor generalization}

Now that we have an estimate of the depolarization for two flavors, we seek its generalization to the real-world situation with three flavors. In general, this requires a completely new analysis~\cite{Doring:2019axc,Capozzi:2020kge}. However, if $\mu$ and $\tau$ flavors are taken to behave identically, the treatment is very simple. In such a case, the three flavor oscillations are treated in a restricted manner -- with the $\nu_e$ oscillating to $\nu_\mu$ and $\nu_\tau$, democratically, and the oscillations between $\nu_\mu$ and $\nu_\tau$ being very efficient. Here, one can guess the effective three-flavor depolarization factor from symmetry considerations alone.

\begin{figure}[t]
	\includegraphics[width=0.75\columnwidth]{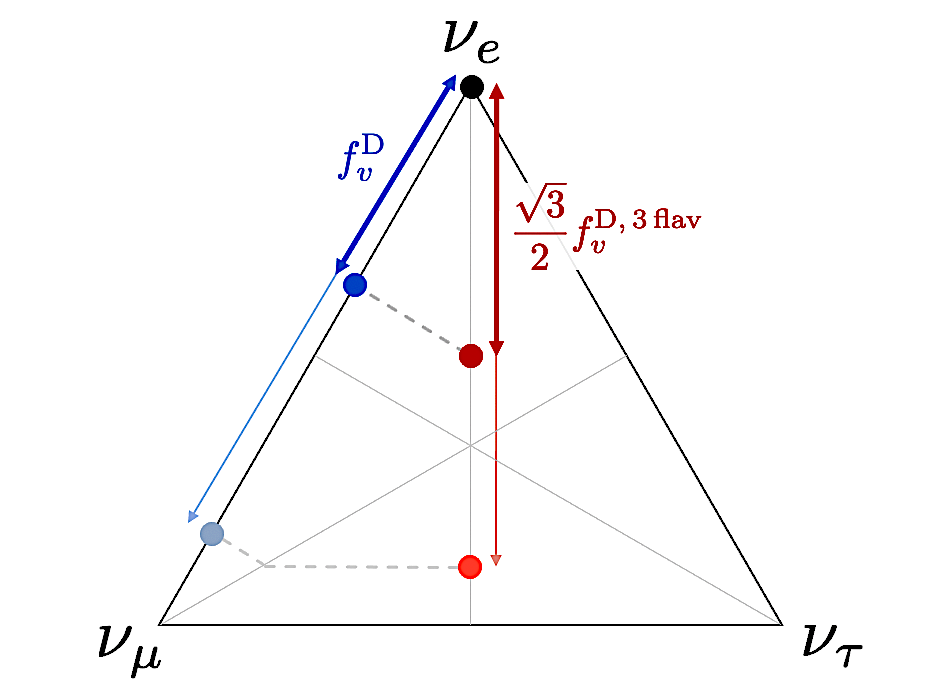}
	\caption{A section of the three-flavor Bloch volume, with the vertices of the unit equilateral triangle corresponding to pure flavor states. The distance of the tip of the Bloch vector (blue dot), away from the top vertex (black dot) along the left edge, is the two-flavor depolarization factor $f^{\rm D}_v$ (blue double arrow). The three-flavor depolarization factor $f^{\rm D,\,3\,flav}_v$, assuming $\mu-\tau$ symmetry, is related to the corresponding distance along the perpendicular bisector (red double arrow, obtained by projecting along the dashed grey lines).}
	\label{fig:3flavfd}
\end{figure}

In Fig.\,\ref{fig:3flavfd}, we show a section of the three-flavor Bloch volume --- the so-called $\hat{\sf e}^{(3)}$-$\hat{\sf e}^{(8)}$ triangle~\cite{Dasgupta:2007ws} --- on which lie the states corresponding to pure flavor states. This region is an equilateral triangle with sides of unit length, with the vertices corresponding to flavor states.  The two-flavor depolarization factor $f^{\rm D}_v$ is the distance of the tip from the top vertex along the left (or right edge). For three flavors, assuming $\mu-\tau$ symmetry, the tip of the Bloch vector lies along the vertical perpendicular bisector.  Note that transverse components of the Bloch vector (i.e., components out of the plane; in analogy to components orthogonal to an edge of the triangle for a two-flavor scenario) get T2-relaxed. The three-flavor depolarization factor $f^{\rm D,\,3\,flav}_v$, to be used in Eq.\eqref{33n}, is then easily recognized as
\begin{align}
f^{\rm D,\,3\,flav}_v=\begin{cases}
\frac{4f^{\rm D}_v}{3} & \text{if ${f^{\rm D}_v}<\frac{1}{2}$}\,,\\
\frac{1+2f^{\rm D}_v}{3} & \text{if ${f^{\rm D}_v}>\frac{1}{2}$}\,,
\end{cases}\label{eq:3flavfd}
\end{align}
in terms of the two-flavor depolarization factor. Note that our analytical estimate of the two-flavor $f^{\rm D}_v$, as in Eq.\,\eqref{29}, stays between $0$ to $1/2$, which corresponds to $f^{\rm D,\,3\,flav}_v$ being in the range $0$ to $2/3$, as one would expect. 

Numerically, one finds  the two-flavor $f^{\rm D}_v$ can sometimes exceed $1/2$. This corresponds to predominant flavor conversion from $\nu_e$ to say $\nu_\mu$, and then partial depolarization. Heres, one expects a similar transition to the third flavor $\nu_\tau$ as well. The combined action projects the Bloch vector as shown by the lighter dashed grey lines. It is easy to see why: if $\nu_e$ almost fully convert to $\nu_\mu$ (while $\nu_\mu$ and $\nu_\tau$ are symmetric), in a three-flavor framework $\nu_e$ has zero survival probability, with equal conversion probability of $1/2$ to both $\nu_\mu$ and $\nu_\tau$.

\section{Prescriptions}\label{Pres}

The takeaway is that we expect depolarization to be the end-state of neutrinos that have undergone fast oscillations. Below, we provide two easily usable set of expressions related to fast oscillated neutrinos. Our intended users are supernova simulators in the first instance, and supernova neutrino phenomenologists for the second.

\subsection{Sub-grid recipe for SN simulations}

In supernova simulations, one computes the neutrino distribution function -- whether in detail or using moments. See, e.g., Refs.~\cite{Glas:2019ijo,Abbar:2018shq,Nagakura:2019sig, DelfanAzari:2019tez,Abbar:2019zoq,Nagakura:2021hyb}. The finite elements for these simulations are about $0.1\,{\rm km}$ in size, and it is inconceivable for the foreseeable future how one could faithfully include fast oscillations occurring on sub-cm scales into these already hugely expensive supernova hydrodynamic calculations.

Our proposal is that one should first identify each `pixel' in the star where fast instabilities can exist. This can be accomplished using a variety of ways, including computationally efficient and increasingly more reliable approximations involving the moments of the neutrino distributions~\mbox{\cite{Dasgupta:2018ulw, Abbar:2020fcl, Nagakura:2021suv}} or simply applying the crossing criterion ~\cite{Morinaga:2021vmc, Dasgupta:2021gfs}. Therein, to obtain an estimate of the effect of flavor oscillations, one should replace the original phase space distributions $F^{\rm ini}$ with the depolarized distributions $F^{\rm depol}$:
\begin{equation}\label{33n}
\begin{split}
F_{v, E}^{\nu_e, \, \rm depol}&=(1-f^{\rm D,\,3\,flav}_v)F_{v, E}^{\nu_e, \, \rm ini}+ f^{\rm D,\,3\,flav}_v\, F_{v, E}^{\nu_{x}, \rm ini}\,, \\
F_{v,E}^{\nu_{x}, \, \rm depol}&=\bigg(1-\frac{1}{2}{f^{\rm D,\,3\,flav}_v}\bigg)F_{v, E}^{\nu_{x}, \rm ini}+  \frac{1}{2}{f^{\rm D,\,3\,flav}_v}F_{v, E}^{\nu_e, \rm ini}\,\\ 
F_{v, E}^{\bar\nu_e, \, \rm depol}&=(1-f^{\rm D,\,3\,flav}_v)F_{v, E}^{\bar\nu_e, \, \rm ini}+ f^{\rm D,\,3\,flav}_v\, F_{v, E}^{\bar\nu_{x}, \rm ini}\,, \\
F_{v,E}^{\bar\nu_{x}, \, \rm depol}&=\bigg(1-\frac{1}{2}{f^{\rm D,\,3\,flav}_v}\bigg)F_{v, E}^{\bar\nu_{x}, \rm ini}+  \frac{1}{2}{f^{\rm D,\,3\,flav}_v}F_{v, E}^{\bar\nu_e, \rm ini}\,, 
\end{split}
\end{equation}
where $x = \mu / \tau$. As fast oscillations are insensitive to neutrino energy $E$, the same $f^{\rm D,\,3\,flav}_v$ applies to neutrinos and antineutrinos.
Note that this does not impose \emph{naive} equalization of all flavors, but a much less extreme mixing consistent with conservation laws. Of course, if perfect depolarization is allowed then the $\nu_e$ distribution becomes $\frac{1}{3}F^{\nu_e,\,{\rm ini}}_v + \frac{2}{3}F^{\nu_{x},\,{\rm ini}}_v$. This is easily recognized as the usual $1:1:1$ mixture of the three flavors.

The main advantage of this sub-grid prescription is that one can avoid performing the expensive fast oscillation calculation, using an analytically pre-computed look-up table instead. Further, it implements a meaningful estimate of the oscillated distributions -- conserving the relevant lepton asymmetry and carrying nontrivial momentum dependence of the degree of depolarization.

\subsection{Depolarized flavor-dependent flux}

To compute the terrestrially observable neutrino fluxes, we need the fluxes at at radius of say about 100 km from the center of the star, where fast oscillations have ceased and one has to then include slower collective effects, MSW transitions, etc. The procedure to include these slower effects are by now well understood. But suppose we only have the undepolarized primary fluxes provided by existing supernova simulations. How can we include an estimate of the depolarization? In general, this is complicated. However, making some symmetry assumptions, a simple estimate is possible.

We assume that the neutrino emission is axially symmetric at each point in the star and that the star is axially symmetric about the axis joining the star and Earth. Thus, the net observable flux from all source regions is simply given by summing over the velocity modes that leave that region in the direction parallel to the axis. The appropriately velocity-weighed depolarization factor is then given by
\begin{equation}\label{3flav2}
\begin{split}
f_0 = \overline{f^{\rm D,\,3\,flav}_v} & = \frac{4}{3} \frac{\int_0^1 {f_v^{\rm D}} dv}{\int_0^1 dv} \\ 
&  = \frac{2}{3}-\frac{7 A}{12 \gamma_0}\,.
\end{split}
\end{equation}
Since we are considering fast oscillations, we approximate the putative neutrino-sphere as an infinite wall. As a result, only the $v >0$ modes can be observed, with $f_v^{\rm D} < 1/2$ always. Note that $\gamma_0$ is the zeroth forward moment of the ELN, cf. Eq.\,\eqref{eq:ELNmom}.  Similarly we can define the $n^{\rm th}$ forward moment of $f_v^{\rm D\,3\,flav}$ in the following way:
\begin{equation}\label{3flav3}
\begin{split}
f_n & = \overline{v^n f^{\rm D,\,3\,flav}_v} \\ 
   & = \frac{4}{3} \left(\frac{1}{2(n+1)}-\frac{A}{4\gamma_0(n+1)}-\frac{3A}{8\gamma_0(n+2)}\right)\,.
\end{split}
\end{equation}
Putting $n = 0$ in Eq.\eqref{3flav3} gives back Eq.\eqref{3flav2}. The total flux per unit energy detected at a distance $r$ from the neutrino sphere of radius $R$ is
\begin{align}\label{36n}
\Phi^{\nu_{\alpha}/\bar{\nu}_{\alpha}}[E, r] \propto \frac{r^2}{R^2}  E^2 \int_{0}^{1} v\,dv \, F^{\nu_{\alpha}/\bar{\nu}_{\alpha}}_{v, E}[r]\,,
\end{align} 
where $\alpha = e, \mu, \tau$. If we consider no oscillation then $F^{\nu_{\alpha}/\bar{\nu}_{\alpha}}_{v, E}[r] = F^{\nu_{\alpha}/\bar{\nu}_{\alpha}, \rm ini}_{v, E}$ and considering only depolarization due to fast oscillations, but neglecting MSW, slow collective or vacuum oscillations, one has $F^{\nu_{\alpha}/\bar{\nu}_{\alpha}}_{v, E}[r] = F^{\nu_{\alpha}/\bar{\nu}_{\alpha}, \rm depol}_{v, E}$. So the ratio of depolarized to the unoscillated flux are given by
\begin{equation}\label{37}
\begin{split}
\frac{\Phi^{\nu_{e}/\bar{\nu}_e}_{\rm dep}[E, r]}{\Phi^{\nu_{e}/\bar{\nu}_e}_{\rm unosc}[E, r]} =  \frac{\int_0^1 v\,dv \, \left(1-f_v^{{\rm D}, \rm 3 \, flav}\right)F_{v, E}^{\nu_e/\bar{\nu}_e, \rm ini}}{\int_0^1 dv \, vF_{v, E}^{\nu_e/\bar{\nu}_e, \rm ini}}+ \\ \frac{ 
\int_0^1 v\,dv	f_v^{{\rm D}, \rm 3 \, flav} F_{v, E}^{\nu_{x}, \rm ini}}{\int_0^1 dv \, vF_{v, E}^{\nu_e/\bar{\nu}_e, \rm ini} }\,,
\end{split}
\end{equation}
\begin{equation}\label{38}
\begin{split}
\frac{\Phi^{{\nu}_{{x}}/\bar{\nu}_x}_{\rm dep}[E, r]}{\Phi^{{\nu}_{x}/\bar{\nu}_x}_{\rm unosc}[E, r]} =  \frac{\int_0^1 v\,dv \, (1-\frac{f_v^{{\rm D}, \rm 3 \, flav}}{2})F_{v, E}^{\nu_{x}/\bar{\nu}_x, \rm ini}}{\int_0^1 v\,dv \, F_{v, E}^{\nu_{x}/\bar{\nu}_x, \rm ini}}+ \\
\frac{\int_0^1 v\,dv \frac{f_v^{{\rm D}, \rm 3 \, flav}}{2} F_{v, E}^{{\nu}_{e}/\bar{\nu}_e, \rm ini} }{\int_0^1 v\,dv \, F_{v, E}^{\nu_{x}/\bar{\nu}_x, \rm ini}}\,.
\end{split}
\end{equation}
For our calculation we assume the following $v$ dependence of $F_{v, E}^{\nu_{\alpha}/\bar{\nu}_{\alpha}, \rm ini}$:
\begin{subequations}\label{41}
\begin{align}
F_{v, E}^{\nu_e, \rm ini} = \sum_n a_n[E] v^n \,, \label{41a} \\
 F_{v, E}^{\bar{\nu}_e, \rm ini} = \sum_n   \bar{a}_n[E] v^n \,, \label{42a} \\
F_{v, E}^{\nu_{x}, \rm ini} = F_{v, E}^{\bar{\nu}_{x}, \rm ini} =  b_n[E] v^n  \,. \label{43a}
\end{align}
\end{subequations}
Using Eqs.\eqref{41a}-\eqref{43a} one can simplify Eqs.\eqref{37}-\eqref{38} as:
\begin{subequations}
	\begin{align}
	\frac{\Phi^{\nu_{e}}_{\rm dep}[E, r]}{\Phi^{\nu_{e}}_{\rm unosc}[E, r]} & = \frac{\sum_n a_n \left( \frac{1}{n+2} - f_{n+1}\right)+b_n f_{n+1}}{\sum_n \frac{a_n}{n+2}} \,, \label{42an}\\
	 \frac{\Phi^{\bar{\nu}_{e}}_{\rm dep}[E, r]}{\Phi^{\bar{\nu}_{e}}_{\rm unosc}[E, r]} & = \frac{\sum_n \bar{a}_n \left( \frac{1}{n+2} - f_{n+1}\right)+b_n f_{n+1}}{\sum_n \frac{\bar{a}_n}{n+2}} \,, \label{42bn}\\
	 \frac{\Phi^{\nu_{x}}_{\rm dep}[E, r]}{\Phi^{\nu_{x}}_{\rm unosc}[E, r]} & = \frac{\sum_n b_n \left( \frac{1}{n+2} - \frac{f_{n+1}}{2}\right)+{a}_n \frac{f_{n+1}}{2}}{\sum_n \frac{b_n}{n+2}} \,, \label{42cn}\\
	 \frac{\Phi^{\bar{\nu}_{x}}_{\rm dep}[E, r]}{\Phi^{\bar{\nu}_{x}}_{\rm unosc}[E, r]} & = \frac{\sum_n b_n \left( \frac{1}{n+2} - \frac{f_{n+1}}{2}\right)+\bar{a}_n \frac{f_{n+1}}{2}}{\sum_n \frac{b_n}{n+2}} \,. \label{42dn}
	\end{align}
\end{subequations}

Thus, knowing the distributions one can compute the coefficients $a_n$, $\bar{a}_n$ and $b_n$,  as well as the moments of the depolarization factor $f_n$. Together, these allow one to compute the depolarized fluxes from the unoscillated fluxes\footnote{Note that  the differential ELN distribution is
\begin{equation}
g_{v, E} = \begin{cases}
 F_{v, E}^{\nu_e, \rm ini}-F_{v, E}^{{\nu}_{x}, \rm ini} , \, E > 0\\
F_{v, E}^{{\nu}_{x}, \rm ini}-F_{v, E}^{\bar{\nu}_e, \rm ini}, \, E<0 
\end{cases}\,.
\end{equation}
Writing $g_{v, E} = \sum_n g_n[E] v^n$, one clearly notices $a_n = g_n+b_n$ and $\bar{a}_n = b_n-g_n$, which can be used to rewrite Eqs.\eqref{42an}-\eqref{42dn} in terms of the ELN and $F^{\nu_x,{\rm ini}}$.}. For multidimensional simulations, one may have more detailed information that allows summing over the momenta appropriately, and the recipe in the previous section is superior in that case. However, if a neutrino phenomenologist wants to approximately readjust the primary fluxes predicted by a supernova simulation to include for potential effects of fast depolarization, the above recipe gives a crude but meaningful estimate.

\section{Summary and Outlook}
\label{Con}
In this paper, we have presented detailed analytical as well as numerical analysis of the late time nonlinear behavior of a dense neutrino gas undergoing fast collective oscillations in the collisionless quantum kinetics approximation. Our study includes time-dependence, but is restricted to one spatial dimension and one nontrivial momentum coordinate that we have taken to be the radial velocity. Unbroken azimuthal symmetry around the radial coordinate is assumed. Under these assumptions, we find the following results:

\begin{enumerate}

\item The evolution of the average flavor content is similar to the motion of a pendulum. However, this pendulum neither preserves its length nor retains its periodic motion, as seen in Fig.\,\ref{fig1lep}. It settles down to a resting point, which is analytically known in terms of the ELN and its moments, cf. Eq.\eqref{26c}, and shown in Fig.\,\ref{fig3mf}.

\item The shrinking of the length of the pendulum and its settling down can be traced to a number of relaxation mechanisms. These fundamentally stem from the quenching of the transverse components of the flavor polarization vectors due to relative dephasing. Such dephasing begins already in the linear regime of flavor growth, as shown in Figs.\,\ref{figT2relax} and \ref{figT2relax2}. However,  the depolarization depends strongly on which velocity modes experience a large transverse Hamiltonian; see Fig.\,\ref{Srelax} and \ref{Srelax2}.

\item In the nonlinear regime, $n$-multipole cascade and $k$-mode mixing lead to spreading of the flavor disturbance in momentum space and position space, respectively, as shown in Fig.\,\ref{figdepol1n}. 

\item The broad results on depolarization and its extent, as well as mixing of velocity multipoles and $k$-modes, are now confirmed by other groups, i.e., Wu et al.~\cite{Wu:2021uvt} and Richers et al.~\cite{Richers:2021nbx,Richers:2021xtf}. The apparent conflicts are resolved, with Wu et al.~\cite{Wu:2021uvt} as shown in Fig.\,\ref{fig10} and with Martin et al.~\cite{Martin:2019gxb,Martin:2021xyl} as shown in Fig.\,\ref{fig9}. The conflicts arose from minor misunderstandings: the former applied our criterion of comparing the Hamiltonian components in a non-standard way (see Fig.\,\ref{fig11}), and the latter didn't show results after spatial averaging at sufficiently late time (see Fig.\,\ref{fig10a}).

\item The flavor content eventually acquires an approximately time-independent character. This is called depolarization. The extent of depolarization is non-uniform over neutrino and antineutrino momentum, as shown in Fig.\,\ref{figdepol1}. In general, it depends on the ELN. This is essentially because the net lepton asymmetry needs to remain conserved.

\item The extent of depolarization, encoded in the depolarization factor, can be predicted -- if the range of fully depolarized modes is assumed. The prediction is based on a series expansion of the final flavor composition, and enforces lepton number conservation. Equation\,\eqref{29} gives an estimate to linear order in $v$, in the two-flavor approximation.

\item The above result is in the two-flavor approximation. Equation\,\eqref{eq:3flavfd} generalizes it to a restricted three-flavor scenario where the initial conditions and evolution of the $\mu$ and $\tau$ flavors are taken to be identical.

\item The depolarized flavor distributions (in Eq.\eqref{33n}) and the depolarized fluxes (in Eqs.\eqref{42an}-\eqref{42dn}) are given in terms of the original distributions (in Eqs.\eqref{41a}-\eqref{43a}) and forward moments $f_n$ of the depolarization factor (in Eq.\eqref{3flav3}). These are approximate but readily usable ingredients for implementation in supernova/nucleosynthesis simulations and for computations of neutrino signals.

\end{enumerate}

Dephasing leads to qualitatively different results than purely coherent evolution. This the fundamental result we hope to have conveyed. Our treatment of depolarization rests on the idea that there is dephasing of many modes.  It is the dephasing assumption that allows going from Eq.\eqref{eom2} to Eq.\eqref{eom5}, allows truncation of the multipole equations, introduces irreversibility, leads to the steady-state solution in Eq.\eqref{26c}, and allows a description of depolarization. While we do not use the truncated or dephased equations for any numerical computations, rather preferring to solve Eq.\eqref{eom2} directly and then averaging the solutions appropriately, the analytical results of the relaxed and truncated multipole equations, e.g., Eq.\eqref{26c}, provide remarkable agreement with the numerical solutions of the full equations at late times.

The reader may see parallels with the ``derivation'' of the Boltzmann equation~\cite{Huang:1987,Kardar:2007}. Hamilton's equations for many interacting particles can be expressed as the BBGKY hierarchy, but there is no way to truncate that hierarchy without \emph{assuming} something more, viz., molecular chaos, coarse graining, etc. These assumptions serve to introduce, \emph{by hand}, the loss of correlation required to explain irreversibility. While the derivation continues to be a matter of discussion, there is no doubt that its end result, i.e., the Boltzmann equation, is extraordinarily useful and describes macroscopic reality much more appropriately than the technically better justified microscopic equations of motion.

We conclude this paper with our outlook for further work on this subject. We believe that an immediate task is to arrive at a better estimate of the range of depolarized modes. Perhaps the answer will lie in devising an improved criterion on the Hamiltonian, or finding the exact depolarization envelop. With that, the problem of computing the depolarized final state of fast oscillating neutrinos would be largely accomplished. It is our belief that this will be important and useful for any practical study accounting for the fast flavor oscillations of neutrinos in supernovae.

\section*{Acknowledgements}
We thank Ian Padilla-Gay, Georg Raffelt, and Irene Tamborra, for  important clarifications regarding their paper, and for helpful suggestions about our manuscript. We also thank H.-T.\,Janka, Meng-Ru Wu and Zewei Xiong for helpful discussions. This work is supported by the Dept. of Atomic Energy (Govt. of India) research project RTI 4002, the Dept. of Science and Technology (Govt. of India) through a Swarnajayanti Fellowship, and by the Max-Planck-Gesellschaft through a Max Planck Partner Group. The numerical computations were performed on the compute clusters, Flock, Leap, Pride, and Raft, at the Dept. of Theoretical Physics (TIFR, Mumbai).

\end{document}